\documentclass[twocolumn,english,aps,prb,reprint, superscriptaddress,showpacs,longbibliography,showkeys, nofootinbib]{revtex4-2}
\usepackage{amsmath,amssymb,bbm,mathrsfs,bm,color,graphicx,comment,xcolor,dsfont,multirow}
\usepackage[colorlinks,citecolor=blue,urlcolor=blue,linkcolor = blue]{hyperref}
\usepackage{enumitem}
\usepackage[mathscr]{euscript}
\usepackage[T1]{fontenc}
\usepackage{siunitx}
\usepackage{placeins}
\usepackage{cancel}
\usepackage{todonotes}

\renewcommand{\vec}[1]{\boldsymbol{#1}}

\begin{document}

\title{Replacement-Type Quantum Gates}

\author{Florian Ginzel}
    \affiliation{Parity Quantum Computing Germany GmbH, Schauenburgerstraße 6, 20095 Hamburg, Germany}

\author{Javad Kazemi}
    \affiliation{Parity Quantum Computing Germany GmbH, Schauenburgerstraße 6, 20095 Hamburg, Germany}

\author{Valentin Torggler}
    \affiliation{Parity Quantum Computing Germany GmbH, Schauenburgerstraße 6, 20095 Hamburg, Germany}

\author{Wolfgang Lechner}
    \affiliation{Parity Quantum Computing Germany GmbH, Schauenburgerstraße 6, 20095 Hamburg, Germany}
    \affiliation{Parity Quantum Computing GmbH, Rennweg 1, Top 314, 6020 Innsbruck, Austria}
    \affiliation{Parity Quantum Computing France SAS, 10 Avenue de Kl{\'e}ber, 75016 Paris, France}
    \affiliation{Institute for Theoretical Physics, University of Innsbruck, 6020 Innsbruck, Austria}

\date{\today}

\begin{abstract}
    We introduce the paradigm of replacement-type quantum gates. This type of gate introduces input qubits, candidate qubits, and output qubits. The candidate qubits are prepared such, that a  displacement conditional on the  input qubit results in the targeted output state. Finally, the circuit continues with the output qubits constructed from the candidate qubits instead of the input qubits, thus the name "replacement-type gate". We present examples of replacement-type $X$ and $\mathrm{CNOT}$ gates realized with spin qubits and with neutral atom qubits with error rates predicted near the threshold of the XZZX surface code. By making use of the extended Hilbert space, including the position of the particles, these gates approximately preserve the innate noise bias of the qubits. The gate preserves the noise bias which motivates advanced quantum computer architectures with quantum error correction.
\end{abstract}

\maketitle

\section{Introduction\label{sec_introduction}}

Quantum computing platforms are at the cusp of advancing beyond laboratory-scale proof-of-principle demonstrations. However, all types of qubits are vulnerable to environmental noise and are therefore error-prone, making quantum error correction (QEC) a necessity before real-world problems can be tackled~\cite{Preskill2018quantumcomputingin,NielsenChuang,Preskill1998}. A large number of generic QEC codes have been proposed, however, the overhead required for successful error correction is widely considered prohibitive for large-scale applications on near-term hardware~\cite{PhysRevA.52.R2493,PhysRevA.54.1098,PhysRevLett.77.793,KITAEV20032,PhysRevA.86.032324,Bravyi2024}.

One strategy for alleviating the requirements for QEC is to rely on specific advantages of the underlying hardware, which is oftentimes described by a less destructive error model than assumed for a general-purpose QEC code~\cite{Jayashankar2022,Wu2025,PhysRevX.13.041013,PhysRevA.109.032433}. Particularly beneficial are biased-noise qubits, where one type of error -- phase errors, modelled by the $\sigma_z$ Pauli matrix, or bit errors, modelled by $\sigma_x$ -- is dominant.

Tailored codes can exploit biased noise to significantly simplify QEC, improve the performance and reduce the requirements~\cite{PhysRevA.78.052331,Wu2025,PhysRevX.9.041053,chamberland2022building,ruiz2025ldpc,putterman2025hardware,PhysRevX.12.021049,PhysRevA.109.032433,PhysRevX.13.041013,Messinger2024}. As particularly impressive example, Clifford-deformed topological codes such as the XZZX surface code and the XYZ color code approach the theoretical maximum threshold error rate of $50\%$ if the noise bias approaches infinity~\cite{BonillaAtaides2021,Roffe2023biastailoredquantum,Tiurev2023correctingnon,PRXQuantum.4.030338}. Intuitively, this can be understood by considering that the vast overhead in QEC serves to protect the quantum state from errors. While a general-purpose code protects against arbitrary errors, knowledge about the noise model may be used to eliminate protection which is not needed and to focus resources on the dominant errors.

An innate noise bias is found in many physical systems, for example semiconductor spin qubits, where the qubit is formed by the spin-$1/2$ of an electron or hole confined to an effectively zero-dimensional quantum dot (QD) by an engineered electrostatic potential~\cite{RevModPhys.95.025003,RevModPhys.79.1217}: Here, the noise is dominated by electrical fluctuations which rarely flip the spin but contribute to dephasing~\cite{PhysRevB.110.235305,Connors2022,PhysRevB.100.165305,RojasArias2024}. Another prominent example is neutral atom qubits, where the electronic states of Rydberg atoms trapped in optical potentials encode quantum information~\cite{ Henriet2020, Morgado2021}. In this case, a highly excited valence electron originating from a certain qubit state mediates strong interaction with nearby atoms, which is then used to engineer entangling gates. Such gates are highly biased towards phase errors since only one of the qubit states is involved in the gate operation under careful engineering of the qubit encoding \cite{PhysRevX.12.021049}. 

The noise bias of these qubit types is not necessarily preserved by gate operations for quantum computing. A phase (bit) error occurring during a rotation of a qubit on the Bloch sphere is partially or entirely translated into a bit (phase) error, resulting in a depolarizing rather than a biased noise. Furthermore, a common decomposition of the $\mathrm{CNOT}_{q_c, q_t}$ gate is $H_{q_t} \mathrm{CZ}_{q_c,q_t} H_{q_t}$ since $\mathrm{CZ}$ is a native gate to many platforms~\cite{Matsumoto2025,Mills2022,Xue2022,Jandura2022,Pagano2022,ion_trap_review,PhysRevX.13.041052}. Here, the Hadamard gates $H_{q_t}$ temporarily swap the $z$- and $x$-bases on the target qubit $q_t$, effectively inverting the noise bias on $q_t$ for the duration of the gate.

Since the commonly noise-biased operations -- (controlled) phase gates, preparation and measurement of certain bases, and idling -- do not provide a universal gate set, composite gates like $\mathrm{CNOT}_{q_c, q_t} = H_{q_t} \mathrm{CZ}_{q_c,q_t} H_{q_t}$ with a more balanced noise model are a necessity~\cite{NielsenChuang}. This challenge is typically tackled by relying on magic state preparation, distillation and injection for performing arbitrary operations~\cite{PhysRevA.78.052331,PhysRevA.92.062309}. The distillation of magic states introduces an additional resource overhead, in particular if multiple distillation rounds are required to reach the target fidelity. It is expected that several hundreds to thousands of raw magic states are required for preparing one high-quality magic state~\cite{PhysRevA.71.022316,PhysRevA.86.032324,Gidney2019efficientmagicstate}. This excessive overhead can be avoided by relying on an intrinsically noise-bias-preserving gate set, which preserves the bias throughout all operations~\cite{PhysRevX.9.041053}.

A noise-bias-preserving gate set with this property is known for bosonic cat qubits \cite{mirrahimi2014dynamically, puri2020bias, chamberland2022building, PhysRevX.9.041053} but remains elusive in general. In fact, a no-go theorem states that a noise-bias-preserving $\mathrm{CNOT}$ gate is impossible in a system with finite Hilbert space~\cite{PhysRevX.9.041053}. Nonetheless, approximately biased-noise gates can be designed if the quantum state is not confined to the Bloch sphere of the qubit's Hilbert space, avoiding the problematic rotations. One example is known for Rydberg atoms, harnessing additional hyperfine states of the atoms~\cite{PhysRevX.12.021049}.

A larger Hilbert space does not necessarily require additional internal states of the qubit. In several quantum computing platforms, the qubits are formed in the internal degrees of freedom of indistinguishable particles confined by an engineered potential, e.g., electrons or holes in QDs and trapped atoms or ions~\cite{ion_trap_review}. These particles can be shuttled -- physically moved to remote sites by displacing the potential minimum of a qubit. Shuttling has gained much attention as a method for mediating interactions between distant qubits and for moving qubits to dedicated readout, manipulation or interaction zones~\cite{Taylor2005,Noiri2022,Kuenne2024,PhysRevB.110.075302,PhysRevX.13.041052,PRXQuantum.4.040313,Lekitsch2017,PhysRevLett.117.220501, Bluvstein2022, Finkelstein2024, Bluvstein2025}.

In this paper, we propose a new paradigm for single- and two-qubit quantum gates which entirely avoid rotations of the qubit on the Bloch sphere. By hybridizing the qubit state with the positional degree of freedom of the particle the Hilbert space is extended beyond the Bloch sphere, allowing an approximately noise-bias-preserving gate. Key element of the gate is a quantum operation which replaces the original qubit with a candidate qubit, conditional on the input state. We present the general framework and discuss possible realizations for both QD spin qubits and Rydberg atom qubits. In both cases our investigation of the error mechanisms shows that the gates can indeed be approximately noise-bias-preserving, and we identify the leading error sources.

This paper is organized as follows. We introduce the general concept of replacement-type gates in Sec.~\ref{sec_general}. Subsequently, in Sec.~\ref{sec_spins_requirements}, we introduce the required primitives for spin-$1/2$ qubits in semiconductor QDs, in Sec.~\ref{sec_spins_implementation} we present concrete examples of replacement-type gates, in Sec.~\ref{sec_spins_errors} we discuss the impact of various error mechanisms on the gates and the noise bias, and in Sec.~\ref{sec_spins_qpt} we estimate the performance. Analogously, a realization with neutral atom qubits is presented in Sec.~\ref{sec_atoms}. In Sec.~\ref{sec_practice} we investigate possible use cases for replacement-type gates in QEC and formulate requirements to achieve an advantage over standard gates. Finally, we summarize our findings and give an outlook in Sec.~\ref{sec_conclusion}.

\section{Replacement-Type Gates\label{sec_general}}

Here, we present the abstract concept underlying our proposed replacement-type quantum gate. The core idea is to directly exploit the positional degree of freedom of a mobile qubit to implement a quantum gate, following the protocol outlined below. We consider a register comprised of multiple sites, where the position of a qubit and the the occupation number of a site are quantum mechanical observables.

\begin{figure}
    \centering
    \includegraphics[width=0.75\columnwidth]{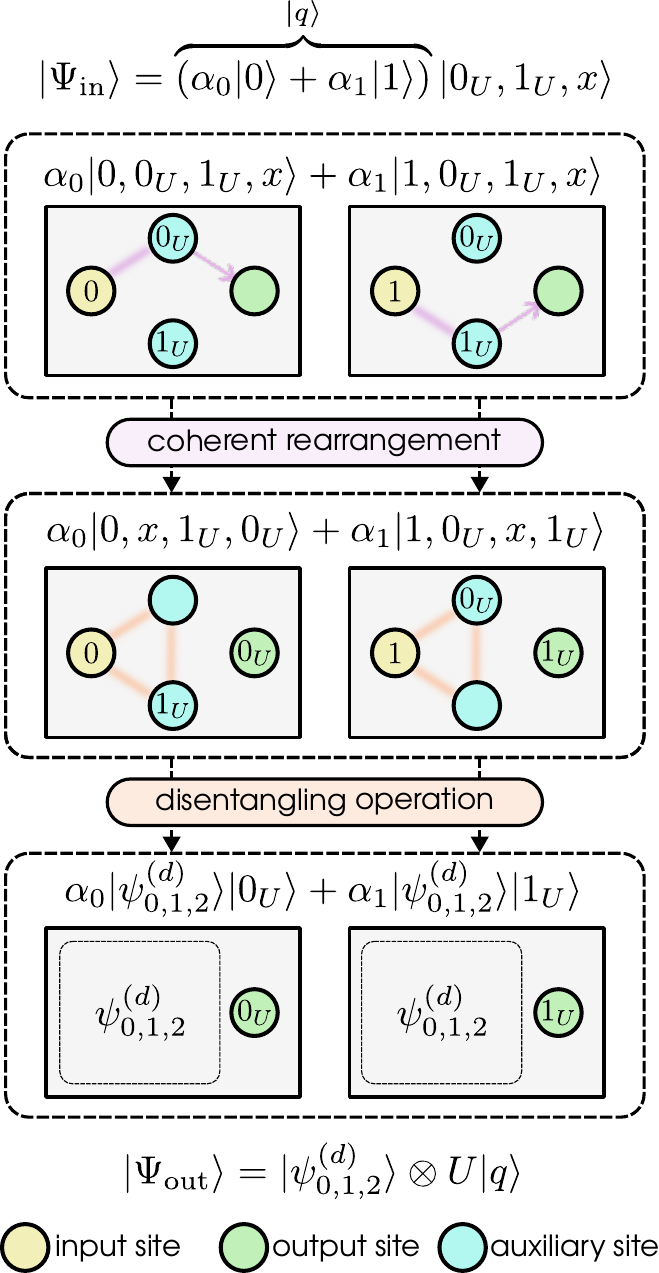}
    \caption{Concept of the replacement-type single-qubit gate $U$. An input qubit has the general qubit state $|q \rangle = \alpha_0 |0 \rangle + \alpha_1 |1 \rangle$ and is placed at the yellow input site. Two candidate qubits are initialized in suitable basis states of the Hilbert space, here in the states $|0_U (1_U) \rangle = U|0(1) \rangle$, and are placed at the blue auxiliary sites, the green output site is initially empty, $|x\rangle$. Then, the register is rearranged coherently in order to transfer one of the candidate qubits into the output site, conditional on the state of the input qubit. This leads to an entangled state of all four qubits and the configuration of the positions. The amplitude $\alpha_0$ ($\alpha_1$) is now associated with state $|0_U\rangle$ ($|1_U \rangle$) at the output site, imposing the effect of the gate U. Disentangling the population of the output site from the rest of the register with suitable measurements leaves a single qubit in the final separable state $|\psi_{0,1,2}^{(d)}\rangle \otimes U|q \rangle$, corresponding to a gate $U$ and a relocation of the qubit state.}
    \label{fig_general}
\end{figure}

To implement a quantum operation $U$ on one or multiple input qubits as replacement-type gate, auxiliary qubits need to be initialized in the states $U|\psi_i\rangle$, where the set of all $|\psi_i\rangle$ forms a basis of the Hilbert space $U$ is acting upon. We refer to these auxiliary qubits as the candidate qubits, they are designated to replace the input qubits at the end of the gate. This is achieved through a suitable rearrangement operation, which coherently moves one candidate qubit per input qubit to a designated output site, conditional on the state of the input qubits. The entangled state of input and candidate qubits created by the rearrangement operation can then be disentangled by an appropriate disentangling operation, which discards all qubits except the population of the output site. The remaining output qubits are kept instead of the input qubits. The procedure is illustrated in Fig.~\ref{fig_general}.

First, we consider a single input qubit in the state $|\psi_\mathrm{in}\rangle = \alpha_0 |0\rangle + \alpha_1 |1\rangle$, to which a quantum gate $U$ should be applied. The basis states are transformed by $U$ as $U|0(1)\rangle = |0_U (1_U)\rangle$ and we assume that it is possible to directly initialize qubits in both states $|0_U\rangle$ and $|1_U\rangle$. This may not be possible for arbitrary states, such that only discrete gates are allowed.

In the first step of the replacement-type gate, we introduce a register with three sites occupied by two candidate qubits in the state $|0_U, 1_U, x\rangle$, where $x$ indicates an empty site. The empty site serves as the output site, which will eventually be populated with a qubit in the targeted state $U|\psi_\mathrm{in}\rangle$. Note that we chose the output site to be an additional site for clarity here, a suitable rearrangement operation could also be configured such that one of the initially populated sites is chosen as output site. The total initial state is
\begin{align}
    |\Psi_\mathrm{in}\rangle = \Big(\alpha_0 |0\rangle + \alpha_1 |1\rangle\Big) |0_U, 1_U, x \rangle.
\end{align}

In the next step, the qubits are rearranged by a unitary operation that maps
\begin{align}
    |0\rangle|0_U, 1_U, x \rangle &\to |0\rangle|x, 1_U, 0_U \rangle,\\
    |1\rangle|0_U, 1_U, x \rangle &\to |1\rangle|0_U, x, 1_U \rangle.
\end{align}
The total state is thus transformed into an intermediate, entangled state
\begin{align}
    |\Psi_\mathrm{mid}\rangle = \alpha_0 |0\rangle|x, 1_U, 0_U \rangle + \alpha_1 |1\rangle|0_U, x, 1_U \rangle.\label{eq_general_state_after_X}
\end{align}
The nature of this operation and the relevant interactions depend on the hardware platform and may not be possible for arbitrary $U$. 

Following the rearrangement operation, the state of the qubit is encoded in a multi-particle state, Eq.~(\ref{eq_general_state_after_X}), where the probability distribution for a measurement of the qubit state at the output site corresponds to a single qubit in the state $U|\psi_\mathrm{in}\rangle$. However, if the qubit is not immediately measured following the gate, it is beneficial to disentangle the qubits again, since a hybrid state of multiple qubits and their position may be more vulnerable to environmental noise than the state of a single qubit.

Such a disentangling operation can be achieved through measurements performed in an appropriate basis. Introducing the superposition states $|\pm\rangle = (|0\rangle \pm |1\rangle)/ \sqrt 2$, the total state can be written as
\begin{align}
    % &\frac{\alpha_0}{\sqrt 2} \left(|+\rangle + |-\rangle\right) |x, 1_U, 0_U \rangle + \frac{\alpha_1}{\sqrt 2} \left(|+\rangle - |-\rangle\right) |1\rangle|0_U, x, 1_U \rangle = \\
    |\Psi_\mathrm{mid}\rangle = & \frac{1}{\sqrt 2} |+\rangle \Big(\alpha_0 |x, 1_U, 0_U \rangle + \alpha_1 |0_U, x, 1_U \rangle \Big) \nonumber\\
    & + \frac{1}{\sqrt 2} |-\rangle \Big(\alpha_0 |x, 1_U, 0_U \rangle - \alpha_1 |0_U, x, 1_U \rangle \Big).
\end{align}

Measuring the state of the original input qubit in the basis $\{|+\rangle, |-\rangle\}$, results in the post-measurement state
\begin{align} 
    \frac{1}{\sqrt 2} \Big( |0\rangle + s |1\rangle \Big) \Big( \alpha_0 |x, 1_U, 0_U \rangle + s \alpha_1 |0_U, x, 1_U \rangle \Big),
\end{align}
where the relative phase $s=(-1)^m$ is conditional on the measurement outcome $m$. Proceeding analogously with a measurement of the hybrid states $(|x, 1_U\rangle \pm |0_U, x\rangle)/ \sqrt 2$ at the initial positions of the candidate qubits results in a separable state. We omit all qubits except the one remaining in the output site, which has the final state
\begin{align}
    |\psi_\mathrm{out}\rangle = \alpha_0 |0_U \rangle \pm \alpha_1 |1_U \rangle = U Z^c |\psi_\mathrm{in}\rangle,
\end{align}
with $Z = |0\rangle\langle 0| - |1\rangle\langle 1|$. Here, $c$ is a classical bit obtained from the two measurement outcomes. Note that performing two separate measurements is not always necessary; if experimental constraints permit, a single measurement can be advantageous in mitigating measurement errors.

The final state is the targeted state $U |\psi_\mathrm{in}\rangle$ up to a phase flip that can be tracked by storing the measurement outcomes. Note that the disentangling operation does not require interaction between the auxiliary qubits and the output qubit, allowing the output qubit to be shuttled to a new location during the measurement. A concrete example for a disentangling operation is discussed in Sec~\ref{sec_spins_disentangle}.

The generalization to multi-qubit gates is straightforward. For an operation $U$ targeting $N$ qubits, the rearrangement operations need to produce $2^N$ configurations of the candidate qubits, corresponding to the action of $U$ on a suitable basis of the multi-qubit Hilbert space. The main challenge of implementing the replacement-type gate lies in designing a feasible rearrangement operation. In the following chapters, we discuss examples based on two different hardware platforms.

We note that any errors occurring during the disentangling operation result in dephasing of the output qubit, regardless of the protocol: The objective of the disentangling operation is to transform a highly entangled state into a separable state,
\begin{align}
    & |\Psi_\mathrm{mid}\rangle = \alpha_0 |\psi_{0,1,2}^{(0)} \rangle | 0_U\rangle + \alpha_1 |\psi_{0,1,2}^{(1)} \rangle | 1_U\rangle \nonumber\\
    & \to |\Psi_\mathrm{out}\rangle = |\psi_{0,1,2}^{(d)} \rangle \left( \alpha_0 | 0_U\rangle + \alpha_1 | 1_U\rangle \right),
\end{align}
where $|\psi_{0,1,2}^{(0/1)} \rangle$ is the state of sites 0, 1 and 2 after the rearrangement operation, labelling them from left to right, and $|\psi_{0,1,2}^{(d)} \rangle$ is their state after perfect disentangling. An error with probability $p_d$ will result in the density matrix
\begin{align}
    \rho_\mathrm{out} = & (1-p_d) |\psi_{0,1,2}^{(d)} \rangle \left( \alpha_0 | 0_U\rangle + \alpha_1 | 1_U\rangle \right)(\mathrm{h.c.}) \nonumber \\
    & + p_d \left(\alpha_0 |e_{0,1,2}^{(0)} \rangle | 0_U\rangle + \alpha_1 |e_{0,1,2}^{(1)} \rangle | 1_U\rangle \right) (\mathrm{h.c.}),
\end{align}
with the erroneous states $|e_{0,1,2}^{(0/1)}\rangle$, $\mathrm{h.c.}$ denotes the hermitian conjugate.

The subsequent discarding or resetting of the auxiliary qubits can be modelled as relaxation to the vacuum or ground state in these sites with unit probability. This leaves the output qubit dephased with probability $p_d$:
\begin{align}
    \rho_\mathrm{out} \to & (1-p_d) \left( \alpha_0 | 0_U\rangle + \alpha_1 | 1_U\rangle \right)\left( \alpha_0^* \langle 0_U | + \alpha_1^* \langle 1_U | \right)\nonumber \\
    & + p_d \left(|\alpha_0|^2 | 0_U\rangle \langle 0_U| + |\alpha_1|^2 | 1_U\rangle \langle 1_U| \right).
\end{align}
Similarly, measurement errors will result in the wrong phase correction being applied to the output qubit. The observation that during the disentangling operation bit errors are translated into phase errors but not vice versa is an important result for considering the noise bias of the total gate.

\section{Realization with Spin Qubits\label{sec_spins}}

\begin{figure}
    \centering
    \includegraphics[width=0.9\columnwidth]{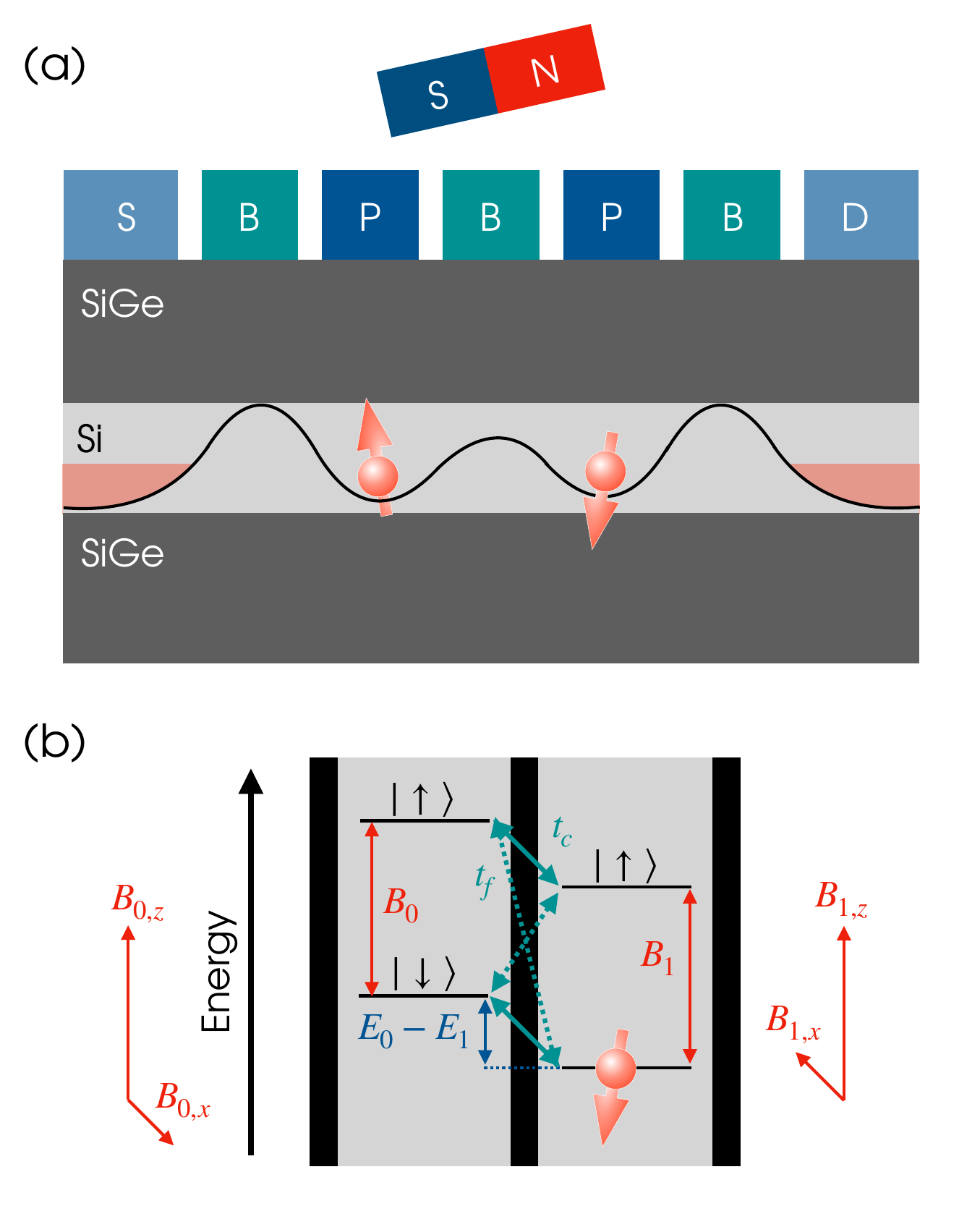}
    \caption{(a) Sketch of two QDs in a Si/SiGe heterostructure. The electrical potential applied to plunger (P) and barrier (B) gates defines potential wells, which can be filled with single electrons or holes by coupling them to source (S) and drain (D) reservoirs. A strong magnetic field inhomogeneity, also referred to as synthetic spin-orbit interaction, is commonly created by a micromagnet.
    (b) Energy levels of two QDs, left ($0$) and right ($1$) occupied by a single spin. In each QD, two spin states $|\!\downarrow\rangle$ and $|\!\uparrow\rangle$ are available, with a splitting corresponding to the total magnetic field $B_{0(1)}$ (red). A transverse magnetic gradient $B_{0,x} - B_{1,x} \neq 0$ leads to non-collinear quantization axes. %, and a longitudinal magnetic gradient $B_{0,z} - B_{1,z}$ results in a different Zeeman splitting.
    The energy detuning between the ground states, $E_0-E_1$ (blue), is set by the electrical potentials applied to the plunger gates. The voltage applied to the barrier gates tunes the tunneling between the QDs (teal) with a spin-conserving ($t_c$) and spin-flipping ($t_f$) matrix element.}
    \label{fig_spins_sketch}
\end{figure}

In this section, we consider qubits encoded in the spins of confined electrons~\cite{PhysRevA.57.120}. The confinement is achieved by creating a two-dimensional electron gas at the interface of a semiconductor heterostructure, such that the conduction band offset provides out-of-plane confinement. In-plane confinement is then achieved by electrostatic gating~\cite{RevModPhys.79.1217}. A sketch of such QDs is depicted in Fig.~\ref{fig_spins_sketch}.

We consider an array of $N$ QDs where only the orbital ground state is available in the few-electron regime. For simplicity, we neglect the valley degree of freedom, assuming a large valley splitting or a material where the conduction band minimum is non-degenerate~\cite{RevModPhys.95.025003,RevModPhys.85.961}. The system can be modelled with an extended Hubbard Hamiltonian~\cite{RevModPhys.95.025003}, which describes all unitary operations in the replacement-type gate,
\begin{align}
    H_\mathrm{spin} =& \sum_j^N \left( E_j n_j + \vec{B}_j \cdot \vec{\sigma}_j + \frac{U_2}{2} n_j (n_j-1) \right) \nonumber\\
    & + \sum_j^{N-1} \sum_{\sigma,\sigma'} \left(t_{j,\sigma,\sigma'} \left( c_j^\sigma \right)^\dagger c_{j+1}^{\sigma'} + \mathrm{h.c.} \right) \nonumber\\
    & + \sum_j^{N-1} U_1 n_j n_{j+1}.\label{eq_spins_hamiltonian}
\end{align}

Here, $\left(c_j^\sigma\right)^{(\dagger)}$ annihilates (creates) an electron with spin $\sigma$ in QD $j$, $n_j = \sum_\sigma \left(c_j^\sigma \right)^\dagger c_j^\sigma$ is the occupation number operator, and $\vec{\sigma}_j$ is the vector of Pauli operators in QD $j$. The first line of Eq.~(\ref{eq_spins_hamiltonian}) includes the electric on-site potential $E_j$, the magnetic field $\vec{B}_j$ (given in energy units) in QD $j$, and the Coulomb repulsion $U_2$ between two electrons in the same QD. An inhomogeneity $\vec{B}_i\neq \vec{B}_j$ may arise from synthetic or intrinsic spin-orbit interaction (SOI). The second line of Eq.~(\ref{eq_spins_hamiltonian}) describes the spin-conserving and spin-flipping tunneling between QDs $j$ and $j+1$ with matrix element $t_{j,\sigma,\sigma'}$, where $\mathrm{h.c.}$ denotes the Hermitian conjugate. The last line finally adds the Coulomb repulsion between the charges in neighboring QDs. Among these parameters, $E_j$ and $t_{j,\sigma,\sigma'}$ are electrically tunable during the experiment. The eigenstates of $H_\mathrm{spin}$ in the example $N=2$ with up to two electrons are illustrated in Fig.~\ref{fig_spins_spectrum}.

In the following, we discuss the physical mechanisms leveraged by the replacement-type gate in Sec.~\ref{sec_spins_requirements}, then we present a realization of $X$ and $\mathrm{CNOT}$ gates in Sec.~\ref{sec_spins_implementation}. Finally, in Sec.~\ref{sec_spins_errors}, we discuss the dominant error mechanisms and ins Sec.~\ref{sec_spins_qpt} we simulate quantum process tomography to investigate the noise bias and to estimate the performance of the gates.

\subsection{Required Operations\label{sec_spins_requirements}}

With Loss--DiVincenzo qubits~\cite{PhysRevA.57.120}, defined by the spin of a single electron, the rearrangement operation can be constructed from well-known primitives. Typically, single-spin rotations in QD $j$ are implemented by effectively realizing an oscillating magnetic field by periodically displacing the electron in an inhomogeneous magnetic field, known as electric dipole spin resonance (EDSR)~\cite{Pioro-Ladriere2008,PhysRevB.74.165319} or by electron spin resonance (ESR) with an oscillating transverse magnetic field $B_{j,x}(t)$~\cite{Koppens2006}. The kinetic exchange is harnessed for coupling spins in adjacent QDs~\cite{Petta2005}, depending on the longitudinal magnetic gradient this is an Ising- or Heisenberg-type interaction% which can be derived from Eq.~(\ref{eq_spins_hamiltonian}) by removing the doubly-occupied states with a Schrieffer--Wolff transformation
~\cite{RevModPhys.95.025003}.

\begin{figure}
    \centering
    \includegraphics[width=0.9\columnwidth]{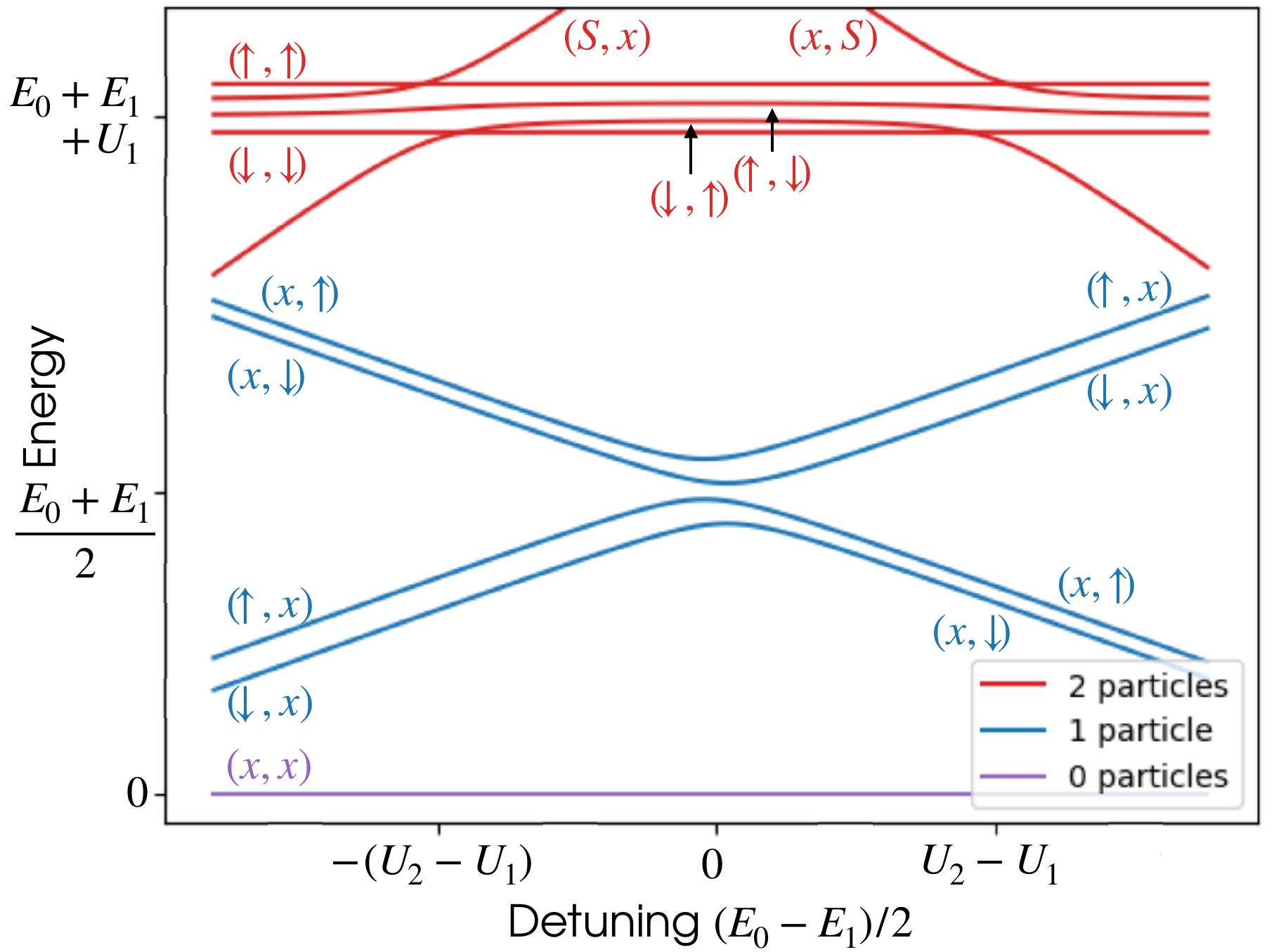}
    \caption{Spectrum of a double QD in an inhomogeneous magnetic field occupied with no (purple), one (blue) or two (red) electrons as a function of the energy detuning. We label the states as $(\sigma_0 , \sigma_1)$ with the spin projection $\sigma_{0(1)}$ in QD $0$($1$), $x$ indicates an empty QD. With a single electron, avoided level crossings (ALCs) are opened at zero detuning by the tunneling matrix elements, far from the ALCs spin and position of the electron are good quantum numbers. The asymmetry is due to the longitudinal magnetic gradient. Adding a second particle leads to an offset in energy due to Coulomb repulsion, the two electrons can occupy a single QD only if their spins form a singlet ($S$). Due to the longitudinal magnetic field gradient the singlet and triplet states decompose into the product states in the regime with one electron per QD.}
    \label{fig_spins_spectrum}
\end{figure}

The replacement-type gate discussed in the following section represents a paradigm shift by leveraging different physical mechanisms. A key element is the conversion of a single-electron spin state into a charge state using Pauli spin blockade (PSB)~\cite{Petta2005,PhysRevLett.103.160503,RevModPhys.79.1217}. Considering two QDs occupied by one electron each, the spin states form a singlet state $|S_{(1,1)}\rangle = \frac{1}{\sqrt{2}} \left[ (c_0^\uparrow)^\dagger (c_1^\downarrow)^\dagger - (c_0^\downarrow)^\dagger (c_1^\uparrow)^\dagger \right] |x\rangle$ and the three triplet states $|T_0\rangle = \frac{1}{\sqrt{2}} \left[ (c_0^\uparrow)^\dagger (c_1^\downarrow)^\dagger + (c_0^\downarrow)^\dagger (c_1^\uparrow)^\dagger \right] |x\rangle$ and $|T_\pm\rangle = (c_0^{\uparrow (\downarrow)} )^\dagger ( c_1^{\uparrow(\downarrow)})^\dagger |x\rangle$, where $|x\rangle$ denotes an empty QD. Due to the Pauli principle, only the singlet state is allowed when both electrons occupy the same QD, $|S_j\rangle = (c_j^{\uparrow} )^\dagger ( c_j^{\downarrow})^\dagger |x\rangle$. Therefore, the triplet states are coupled to the states with double-occupation only via the typically weak spin-flip tunneling.

In the presence of a longitudinal magnetic field gradient $B_{0,z}\neq B_{1,z}$, the states $|S_{(1,1)}\rangle$ and $|T_0\rangle$ are no eigenstates anymore and rearrange themselves into the product states $(c_0^{\uparrow} )^\dagger ( c_1^{\downarrow})^\dagger |x\rangle$ and $(c_0^{\downarrow} )^\dagger ( c_1^{\uparrow})^\dagger |x\rangle$, one of which is coupled to the $|S_j\rangle$ singlet while the other one remains decoupled, as depicted in Fig.~\ref{fig_spins_spectrum}~\cite{RevModPhys.95.025003,RevModPhys.79.1217}. Throughout this section we assume a gradient such that $(c_0^{\downarrow} )^\dagger ( c_1^{\uparrow})^\dagger |x\rangle$ is coupled to $|S_0\rangle$

The inverse process of PSB, splitting a $|S_j\rangle$ singlet into two QDs in the presence of a longitudinal magnetic field gradient, allows for the preparation of a product state of two spins. This technique is commonly used for spin qubit initialization.  The $|S_j\rangle$ singlet can be prepared by filling the QD with two electrons from a reservoir.

Spin-to-charge conversion can in principle also be achieved with a single electron, by precisely driving Landau--Zener transitions which emerge due to the SOI~\cite{PhysRevB.102.195418}. However, since this option is expected to be challenging in practice, we do not consider it further.

We further require the use of different shuttling techniques for manipulating the charge distribution~\cite{PRXQuantum.4.020305}: Conveyor mode shuttling where the electrostatic potential $E_j(t)$ at all sites is manipulated such that the entire potential landscape is uniformly displaced~\cite{Taylor2005,Kuenne2024}, and bucket brigade shuttling. In bucket brigade shuttling, the detuning between two neighboring QDs, $E_j(t) - E_{j+1}(t)$, is swept though the avoided level crossing (ALC) such that a single electron tunnels from one QD to the next, which allows coherent spin transfer over short distances~\cite{PhysRevB.102.125406,PhysRevB.102.195418}. In Fig.~\ref{fig_spins_spectrum}, bucket brigade shuttling corresponds to an adiabatic sweep through the ALCs of the blue single-particle states.

It has been shown that the Coulomb repulsion between neighboring QDs, $U_1$, can facilitate the displacement of a spin qubit conditional on the charge distribution of its surroundings~\cite{vanDiepen2021}. Here, we envision an energy-selective bucket brigade tunneling, where the detuning $\varepsilon (t) = E_j(t) - E_{j+1}(t)$ between QDs $j$ and $j+1$ is swept through the ALCs in the single-electron regime such that tunneling occurs if the target QD is empty but no tunneling occurs if the target QD is occupied. As can be seen in Fig.~\ref{fig_spins_spectrum}, this form of energy-selective tunneling is realistic if $U_2 - U_1$ is sufficiently large compared to the tunneling matrix elements and the ALCs with one and two particles are well separated along the detuning axis. For a high-fidelity charge transition, the detuning at the final time $t_f$ must be sufficiently far from the single-particle ALC, $|\varepsilon (t_f)| \gg 2|t_{1,\sigma,\sigma}|$, furthermore, for the charging energy to suppress movement in the case of two electrons the final detuning must be $|\varepsilon (t_f)| \ll U_2 - U_1$.

Finally, a measurement of the charge distribution will be required for the disentangling operation. A rapid and accurate measurement of the total occupation of a single QD is routinely obtained by means of a radio-frequency single electron transistor in the vicinity~\cite{Fattal2025,Connors2022,PhysRevApplied.13.024019,Takeda2016}. Another method for detecting changes in the charge occupation is dispersive gate sensing, where the reflection of a microwave off one of the voltage gates of the QDs directly is monitored~\cite{PhysRevLett.110.046805}. We emphasize that during the replacement-type gate no charge sensing must occur, otherwise the unintended (non-selective) measurement will lead to gate errors.

\subsection{Implementations of Replacement-Type Gates\label{sec_spins_implementation}}

In this section we present concrete examples of replacement-type gates: The rearrangement operation for an $X$-gate in Sec.~\ref{sec_spins_X}, a related disentangling operation in Sec.~\ref{sec_spins_disentangle}, and a $\mathrm{CNOT}$ gate in Sec.~\ref{sec_spins_CNOT}.

\subsubsection{Single-Qubit $X$ Gate\label{sec_spins_X}}

\begin{figure}
    \centering
    \includegraphics[width=\columnwidth]{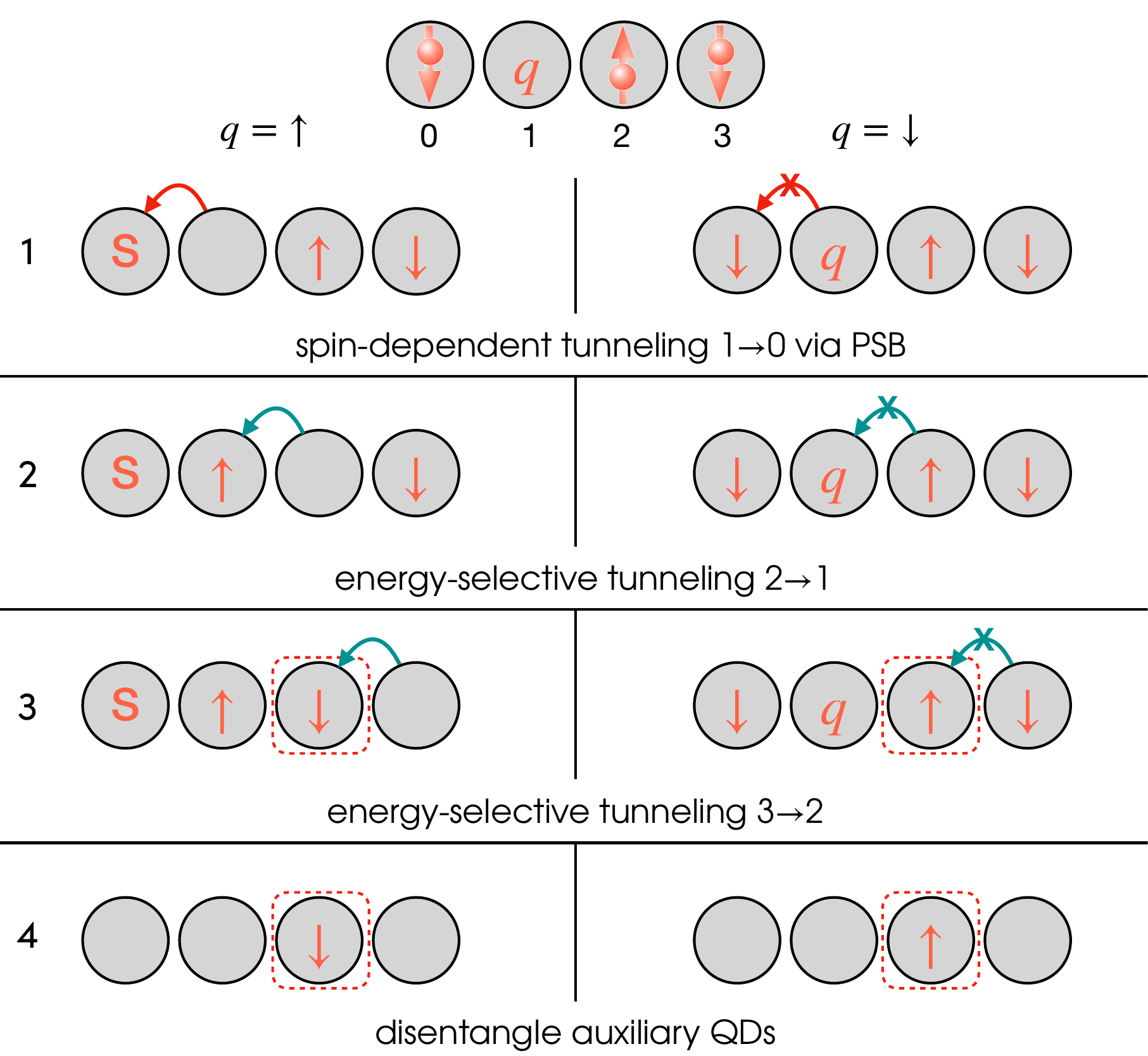}
    \caption{Protocol for a replacement-type $X$ gate with spin qubits. The initial configuration of the array of four QDs is depicted in the top, $q$ indicates the input qubit, $S$ the two-qubit singlet state. Spin-dependent (red arrows) and energy-selective (teal arrows) tunneling are used for manipulating the register, lines 1-4 show the configuration after each step of the protocol. The two columns correspond to the classically distinguishable qubit states $q= \uparrow$ (right) and $q=\downarrow$ (left), in general, a superposition of the columns is obtained. A crossed-out arrow indicates suppressed tunneling. The final output qubit is found at the output site QD 2, highlighted in red. Step 4 is explained in more detail in Sec.~\ref{sec_spins_disentangle} and Fig.~\ref{fig_spins_disentangle}.}
    \label{fig_spins_X}
\end{figure}

The design of the single-qubit gate follows the principles outlined in Sec.~\ref{sec_general} and the protocol is illustrated in Fig.~\ref{fig_spins_X}. A linear array of four QDs labeled 0 to 3 is required, where a specific arrangement of input and candidate qubits is prepared, using shuttling. We place the input qubit $q$ in QD 1 and the two candidate qubits are placed in QDs 2 and 3 in the the states $|\!\uparrow_2\rangle$ and $|\!\downarrow_3\rangle$. Furthermore, a reference spin $|\!\downarrow_0\rangle$ in QD 0 is required for PSB.

In the first step, spin-to-charge conversion via PSB is performed on QDs 0 and 1, such that the reference spin and the input qubit form a singlet state $|S_0\rangle$ in QD 0 if the input qubit was in state $|\!\uparrow_1\rangle$ and movement is suppressed if the input qubit was in state $|\!\downarrow_1\rangle$. In the second step, the energy-selective tunneling introduced above is peformed for the transition from QD 2 to QD 1, then, in the third step, the energy-selective tunneling is repeated for the transition from QD 3 to QD 2.

QD 2 is declared the output site. If the input qubit was in state $|\!\downarrow_1\rangle$, all movements were suppressed and the original candidate qubit in state $|\!\uparrow_2\rangle$ is still found in QD 2. However, if the input qubit was in state ${|\!\uparrow_1\rangle}$, both candidate qubits are shifted to the left by one site and QD 2 is populated by an electron in state $|\!\downarrow_2\rangle$. In general, the input qubit is in a superposition state, therefore a corresponding superposition of the final configurations is obtained. Finally, the population of QDs 0, 1, and 3 is discarded as described in Sec.~\ref{sec_spins_disentangle} and the output qubit replaces the input qubit.

During every step of the operation, the basis states acquire a phase shift, the precise value depends on the microscopic parameters of the Hamiltonian and the protocol for each step. If uncompensated, the total operation has the effect of a gate sequence $R_z(\phi) X$, where $R_z(\phi)$ is a rotation around the $z$ axis of the Bloch sphere by an angle $\phi$, removing the phases with an $R_z(-\phi)$ rotation results in an $X$ gate. We note that only baseband controls are required for the implementation of rearrangement operation, which is beneficial to mitigate adverse heating effects~\cite{PhysRevX.13.041015,Unseld2024}.

\subsubsection{Disentangling Operation\label{sec_spins_disentangle}}

\begin{figure}
    \centering
    \includegraphics[width=\columnwidth]{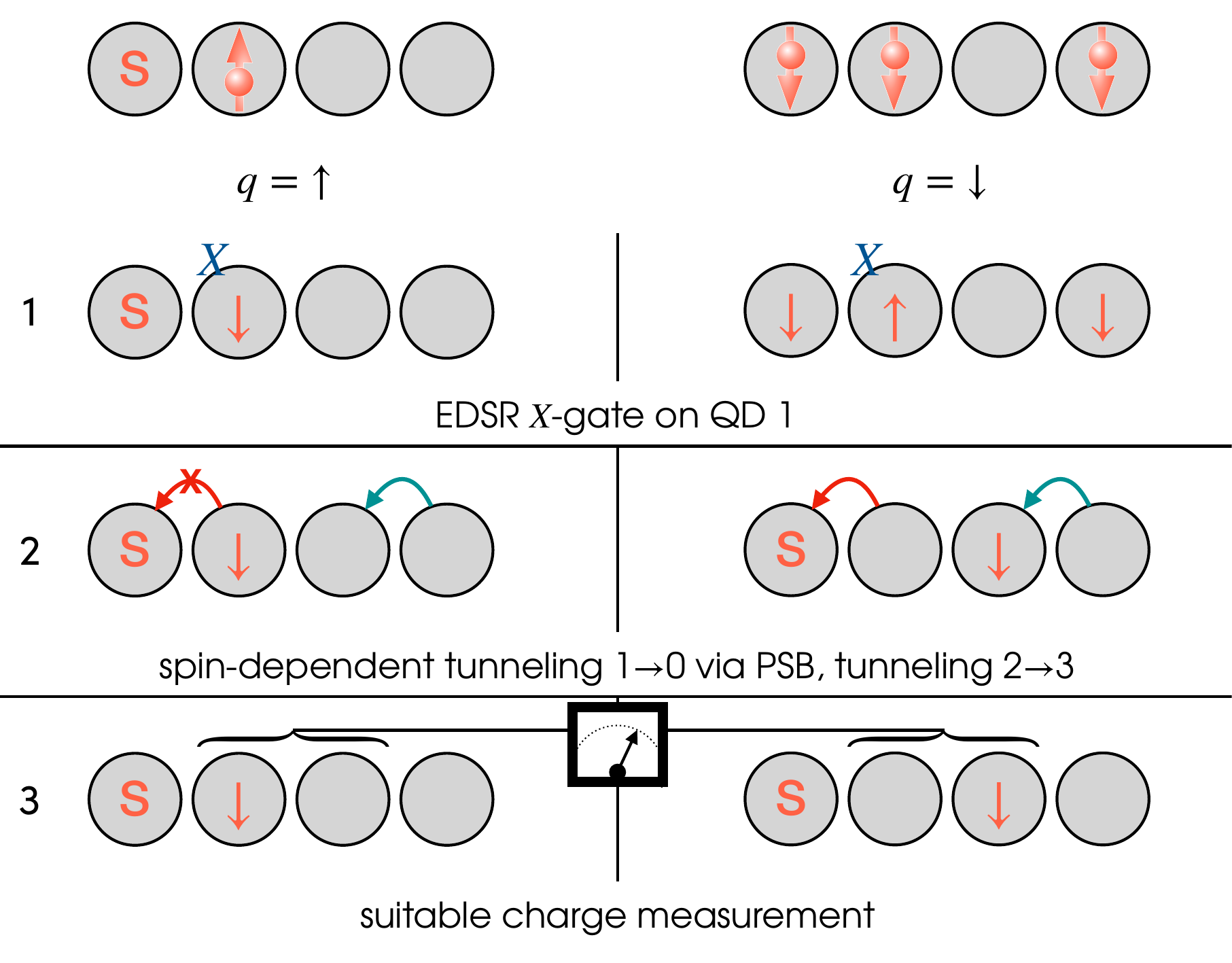}
    \caption{Example of a disentangling operation for spin qubits following an $X$ gate. The top row shows the final configuration of the QD array after the rearrangement from Fig.~\ref{fig_spins_X} for both input states, with the output qubit from QD 2 shuttled away. Steps 1 and 2 prepare a configuration where a single measurement is sufficient for the disentangling, the figure shows the configuration after each step. In step 3, the charge of QDs 1 and 2 is measured in the basis defined by $\left( |1_1\rangle |x_2\rangle \pm |x_1\rangle |1_2\rangle \right)/\sqrt{2}$, where $1(x)$ indicates a QD occupied by one (zero) electrons. A measurement of the total charge occupation in QD 1 or 2 would be equivalent to measuring the output qubit.}
    \label{fig_spins_disentangle}
\end{figure}

As conclusion of the $X$ gate presented in the previous section, the additional electrons need to be disentangled from the output qubit, otherwise enhanced dephasing is expected due to the vulnerability of the charge state against charge noise. One example of a disentangling operation is outlined in Fig.~\ref{fig_spins_disentangle}. The output qubit is not required for the disentangling operation, therefore we assume it is shuttled to a new destination while the disentangling commences, freeing up QD 2.

In the first step, the spin from QD 1 is flipped by an $X$ gate performed with EDSR, which may be exploited to apply the phase compensation as virtual $R_z(-\phi)$ rotation. Subsequently, another PSB operation in QDs 0 and 1 is executed, such that a singlet is formed in QD 0 also in the case of $q = \downarrow$ as input state, while the energy offset of the three charges in the case $q = \uparrow$ suppresses all movement. Simultaneously to the PSB, the population from QD 3 is shuttled to QD 2, both bucket brigade or conveyor mode shuttling are viable options. Now, the two configurations differ only by the position of a single charge.

As discussed in Sec.~\ref{sec_general}, a charge measurement in the basis $\{\frac{1}{\sqrt{2}}\left(|1_1\rangle |x_2\rangle + |x_1\rangle |1_2\rangle \right), \frac{1}{\sqrt{2}}\left(|1_1\rangle |x_2\rangle - |x_1\rangle |1_2\rangle \right)\}$, where $1(x)$ indicates a QD occupied by one (zero) electron, is sufficient for disentangling the auxiliary qubits from the output qubit. This can be accomplished by adiabatically tuning the two QDs to resonance and measuring the population in one of the dots. The auxiliary qubits can then be discarded into a reservoir or reset. Due to the accumulation of phases during the first two steps and the non-deterministic outcome of the measurement, additional phases may be added to the output qubit, which can be removed with a physical $R_z$ rotation.

We emphasize that this protocol is not unique. Other measurements may be favourable for the disentangling, depending on the chip layout, the measurement time compared to the duration of the charge and spin operations, and the measurement fidelity. In Sec.~\ref{sec_general} it was shown that all errors on the discarded qubits translate into dephasing of the output qubit, thus operations which are not noise biased may be used, including the conventional quantum gates.

\subsubsection{Two-Qubit $\mathrm{CNOT}$ Gate\label{sec_spins_CNOT}}

\begin{figure*}
    \centering
    \includegraphics[width=\textwidth]{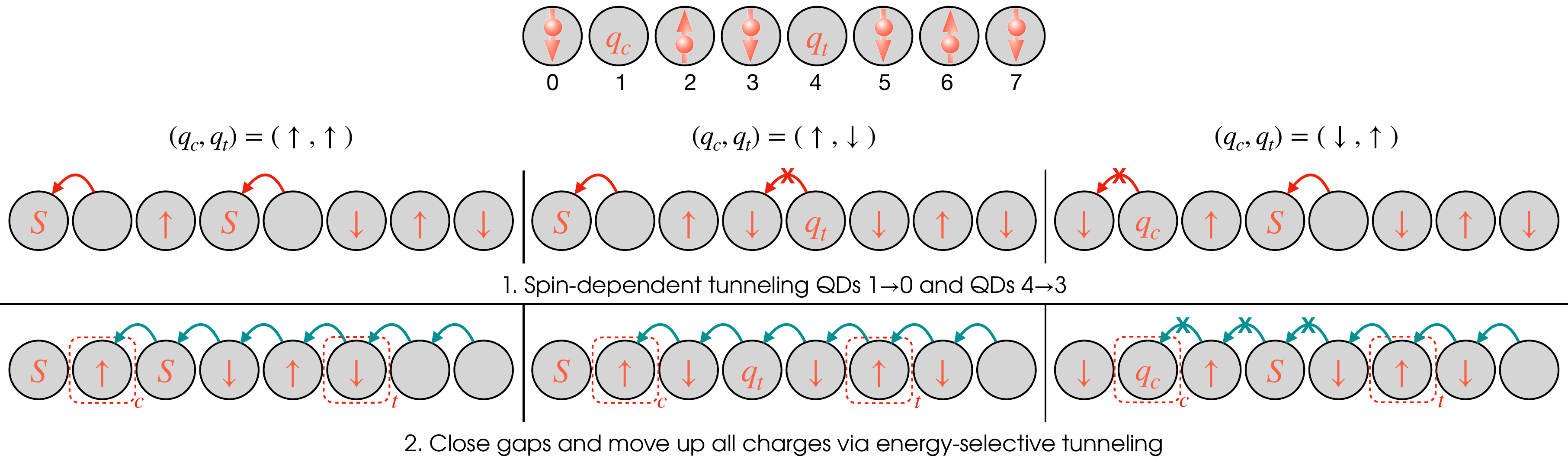}
    \caption{Protocol for a replacement-type $\mathrm{CNOT}$ gate for spin qubits. The top row shows the initial configuration of the input, candidate, and reference qubits, the following rows show the possible configurations after the described operation. The label $q_{c(t)}$ indicates the control (target) qubit. The columns correspond to different two-qubit basis states $(q_c, q_t)$. Output sites for the new control ($c$) and target ($t$) qubit are highlighted in the bottom row. The desired final state is prepared by two parallel spin-dependent tunneling events, followed by moving up all qubits from left to right. In the case $(\downarrow, \downarrow)$, which is not shown, all movement is suppressed.}
    \label{fig_spins_CNOT}
\end{figure*}

A $\mathrm{CNOT}$ gate, up to phase corrections, can be designed following the same principles as the $X$ gate. With two input qubits, candidate qubits to distinguish the four two-qubit basis vectors need to be prepared, doubling the required resources. One possible realization is shown in Fig.~\ref{fig_spins_CNOT}. The replacement-type $\mathrm{CNOT}$ gate uses the same techniques introduced in Sec.~\ref{sec_spins_requirements}.

In the first quantum dot, QD 0, a reference spin for the control qubit in state $|\!\downarrow_0\rangle$ is prepared, followed by the control qubit $q_c$ in QD 1. Analogously, in QD 3 a reference spin $|\!\downarrow_3\rangle$ for the target qubit is prepared, followed by the target qubit $q_t$ in QD 4. The first step of the $\mathrm{CNOT}$ gate uses spin-selective tunneling from QD 1 to QD 0 and, in parallel, from QD 4 to QD 3. The PSB operations create a charge distribution conditional on $(q_c, q_t)$.

One candidate qubit is prepared in state $|\!\uparrow_2\rangle$ in QD 2 and three further candidates are prepared in the state $|\!\downarrow_5, \uparrow_6, \downarrow_7\rangle$ in QDs 5, 6 and 7. After the first step, all charges are moved up towards lower indexed QDs via energy-selective tunneling to close all gaps. Depending on $(q_c, q_t)$, the candidate qubits are moved up by two sites, one site, or all displacements are suppressed.

Finally, QD 1 is declared the output site for the control qubit, it thus is replaced by a superposition of the candidate qubit from QD 2 and the amplitude of $q_c$ which was blockaded by the PSB. For the target qubit, QD 5 is chosen as output site. If the spin projections of $q_c$ and $q_t$ are anti-aligned, the charges are moved up by one site and $|\!\uparrow_5\rangle$ is found, otherwise  the spin state is $|\!\downarrow_5\rangle$. Therefore, the replacement-type gate reproduces the truth table of a $\mathrm{CNOT}$ gate, although, as with the single-qubit gates, each basis state acquires a unique phase during the time evolution, which may need to be compensated. Again, the gate should be followed by a disentangling operation, entailing a measurement of the charge configuration in a suitable basis.

We note that closing the gaps requires the energy-selective displacement of a two-electron singlet in the case $(q_c, q_t) = {(\uparrow, \uparrow)}$. This may be experimentally challenging but can be circumvented by separating the double-occupied charge state before the tunneling into a laterally adjacent empty QD, if the array is not strictly linear. Alternatively, a sequence of three PSB events may be used, each immediately followed by a chain of energy-selective tunneling events: from QD 1 to QD 0, then from QD 3 to QD 2, then from QD 4 to QD 3. This way, the singlet state is created in its final position, however, the duration of the gate is increased, and the final PSB step needs to be energy-selective such that tunneling is suppressed if QD 2 is double-occupied, or another qubit $|\!\downarrow_8\rangle$ needs to be appended.

Entangling a fresh auxiliary qubit with an existing qubit is possible with considerably less overhead than the $\mathrm{CNOT}$ gate between two qubits in arbitrary states. This is achieved with a modification of the $X$ gate protocol from Fig.~\ref{fig_spins_disentangle}: The populations of both QDs 1 and 2 are chosen as output, only the population of QDs 0 and 3 are disentangled with a suitable measurement. If the population in QD 1 is kept instead of the original qubit $q$ and the population of QD 2 is understood as a new qubit, the effect of the operation is equivalent to the gate $\mathrm{CNOT}_{1\to 2} |q_1\rangle |0_2\rangle$, up to phase corrections. The auxiliary qubit from QD 2 could then be used as reference or candidate in a replacement-type gate on another qubit in a form of entanglement swapping. This may facilitate a more efficient implementation of the syndrome-measurements in quantum error correction compared to the general replacement-type $\mathrm{CNOT}$ gate.

\subsection{Discussion of Error Mechanisms\label{sec_spins_errors}}

The replacement-type gates for spin qubits primarily rely on two physical mechanisms: Pauli spin blockade, which can have an infidelity ranging from $7\times 10^{-3}$ to $10^{-3}$ with a duration from $\SI{300}{\nano\second}$~\cite{PhysRevApplied.18.054090,Kuenne2024} down to $\SI{50}{\nano\second}$~\cite{Matsumoto2025}, and energy-selective tunneling, which can have an infidelity ranging from $10^{-3}$ to $< 10^{-5}$ with a duration of $\lesssim \SI{10}{\nano\second}$~\cite{PhysRevB.111.115305,Ventura2024,PRXQuantum.4.030303}. Thus, the duration of the rearrangement operation is mainly determined by the PSB and comparable to conventional gates in shuttling-based architectures~\cite{Matsumoto2025}. The same is true for the disentangling operation, which includes a measurement of the charge state, which typically has an integration time of $\gtrsim \SI{100}{\nano\second}$~\cite{PhysRevApplied.13.024019}. However, we note that while duration of the disentangling is relevant for the dephasing of the output spin qubit, it should not be counted towards the duration of the gate: Since the output qubit is not directly involved, it can be shuttled to a new destination or even be involved in further Clifford operations while the disentangling commences.

During the rearrangement, a number of errors can lead to a candidate qubit with the incorrect spin state (bit error), or even no qubit at all in the output site (charge error).

A major source of bit errors is the lifting of the spin blockade, allowing other two-particle states than the intended one to be merged into a single quantum dot. This can be caused by spin-flip processes, $t_{i,j,\uparrow,\downarrow}$, $t_{i,j,\downarrow,\uparrow}$ in Eq.~(\ref{eq_spins_hamiltonian}), which may emerge due to synthetic or intrinsic SOI~\cite{Fujisawa2002,PhysRevB.80.041301}. Furthermore, a low-lying orbital state, the excited valley state for electrons in silicon or bilayer graphene, in particular, can lead to an available triplet state in the double-occupied dot~\cite{PhysRevB.82.155312,PhysRevB.82.155424,PhysRevB.80.201404}. Relaxation of the excited spin state before or during the tunneling process will also result in an error, but is highly unlikely due to the long spin lifetime. Finally, a non-adiabatic sweep of the detuning leads to Landau--Zener (LZ) transitions between the instantaneous eigenstates, resulting in the undesired outcome where the population of an unblockaded state remains in its charge configuration~\cite{Wittig2005,PhysRevA.53.4288}.

LZ transitions are a leading error mechanism of the energy-selective tunneling with a single electron as well. A non-adiabatic sweep with respect to the spin-conserving tunneling can result in cases where the electron is left behind, potentially in combination with a spin flip or valley excitation~\cite{PhysRevB.102.195418}. Of particular relevance is the regime $(B_i + B_j) / 2 > 2 |t_{i,j,\sigma,\sigma}|$ with a non-zero spin-flip tunneling, where different paths of adiabatic and non-adiabatic transitions through the sequence of ALCs lead to the same outcome. Depending on the relative phase accumulated between the ALCs constructive or destructive interference between these paths may emerge~\cite{PhysRevB.101.035303,PhysRevB.102.195418}.

The energy selective tunneling may suffer from leakage, allowing tunneling also for the states with two electrons in the two QDs if the charging energy $U_2 - U_1 \not\gg |t_{i,j,\sigma,\sigma}|$. This can be understood from Fig.~\ref{fig_spins_spectrum}: Traversing the single-electron ALCs may bring the system in the vicinity of the two-electron ALCs, where the single- and double-occupied states hybridize.

Further imperfections which are reflected as bit errors arise from the erroneous preparation of the reference and candidate qubits. As discussed in Sec.~\ref{sec_general}, errors during the disentangling operation, including measurement errors, result in phase errors.

Throughout all operations, the qubits experience dephasing (phase errors) due to low-frequency fluctuations of the energy levels caused by electric charge noise, the coherences of the charge state are particularly vulnerable~\cite{RevModPhys.79.1217,RevModPhys.95.025003}. The rapid dephasing due to charge noise favors fast detuning sweeps during the gate, therefore, a trade-off between adiabaticity and dephasing is anticipated~\cite{PRXQuantum.4.020305}. Furthermore, even in the regime of an adiabatic detuning sweep for the energy-selective tunneling, LZ transitions induced by charge noise can cause the electron to be left behind, fundamentally limiting the noise bias of the gate~\cite{PhysRevB.101.035303,PhysRevB.104.075439}.

We note that at no point the qubit state is manipulated by a spin rotation, neither are the basis vectors temporarily relabelled. This eliminates the direct translation of the dominant phase errors stemming from charge noise into bit errors. We further note that the mechanisms for bit and charge errors are largely insensitive to the phase of the spins. It is possible for phase errors to impact the spin projection and charge distribution if the energy-selective tunneling is performed in a non-adiabatic regime where the probability amplitudes for different paths in the level diagram can interfere due to LZ transitions. However, this interference can be strongly suppressed by choosing $2 |t_{i,j,\sigma,\sigma}| > (B_i + B_j) / 2$ and a sufficiently adiabatic level velocity of the detuning sweep~\cite{PhysRevB.101.035303,PhysRevB.102.195418}.

The phase of the spin state can in general impact the PSB: The singlet state $|S_{(1,1)}\rangle$ is mapped to the triplet state $|T_0\rangle$ and vice versa by a phase flip on one of the spins. In our proposed protocol, however, a longitudinal magnetic gradient is used, in the presence of which the singlet and triplet states are no eigenstates anymore. Instead, the relevant two-spin states are $({|S_{(1,1)}\rangle \pm |T_0\rangle)/\sqrt{2}}$, for which a phase flip results in a global phase. Therefore, the translation of phase to bit errors in the PSB can be suppressed by choosing a sufficiently strong magnetic field gradient.

\subsection{Performance Analysis\label{sec_spins_qpt}}

From the discussion of the error mechanisms and the translation of errors, we expect that bit errors of the replacement-type gate are strongly suppressed if the detuning sweeps are sufficiently adiabatic and if the singlet-triplet splitting and on-site charging energy are sufficiently large. Therefore, we expect the replacement-type gate to be strongly noise-biased, dominated by phase errors. We confirm the noise bias by simulating quantum process tomography~\cite{Chuang01111997,PhysRevLett.78.390,PhysRevLett.93.080502} for both the replacement-type $X$ and $\mathrm{CNOT}$ gates. The disentangling operation, which contributes only phase errors, is not included in the simulation in order to study the error of the rearrangement operation alone.

The simulation is based on the Hamiltonian from Eq.~(\ref{eq_spins_hamiltonian}), including the single-occupied states with both spin projections for all QDs, the empty state for all QDs except the leftmost one and the double-occupied singlet state and the lowest triplet state for all QDs except the rightmost one. We assume that the triplet state lies $E_\mathrm{ST} = \SI{200}{\micro\electronvolt}$ above the singlet.

We extract the parameters of electron spin qubits in Si/SiGe from recent literature~\cite{Takeda2024,PhysRevB.111.115305,PhysRevB.102.125406,PRXQuantum.4.030303,PhysRevLett.110.086804}.
For the PSB, we follow the procedure of Ref.~\cite{PhysRevLett.110.086804}, slowing down the detuning sweep to a level velocity of $\SI{100}{\micro\electronvolt\per\nano\second}$ near the ALC and using a fast ramp with a level velocity of $\SI{20}{\milli\electronvolt\per\nano\second}$ away from the sensitive spots, in order to shorten the duration. With a spin-conserving tunnel coupling $t_{i,j,\sigma,\sigma} = \SI{25}{\mu\electronvolt}$ and a spin-flip tunneling $t_{i,j,\sigma,\sigma'} = \SI{0.25}{\mu\electronvolt}$, $\sigma \neq \sigma'$, this allows for a highly adiabatic and rapid PSB. We further assume a charging energy $U_2 - U_1 = \SI{5}{\milli\electronvolt}$ and a longitudinal magnetic field gradient $B_j - B_{j+1} = \SI{10}{\micro\electronvolt}$. As non-unitary sources of error we assume imperfect initialization with success probability $p_i = 1 - 10^{-3}$, dephasing of the spin state with $T_2 = \SI{20}{\micro\second}$ and the charge state with $T_{2,c} = \SI{25}{\nano\second}$, and a spin relaxation time $T_1 = \SI{1}{\second}$.

\begin{figure}
    \centering
    \includegraphics[width=0.98\columnwidth]{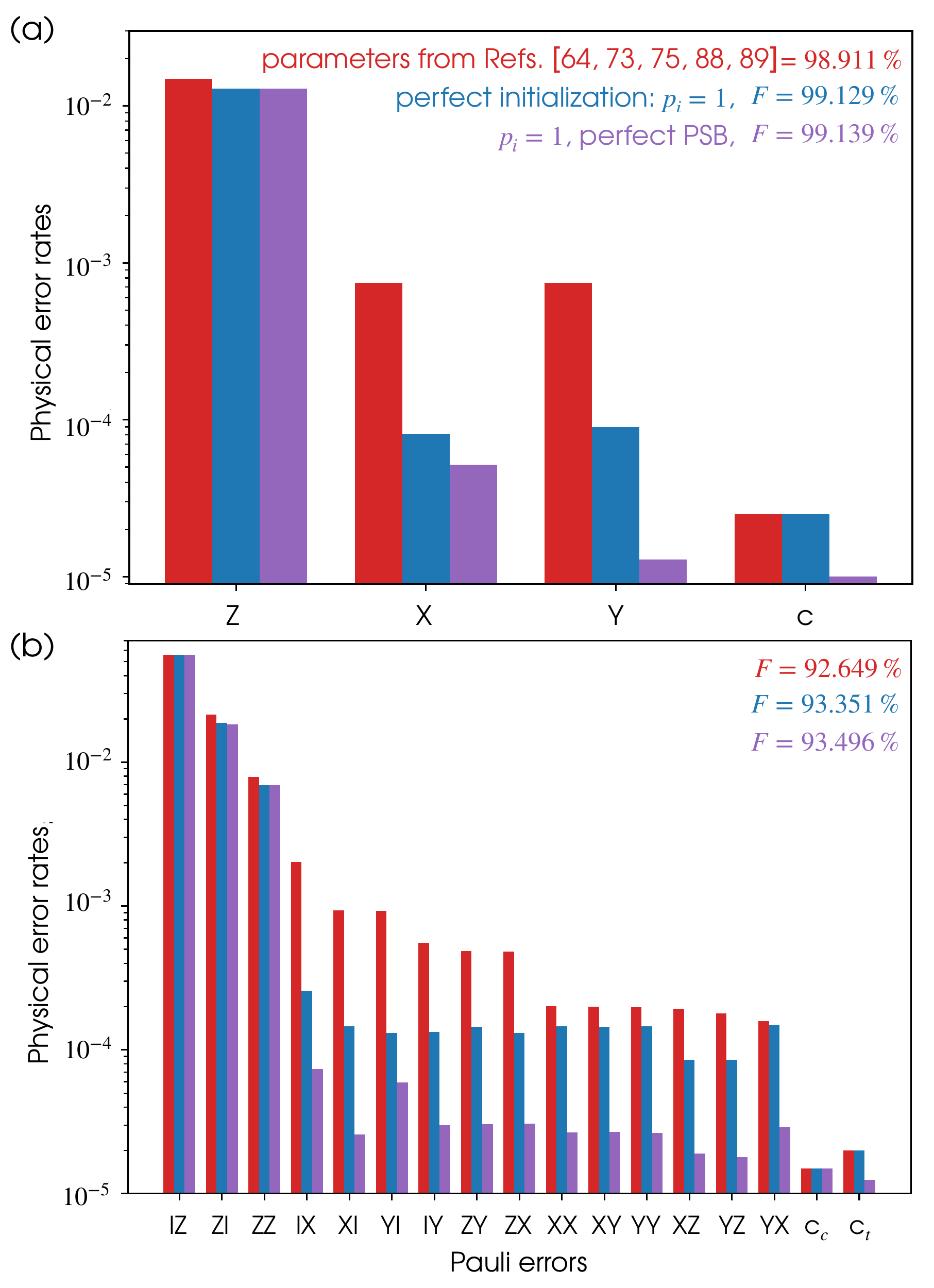}
    \caption{Quantum process tomography of a simulated replacement-type $X$ (a) and $\mathrm{CNOT}$ gate (b) with spin qubits, not including the disentangling operation. The bar diagram illustrates the probability of the Pauli and charge ($c$) errors on the involved qubits, the fidelity $F$ for each case is shown in the top right corner. In red the parameters discussed in the main text are used, in blue a perfect initialization is assumed, $p_i = 1$, and in purple the PSB is assumed to be error-free in addition to $p_i = 1$. Phase errors dominate throughout, initialization and PSB imperfections are a major source of bit errors but contribute hardly to phase errors. Charge errors occur with a very low probability.}
    \label{fig_spins_sim}
\end{figure}

In Fig.~\ref{fig_spins_sim}(a) we show simulated quantum process tomography for the replacement-type $X$ gate. With the realistic parameters, bit errors occur with a probability one order of magnitude lower than phase errors, and are mostly caused by erroneous initialization of the reference and candidate qubits. Assuming a perfect initialization of all qubits, $p_i = 1$, the probability of bit errors is further suppressed by one order of magnitude. Further assuming a perfect PSB, slightly reduces $x$-errors and suppresses charge errors, which are generally on a very low level. The remaining errors are due to dephasing and imperfections in the energy-selective tunneling. The effect of spin relaxation is negligible.

In Fig.~\ref{fig_spins_sim}(b), the same is shown for the replacement-type $\mathrm{CNOT}$, where a similar behavior is observed. Here, the effect of the bit errors related to PSB is stronger compared to the $X$ gate, which is explained by the use of multiple PSB operations in the gate. Again, charge errors occur with a very low probability, slightly more likely on the target ($c_t$), which is moved considerably more than the control ($c_c$) qubit. Note that for the $\mathrm{CNOT}$ gate we assumed a larger tunneling in order to accommodate for $2 |t_{i,j,\sigma,\sigma}| > (B_i + B_j) / 2$ throughout the longer QD array.

The results from quantum process tomography illustrate that the noise bias of the rearrangement operation can approach the ratio known for idling spin qubits, although a reduction of the errors of the spin initialization and spin-selective tunneling is required. We note that both techniques are based on PSB, where the fidelity can be enhanced by optimal control techniques~\cite{PhysRevApplied.18.054090,Ventura2024}, by optimizing the confinement potential, and by engineering the valley splitting. The latter can be achieved by optimizing the width of the quantum well and the interfaces~\cite{Lima_2023,Stehouwer2025} or by introducing germanium into the silicon quantum well~\cite{PhysRevB.104.085406,McJunkin2022,PhysRevB.106.085304,PaqueletWuetz2022}.

The rearrangement is followed by a disentangling operation, which contributes only phase errors, as discussed in Sec.~\ref{sec_general}. We assume the same experimental parameters as above, EDSR gates with an error probability of $10^{-4}$~\cite{Wu2025highfidelitygate} and  a charge readout error of $10^{-3}$ at an integration time of $\SI{300}{\nano\second}$~\cite{PhysRevApplied.13.024019}. We estimate a fidelity of $99.6\%$ ($99.7\%$) for the parameters from recent literature (with perfect PSB). The fidelity of the disentangling is so high compared to the rearrangement because the quantum information is stored in a robust spin state instead of a vulnerable hybrid state, and experiences only little dephasing during the rapid operation.
% F_disentangle = F_EDSR * F_PSB * F_hopping * F_charge_measurement * F_idling_spin(~ 600 ns)
% = 0.9999 * 0.999 * 0.999 * 0.999 * exp[-(0.6 / 20)^2]

The total fidelity of the replacement-type gates may appear low in comparison to conventional gates. Note however, that, for example, the XZZX surface code with minimum-weight perfect-matching for fault tolerant decoding has a threshold error rate of $\approx 0.078$ at a noise bias of $10^2$ and $0.085$ at a noise bias for $10^3$~\cite{BonillaAtaides2021}. Thus, we expect replacement-type gates around the threshold for fault tolerance to be experimentally feasible with only slightly optimistic assumptions. Any improvement of the noise bias by further suppressing the errors of the initialization, PSB or energy-selective tunneling will further improve the performance of the QEC code.

\begin{figure}
    \centering
    \includegraphics[width=0.95\columnwidth]{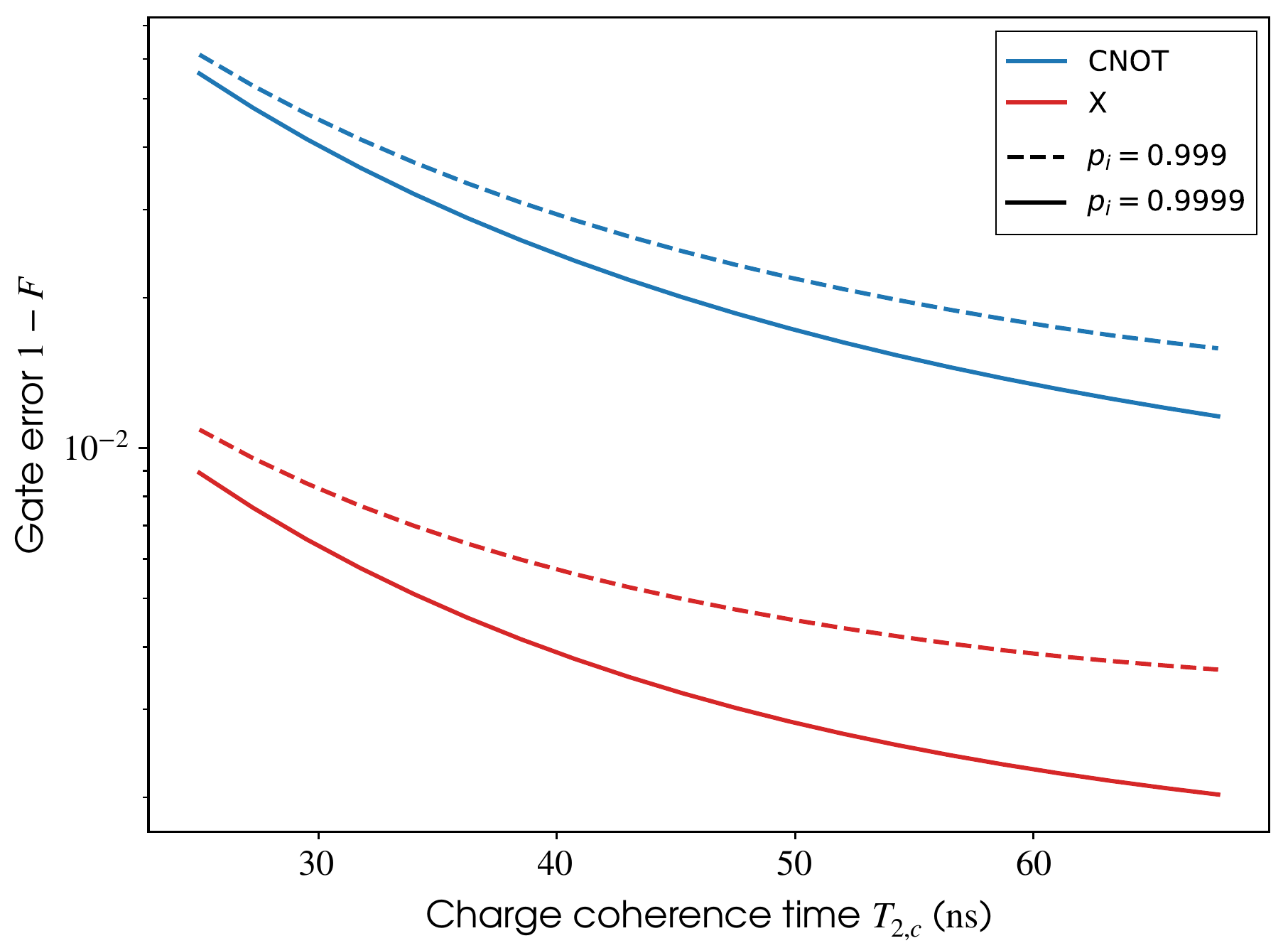}
    \caption{Infidelity $1-F$ of replacement-type $X$ (red) and $\mathrm{CNOT}$ (blue) gates with spin qubits as a function of the decoherence time $T_{2,c}$ of the charge state. Solid lines correspond to a probability for correct initialization of $p_i = 0.9999$ and dashed lines to $p_i = 0.999$, the other parameters are identical to Fig.~\ref{fig_spins_sim}, where $T_{2,c} = \SI{25}{\nano\second}$ was used. Extending the coherence to $T_{2,c} = \SI{70}{\nano\second}$ allows a $\mathrm{CNOT}$ fidelity as high as $\approx 0.989$ and an $X$ gate fidelity of $\approx 0.998$.}
    \label{fig_spins_T2}
\end{figure}

In our model, the fidelity of the replacement-type gate is limited by dephasing, in particular of the hybrid spin-charge state. In Fig.~\ref{fig_spins_T2} we investigate the dependence of the infidelity on the charge coherence time $T_{2,c}$ of the charge state. We find that with a sufficiently high initialization fidelity and with an extended $T_{2,c}$, a $\mathrm{CNOT}$ gate infidelity $\lesssim 10^{-2}$ is possible. Extending $T_{2,c}$ requires suppression of the environmental charge noise, which is challenging in practice. 

Conventional dynamical decoupling (DD) techniques, which are widespread for extending the coherence time, cannot be interspersed into the rearrangement or disentangling operations as they would compromise the noise bias. However, since the replacement-type gate relies on moving qubits, a similar effect is expected from motional narrowing~\cite{PRXQuantum.4.020305,Struck2023,PRXQuantum.2.030331}. By averaging out quasistatic but spatially varying contributions of the magnetic and electric noise it was observed that the coherence time of moving electron spin qubits can be increased by a factor up to $\approx 4$ in natural silicon~\cite{Struck2023} or $\approx 10$ in GaAs~\cite{PRXQuantum.2.030331}. To fully leverage motional narrowing, the replacement-type gate can be performed inside the moving QDs of a conveyor mode shuttling lane.

Furthermore, recent research is advancing the understanding of the microscopic origin of charge noise such that a noise reduction through adaptations of the device design is conceivable~\cite{PhysRevB.110.235305,Connors2022,PhysRevB.100.165305,RojasArias2024} and reduction strategies can be developed~\cite{PaqueletWuetz2023,Unseld2024,Choi2024}. Another direction for future improvements are modifications of the rearrangement operation which increase the admixture of the spin states to the hybrid spin-charge basis of the intermediate steps. Since the coherence time of the spin alone is many times longer than the gate time, this will reduce the vulnerability to charge noise~\cite{RevModPhys.95.025003}.

We emphasize that the parameters of the Hamiltonian here are not specifically optimized for the performance of the replacement-type gate. We expect that significant improvements can be achieved by finding an optimal regime for the magnetic field gradient which impacts PSB, the energy-selective tunneling, and the susceptibility of the spin state to charge noise. Uniformly displacing the entire register between regions of different gradients with conveyor mode shuttling may even allow tailoring its value to each operation. Optimizing the shape of the detuning ramps with optimal control theory is expected to bring further enhancement.

Finally, the question arises whether the noise bias is still preserved if the coherence times are improved by reducing the coupling to charge noise. We expect this to be the case since reducing the dephasing due to charge noise allows for more adiabatic detuning sweeps: $X$, $Y$, and charge errors stemming from LZ transitions can be further suppressed by readjusting the trade-off of between dephasing and non-adiabatic excitation. Furthermore, the LZ transitions triggered by the electric fluctuations represent the minimal $X$ and $Y$ errors added in addition to the statistics of an idle spin~\cite{PhysRevB.101.035303,PhysRevB.104.075439}. Thus, suppressing charge noise can even allow for a better approximation of the intrinsic noise bias, provided sufficient control over the detuning sweeps. Finally we note that the errors during the disentangling will always be phase errors, which naturally ensures a noise bias even if the error channel of the rearrangement alone is more balanced.

\section{Realization with Neutral Atom Qubits\label{sec_atoms}}

The idea of the replacement-type gates can be extended to other qubit platforms, such as atoms or molecules. For concreteness, we here focus on neutral atom qubits in optical tweezers that have shown significant progress for quantum computational tasks due to their versatile properties, including scalability in qubit count, and long-range interactions mediated by mid-circuit shuttling. In our proposed setup, the atoms are trapped using optical tweezers created by focused laser beams, and their internal states are manipulated through additional laser-driven transitions. Exciting these atoms to Rydberg states, highly excited electronic states with large principal quantum numbers, not only enables strong, tunable interactions for implementing entangling gates, but also facilitates controlled tunneling. This controlled tunneling plays a key role in realizing replacement-type gates via occupation-dependent tunneling and state-dependent tunneling. 

For concreteness, we focus on the fermionic isotope of Alkaline-earth(-like) atoms possessing large nuclear spin $I$ (such as $\mathrm{^{87}Sr}$ with $I=9/2$ and $\mathrm{^{173}Yb}$ with $I=5/2$) where quantum information is encoded in the nuclear spin of the low-lying electronic states \cite{Daley2011, Cooper2018}.
The electronic ground state $\mathrm{^1S_0}$ and the metastable clock state $\mathrm{^3P_0}$ are of particular interest as they feature a rich subspace due to Zeeman splitting.
These hyperfine sublevels $|m_F\rangle$ are dominantly characterized by the nuclear spin $F=I$ as the total electronic angular momentum $J$ is zero. Consequently, the nuclear spin is largely decoupled from magnetic-field noise and photon scattering, leading to remarkably long coherence times. Experiments with fermionic strontium and ytterbium isotopes have demonstrated relaxation times exceeding the vacuum-limited lifetimes of tens to hundreds of seconds, and coherence times ranging from several seconds up to minutes for different qubit encodings \cite{Barnes2022,Jenkins2022,Ma2022, Yang2025}. This long-lived and well-controlled manifold of nuclear-spin states provides a natural platform for diverse qubit and qudit encodings \cite{Lis2023} and enables versatile simulations of complex Hamiltonians, including lattice gauge theories \cite{Gonzalez2022,Zache2023}.

We utilize the Rydberg manifold $\mathrm{^3S_1}$ passing through the metastable excited manifold $\mathrm{^3P_0}$ to activate interactions between atoms. The strong Van der Waals (VdW) interaction between nearby atoms is the main resource to implement entangling gates, such as controlled-$Z$ gates, via the so-called Rydberg Blockade mechanism. Moreover, the Rydberg manifold $\mathrm{^3S_1}$, in contrast to the ground and metastable manifold, possesses non-zero $J=1$ and therefore strong hyperfine coupling. Consequently, a magnetic field strongly splits the Zeeman sublevels of the Rydberg manifold. This allows selective addressing of the sublevels, which is crucial for our Rydberg-based gate operations. The following will explain how this level structure enables the conditional operations required for the replacement-type gates.

\begin{figure*}[ht!]
    \centering
    \includegraphics[width=0.9\textwidth]{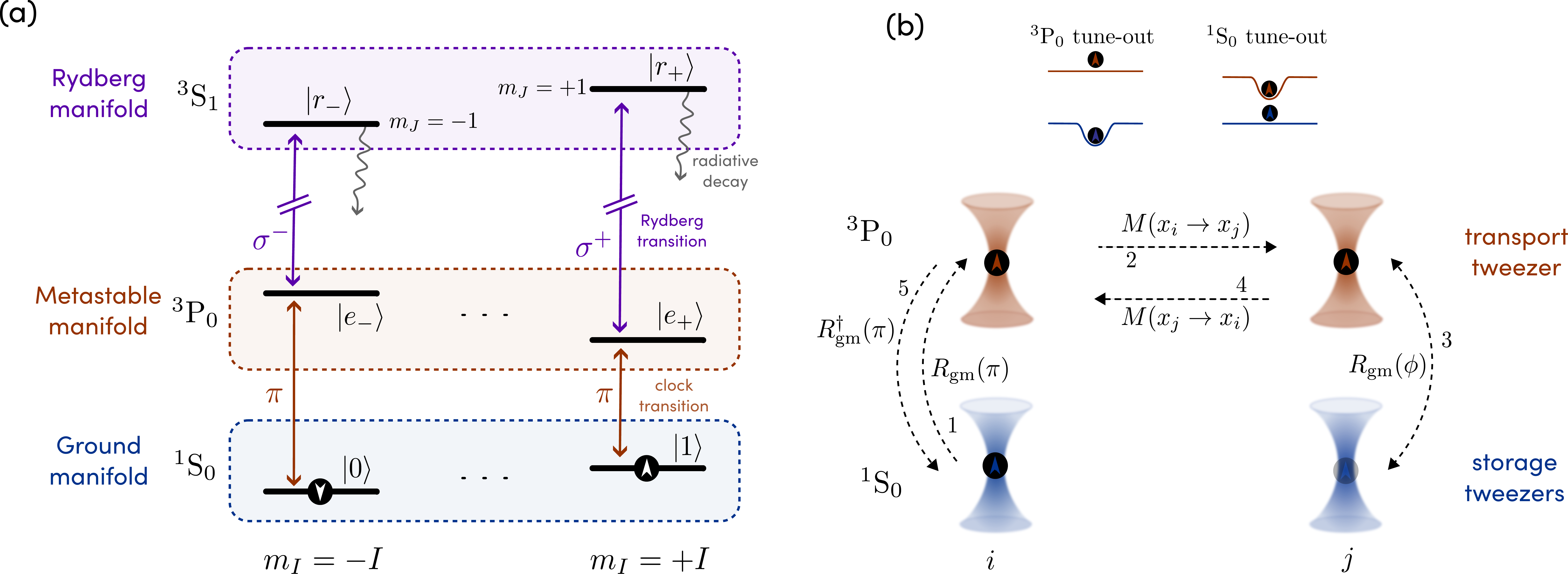}
    \caption{(a) Qubit states are encoded in the stretched nuclear spin states located at the edges of the total magnetic quantum number within the ground manifold, specifically $|0\rangle \equiv |\mathrm{^1S_0}, F = -m_F = I\rangle$ and $|1\rangle \equiv |\mathrm{^1S_0}, F = m_F = I\rangle$, which are interchangeably referred to as $\left|\downarrow\right\rangle$ and $\left|\uparrow\right\rangle$, respectively. A dual-tone optical $\pi$-pulse addresses the clock transition to the metastable manifold $^3\mathrm{P}_0$ for both qubit levels, namely $|0\rangle \leftrightarrow |e_{-}\rangle$ and $|1\rangle \leftrightarrow |e_{+}\rangle$. The metastable state serves two purposes: It enables transport of the atom via tweezers that selectively trap $^3\mathrm{P}_0$ while being tuned out for $^1\mathrm{S}_0$, and it acts as an intermediate state to Rydberg excitation for entangling gates. With stretched Rydberg states defined as $|r_{\mp}\rangle \equiv |\mathrm{^3S_1}, F = \mp m_F = I + 1\rangle$, the required Rydberg coupling is achievable via $\sigma^{\mp}$-polarized UV light.
    (b) Shuttling protocol utilizing state-selective trapping: Two static storage tweezers positioned at sites $i$ and $j$ (blue) confine atoms in the ground state while being tuned out for $^3\mathrm{P}_0$, and a mobile transport tweezer (orange) shuttles atoms in $^3\mathrm{P}_0$ between the two. As an example, suppose the tweezer at site $i$ holds an atom in $|0\rangle$, and site $j$ is initially unoccupied. The transport tweezer is then turned on at site $i$, followed by a $\pi$ pulse $R_{\mathrm{gm}}(\pi)$ that maps $|0\rangle \rightarrow |e_-\rangle$, loading the atom into the mobile tweezer. After moving to site $j$, a $R_{\mathrm{gm}}(\phi)$ operation redistributes population into the storage tweezer at site $j$, depending on the angle $\phi$. The transport tweezer returns to site $i$, and a final $R_{\mathrm{gm}}^\dagger(\pi)$ operation transfers the remaining population back to the ground state, creating a spatial superposition of $|0\rangle$ or vacuum. This sequence effectively realizes a hopping of a particular qubit state between sites $i$ and $j$.}
    \label{state_dependent_shuttling}
\end{figure*}

The details of our scheme are illustrated in Fig.~\ref{state_dependent_shuttling}. The atomic level structure comprises a ground manifold in which the qubit is encoded in the stretched nuclear spin states with $|0\rangle$ and $|1\rangle$, corresponding to $ |\mathrm{^1S_0}, F = \mp m_F = I\rangle$, respectively. Coherent population transfer to the metastable clock state $\mathrm{^3P_0}$ is achieved using a two-tone $\pi$-pulse that simultaneously drives the clock transitions $|0\rangle \rightarrow |e_+\rangle$ and $|1\rangle \rightarrow |e_-\rangle$, where $ |e_{\mp}\rangle \equiv |\mathrm{^3P_0}, F =\mp m_F = I\rangle$. Each frequency component of the pulse is tuned to be resonant with its respective $m_F\rightarrow m_F^\prime$ transition, ensuring state-selective excitation without cross-talk. 

Excitation from the metastable to the Rydberg manifold is performed via another dual-tone laser field that couples the states $ |e_{\mp}\rangle $ to the stretched Rydberg states $ |r_{\mp}\rangle \equiv |\mathrm{^3S_1}, F = \mp m_F = I + 1 \rangle $. 
These transitions are selected to obey electric dipole selection rules and are driven with matched Rabi frequencies, maintaining symmetry between the two qubit excitation pathways. This configuration enables coherent and independent manipulation of the two qubit states while significantly suppressing undesired coupling between the subspaces $\{0, e_-, r_-\}$ and $\{1, e_+, r_+\}$, which lie at the extremal values of the $m_F$ manifold. The isolation of these stretched states minimizes bit-flip errors and is a key design feature that supports the implementation of noise-bias-preserving replacement-type gates.

\subsection{Required Operations\label{sec_atoms_requirements}}
The gate protocols rely on two key primitives: state-dependent tunneling and state-dependent Rydberg interactions. Tunneling is implemented using two types of optical tweezers: (i) deep storage tweezers, which localize atoms in the ground state at fixed lattice sites, and (ii) transport tweezers, which enable controlled shuttling of atoms between sites. For Rydberg-mediated interactions, site-resolved, multi-tone laser fields are used to selectively excite atoms depending on their internal state, facilitating controlled, state-dependent coupling between atoms.

\subsubsection{State-dependent trapping}
Optical tweezers confine atoms through the AC Stark effect, where a red-detuned laser induces a potential well that draws atoms toward regions of maximum intensity~\cite{Grimm2000}. Since the AC Stark shift depends on the atomic polarizability, internal states with differing polarizabilities experience distinct trapping potentials, a phenomenon known as state-dependent trapping. This effect is particularly pronounced in alkaline-earth-like atoms, where the polarizabilities of the \( \mathrm{^1S_0} \) ground state and the metastable \( \mathrm{^3P_0} \) state differ significantly across the optical spectrum, owing to their distinct electronic structures and resonance features~\cite{Safronova2015, Heinz2020}. Consequently, specific laser wavelengths can be selected such that one state is strongly trapped while the other remains only weakly confined or untrapped. Detailed analyses of suitable wavelengths for \( \mathrm{^{87}Sr} \) and \( \mathrm{^{173}Yb} \) are available in Ref.~\cite{Pagano2019}. Unlike alkali atoms, the narrow linewidth of the clock transition \( \mathrm{^1S_0} \to \mathrm{^3P_0} \) in these species significantly suppresses spontaneous emission to the ground state manifold.

In what follows, we describe a shuttle gate mechanism based on state-dependent trapping, originally proposed in Refs.~\cite{Gonzalez2023, Zache2023}. In this scheme, a transport tweezer, resonant with the metastable \( \mathrm{^3P_0} \) manifold, is dynamically moved between two sites defined by storage tweezers, which are tuned to trap atoms in the \( \mathrm{^1S_0} \) ground manifold. This allows for coherent spatial transport of atoms conditional on their internal quantum state. It is important to point out that despite the high accuracy achievable with these shuttling operations, they are considerably time-consuming relative to other coherent processes. Current implementations typically require $10$-$\SI{100}{\micro \second}$ for single-atom transport over micrometer distances and up to $10$-$\SI{50}{\milli\second}$ for large-scale rearrangement of atomic arrays. Nevertheless, recent advances in optical-tweezers engineering and optimized control protocols are pushing these limits toward the $\SI{1}{\micro\meter\per\micro\second}$ speed regime~\cite{Lu2025, Hwang2025, Pagano2024}.

The qubit is initially prepared in an arbitrary superposition of the ground manifold states \( |0\rangle \) and \( |1\rangle \), confined by a static optical tweezer at position \( x_i \). To enable state-dependent transport, a second (transport) tweezer is introduced and aligned with the storage tweezer at \( x_i \). A resonant \( \pi \)-pulse drives the population in \( |1\rangle \) to the excited state \( |e_+\rangle \), leaving \( |0\rangle \) unchanged. At this point, the atom experiences a combined potential from both tweezers. The transport tweezer is then spatially displaced, separating the trapping potentials and mapping the internal superposition onto a spatial superposition. The component in \( |e_+\rangle \) is transported to a new location \( x_j \), denoted by the operation \( M_{x_i \rightarrow x_j} \). At this position, an X-rotation \( R_{\mathrm{gm}}^\uparrow(\phi) \) is applied between \( |1\rangle \) and \( |e_+\rangle \), followed by returning the transport tweezer to its original location \( x_i \) (\( M_{x_j \rightarrow x_i} \)). Finally, a de-excitation \( \pi \)-pulse \( (R_{\mathrm{gm}}^\uparrow)^\dagger(\pi) \) maps \( |e_+\rangle \) back to \( |1\rangle \), completing the sequence.

In summary, the shuttle gate sequence described above implements the following unitary operation:
\begin{equation}
    T_{i,j}^\alpha (\phi) = (R_{\mathrm{gm}}^\alpha)^\dagger(\pi) \, M_{x_i \to x_j} \, R_{\mathrm{gm}}^\alpha(\phi) \, M_{x_j \to x_i} \, R_{\mathrm{gm}}^\alpha(\pi),
\end{equation}
where \( \alpha = \uparrow \) corresponds to the case where the population in \( |1\rangle \) is excited to the metastable state \( |e_+\rangle \), as in the example provided. More generally, \( \alpha \) can take the value \( \uparrow \) or \( \downarrow \), depending on whether the metastable state \( |e_+ \rangle \) or \( |e_- \rangle \) is used to shuttle the population from \( |1\rangle \) or \( |0\rangle \), respectively. This sequence realizes a tunneling unitary of the form:
\begin{equation}
    T_{i,j}^\alpha (\phi) = e^{-i \frac{\phi}{2} \left((c_{i}^\alpha)^\dagger c_{j}^\alpha + \mathrm{H.c.}\right)},
\end{equation}
where \( c_i^\alpha \) is the fermionic annihilation operator associated with internal state \( \alpha \) at site \( i \). The parameter \( \phi \), set by the rotation angle in the third step of the sequence, determines the strength of the coherent tunneling between sites \( i \) and \( j \). It is worth noting that the shuttle-gate duration is mainly limited by the atom transport time, which scales with the transport distance, and by the clock-transition \( \pi \)-pulses, set by the available Rabi frequency. With current techniques, the total operation time can be reduced to a few hundred microseconds.

\subsubsection{State-Dependent Interactions}

An essential building block for entangling operations is the use of VdW interactions. These interactions arise when atoms are excited to high-lying Rydberg states of identical parity, enabling strong VdW coupling. When two atoms are positioned within the so-called Rydberg blockade radius, simultaneous excitation to the Rydberg state becomes energetically forbidden. This Rydberg blockade mechanism is widely used to implement entangling gates in neutral atom platforms. 

In alkaline-earth-like atoms, direct one-photon excitation to the Rydberg manifold is feasible using UV laser light. The excitation can be shaped via phase modulation to implement a two-body interaction of the form \( e^{-i \theta n_i n_j} \), where \( n_i \) and \( n_j \) represent the occupation number operators of the Rydberg states on atoms \( i \) and \( j \), respectively~\cite{Jandura2022}. In this work, we set \( \theta = \pi \) to realize a maximally entangling phase gate. 

We further assume that Rydberg excitation is state-selective: Depending on the internal states \( \alpha \) and \( \beta \) of atoms \( i \) and \( j \), respectively, the entangling gate is described by
\begin{equation}\label{eq_CZ}
    E_{i,j}^{\alpha, \beta} = e^{-i \pi\;n_i^\alpha n_j^\beta},
\end{equation}
where \( n_i^\alpha \) and \( n_j^\beta \) denote the occupation numbers of states \( \alpha \) and \( \beta \), with \( \alpha, \beta \in \{\uparrow, \downarrow\} \), corresponding to Rydberg states \( |r_+\rangle \) and \( |r_-\rangle \), respectively. For example, when both atoms are excited to the \( |r_+ \rangle \) state, the interaction becomes:
$E_{i,j}^{\uparrow,\uparrow} = e^{-i \pi\;n_i^\uparrow n_j^\uparrow}$.
Alternatively, if atom \( i \) is driven to \( |r_+ \rangle \) and atom \( j \) to \( |r_- \rangle \) via independently addressable lasers, the resulting interaction is:
$E_{i,j}^{\uparrow,\downarrow} = e^{-i \pi\;n_i^\uparrow n_j^\downarrow}$.

Moreover, VdW interactions can be harnessed to realize multi-qubit entangling gates. This can be achieved by positioning atoms within the blockade radius and employing optimal control techniques to shape the driving fields for implementing a desired unitary operation (see, for example, \cite{Evered2023,Cao2024,Kazemi2025}.
Extending the procedure for the two-qubit entangling gate in Eq.~\eqref{eq_CZ}, state-selective excitation to the Rydberg manifold also enables the implementation of an internal-state-dependent three-qubit entangling gate, which takes the form
\begin{equation}
    E_{i,j,k}^{\alpha, \beta, \gamma} = e^{-i \pi\;n_i^\alpha n_j^\beta n_k^\gamma}.
\end{equation}
In the following section, we will utilize these primitive operations to construct replacement-type gate structures.

A practical challenge in Rydberg-based gates is spontaneous decay from the Rydberg manifold, as the Rydberg lifetimes are comparable to typical gate durations. This decay constitutes a major source of noise~\cite{Pagano2022}. To mitigate its effect and preserve the noise bias in our gate design, we utilize stretched Rydberg states. Due to selection rules, transitions from \( |r_+ \rangle \) to states with negative magnetic quantum number \( m_I \) are strongly suppressed, and similarly for \( |r_- \rangle \) with positive \( m_I \), thereby reducing undesired decay channels. A similar approach for alkali atoms is introduced in Ref.~\cite{PhysRevX.12.021049}.

We expect that the gate fidelity for both two- and three-qubit gates can reach $99.9\%$ for interatomic separations of $\leq \SI{3}{\micro\meter}$. Such high fidelities become feasible through precise pulse shaping and advanced optimal-control techniques that mitigate motional and noise-induced effects. Furthermore, by increasing the Rabi frequency to approximately $\SI{20}{\mega\hertz}$, the total gate duration can be shortened to about $\SI{200}{\nano\second}$~\cite{Kazemi2025}.

\subsubsection{Occupation-Controlled Tunneling Gates \label{sec_controlled_tunneling}}
Occupation-controlled tunneling gates can be engineered by combining state-dependent tunneling with state-dependent entangling operations. These composite gates enable conditional tunneling based on the occupation of one or more control sites. The gate constructions are adapted from Ref.~\cite{Gonzalez2023}.
Note that in this work, we focus only on atom trapping using optical tweezers for brevity. However, similar gate operations can also be realized in optical lattice platforms for example, by tilting the lattice and leveraging collisional interactions, as demonstrated in \cite{Sharma2021}.

\paragraph{Single-Site Control}
The gate $C_i^\alpha T_{j,k}^\beta$ implements a tunneling operation for an atom in internal state $\beta$, interpreted as $|0\rangle$ for $\beta = \downarrow$ and $|1\rangle$ for $\beta = \uparrow$, from site $j$ to site $k$, conditioned on the atom at site $i$ being in internal state $\alpha$. This gate can be realized by the following sequence:
\begin{equation}
    C_i^\alpha T_{j,k}^\beta = T_{j,k}^\beta(\pi/2)\, E_{i,j}^{\alpha, \beta}\, T_{j,k}^\beta(-\pi/2)\, E_{i,j}^{\alpha, \beta}.
\end{equation}

In this sequence, the entangling gates act to modify the effective sign of the intermediate tunneling operation when the control condition is satisfied. The overall effect of the sequence is to apply a tunneling operator conditioned on the occupation of site $i$, which can be expressed more compactly as:
\begin{equation}
    C_i^\alpha T_{j,k}^\beta = e^{-i \frac{\pi}{2} \left((c_j^\beta)^\dagger n_i^\alpha c_k^\beta + \mathrm{H.c.}\right)}.
\end{equation}

\paragraph{Two-Site Control}
The conditional logic can be extended to cases involving two control sites. The gate $C_{i,j}^{\alpha, \beta} T_{i,j}^\gamma$ enables tunneling of a particle in internal state $\gamma$ between sites $i$ and $j$, but only if site $i$ is occupied by an atom in state $\alpha$ and site $j$ by an atom in state $\beta$. The gate is constructed via the sequence:
\begin{equation}\label{eq_2CT}
    C_{i,j}^{\alpha, \beta} T_{i,j}^\gamma = T_{i,j}^\gamma(\pi/2)\, E_{i,j}^{\alpha, \beta}\, T_{i,j}^\gamma(-\pi/2)\, E_{i,j}^{\alpha, \beta}.
\end{equation}

This structure allows the tunneling to occur only if both control site conditions are satisfied. In contrast, one may also construct a gate that triggers tunneling in the absence of a specific two-site configuration. The gate $C_{i,j}^{\neg(\alpha, \beta)} T_{i,j}^\gamma$ permits tunneling only if site $i$ is not in state $\alpha$ or site $j$ is not in state $\beta$. This is implemented by:
\begin{equation}
    C_{i,j}^{\neg(\alpha, \beta)} T_{i,j}^\gamma = T_{i,j}^\gamma(\pi/2)\, E_{i,j}^{\alpha, \beta}\, T_{i,j}^\gamma(\pi/2)\, E_{i,j}^{\alpha, \beta}.
\end{equation}
Here, unlike in Eq.~\eqref{eq_2CT}, both tunneling gates carry the same phase, thereby altering the constructive interference condition based on the control occupation.

\paragraph{Generalization to Non-Local Control}
This conditional tunneling framework can be generalized further to allow for spatial separation between the control and tunneling regions. Specifically, the tunneling of an atom between two arbitrary sites $k$ and $l$ can be conditioned on the internal states of two different control sites $i$ and $j$. This is made possible using the three-body, internal-state-dependent entangling gate introduced earlier. The corresponding gate construction is:
\begin{equation}\label{eq_3CT}
    C_{i,j}^{\alpha, \beta} T_{k,l}^\gamma = T_{k,l}^\gamma(\pi/2)\, E_{i,j,k}^{\alpha, \beta, \gamma}\, T_{k,l}^\gamma(-\pi/2)\, E_{i,j,k}^{\alpha, \beta, \gamma}.
\end{equation}

This generalization greatly enhances the flexibility of occupation-controlled dynamics and enables the design of more complex gate architectures in neutral atom platforms \cite{Gonzalez2022, Schuckert2024}.

\subsection{Implementations of Replacement-Type Gates\label{sec_atoms_implementation}}
With the set of primitive operations introduced in the previous section, we are now equipped to provide concrete protocols for realizing replacement-type gates with neutral-atom qubits. These protocols demonstrate how approximately noise bias-preserving single- and two-qubit gates can be constructed using the available gate primitives.
In the following, we outline explicit sequences for the implementation of an $X$ gate and a $\mathrm{CNOT}$ gate, focusing on their realization via spatially-resolved atomic manipulation in optical tweezers.

\subsubsection{Single-Qubit $X$ Gate}

\begin{figure}
    \centering
    \includegraphics[width=0.49\textwidth]{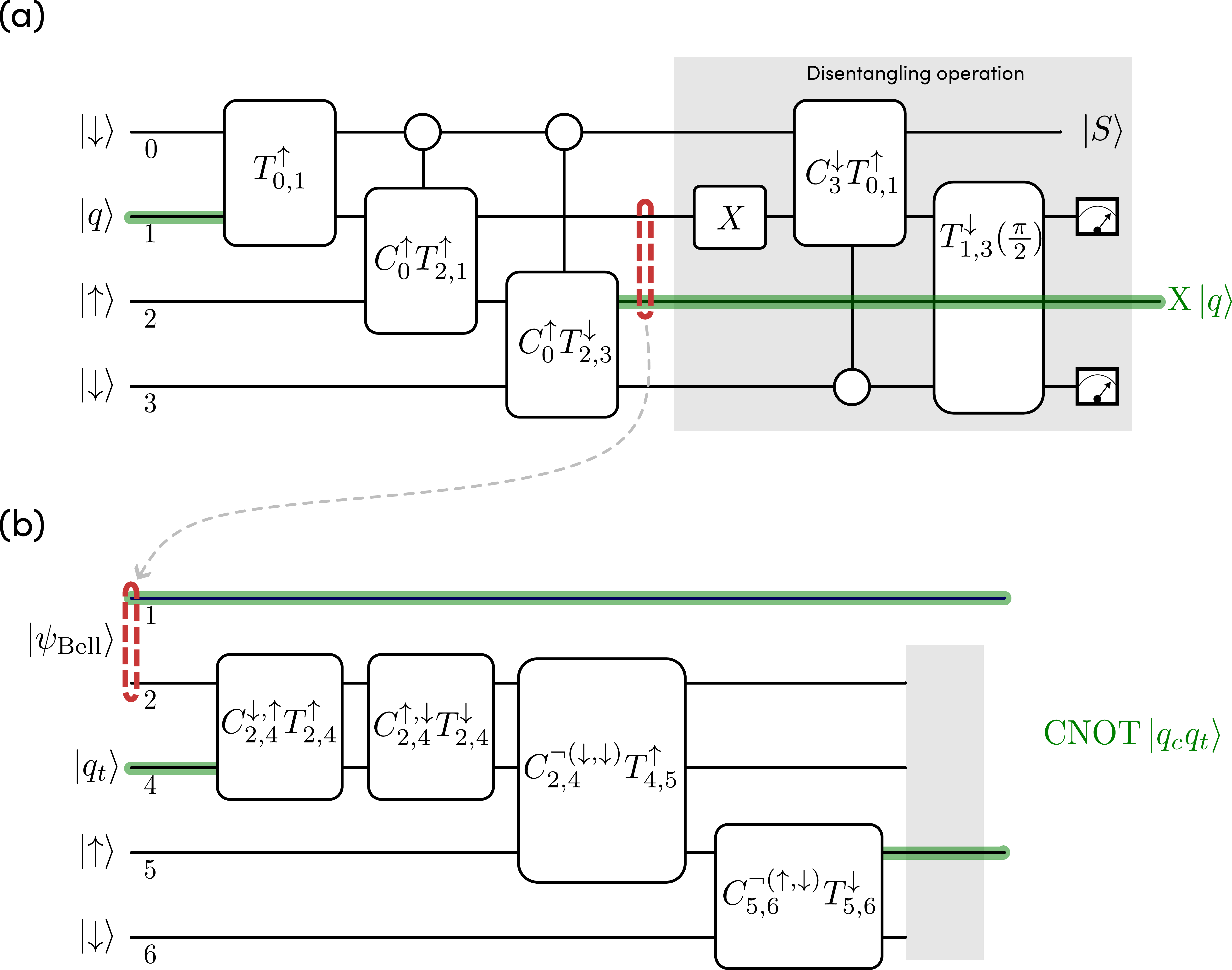}
    \caption{(a) A tunneling-assisted protocol for the implementation of the noise-biased $X$ gate on a neutral atom platform. Here, the occupation-controlled tunneling is performed via state-dependent tunneling and Rydberg entangling gates. The first three gates convert an input qubit at site 1 into the desired state $X|q\rangle$ at site 2, and the remaining gates (gray box) disentangle the auxiliary qubits 0, 1, and 3 from the output qubit. (b) The protocol for the $\mathrm{CNOT}$ gate based on the Bell state from the previous circuit at sites 1 and 2 (red dashed line). Here, sites 1 and 4 represent the control and the target qubits respectively, and sites 5 and 6 represent additional candidate qubits. The depicted four conditional tunneling gates generate the desired state $\mathrm{CNOT}|q_cq_t\rangle$ at sites 1 and 5, and the remaining qubits can then be discarded after a subsequent disentangling operation.}
    \label{fig_X_rydberg}
\end{figure}

A protocol for realizing the $X$ gate is shown in Fig.~\ref{fig_X_rydberg}(a). In analogy to the protocol used for spin-based qubit platforms in Secs.~\ref{sec_spins_X}, \ref{sec_spins_disentangle} we employ four storage tweezers loaded as $|\Psi_\mathrm{in} \rangle = \left |\downarrow_0,q_1,\uparrow_2, \downarrow_3\right\rangle$ representing the reference, input, and two candidate qubits, respectively. Here, $|q\rangle=\alpha_\downarrow\left |\downarrow\right\rangle + \alpha_\uparrow\left |\uparrow\right\rangle$ is an arbitrary state of the input qubit. The bias-preserving $X$ gate is implemented through a sequence of conditional tunneling steps that selectively merge the spatially separated candidate states into an output qubit, generating the desired output qubit $X|q\rangle$.

The procedure relies on occupation-controlled tunneling of internal states, and proceeds as follows: First, a state-dependent tunneling operation $T_{0,1}^\uparrow$ is applied, enabling coherent transfer of the $\left | \uparrow\right\rangle$ state between sites 0 and 1 via a transport tweezer. This step selectively transfers the $\uparrow$ component of the input qubit to site 0. This prepares the system for controlled routing based on the initial qubit state.
Next, a controlled tunneling gate $C_0^\uparrow T_{2,1}^\uparrow$ is applied, enabling the $\left |\uparrow\right\rangle$ state to tunnel from site 2 to site 1 if and only if site 0 contains an atom in the $\left |\uparrow\right\rangle$ state. Otherwise, the candidate qubit at site 2 remains unaffected, effectively encoding a bit-flipped version of the $\downarrow$ component of the input qubit.
A similar conditional operation $C_0^\uparrow T_{2,3}^\downarrow$ is applied to transfer the $\left |\downarrow\right\rangle$ state from site 3 to site 2, again conditioned on the presence of a $\left |\uparrow\right\rangle$ atom at site 0. This step completes the state transformation necessary to enact the $X$ gate behavior at site 2. 

At this stage, auxiliary atoms located at sites 0, 1, and 3 are no longer needed and are discarded. Only the output atom, located at site 2, carries the desired state. Disentangling the output qubit from the rest requires extra gates, while keeping the output qubit intact. One possible disentangling operation contains a sequence of three primitive gates. First, an $X$ gate is applied on qubit 1, which is then followed by a controlled tunneling gate $C_3^\downarrow T_{0,1}^{\uparrow}$ resulting in the formation of a singlet state in site 0, i.e. $|S_0\rangle = (c_0^{\uparrow} )^\dagger ( c_0^{\downarrow})^\dagger |x\rangle$, and $\left |\downarrow\right\rangle$ state in site 2 or 3. Note that the $X$ gate here does not need to preserve the noise bias, as discussed in Sec.~\ref{sec_general}. Now, applying $T_{1,3}^{\downarrow}(\frac{\pi}{2})$ makes an equally weighted superposition of a single spin-down atom shared between sites 1 and 3. Finally, by performing only a single spin-down measurement in sites 1 or 3, the auxiliary qubits are completely disentangled from the output qubit and can be safely discarded or reused for other computational tasks.

This protocol leverages spatial degrees of freedom in addition to internal states to encode and manipulate quantum states, offering an alternative route toward bias-preserving gate implementation using neutral-atom platforms. Incorporating other degrees of freedom, such as the motional states, can also be used as a resource to implement or enhance the replacement-type gates by coherent conversion between electronic and motional states, see for example Ref.~\cite{Shaw2025}.

\subsubsection{Two-Qubit $\mathrm{CNOT}$ Gate}

The $\mathrm{CNOT}$ gate, a key two-qubit entangling operation, requires a more complex implementation compared to the single-qubit \( X \) gate. Beyond the basic \( X \) gate sequence described earlier, an additional entangling step is needed to ensure that the quantum state of a target qubit is flipped conditionally, based on the state of the control qubit. To achieve this, we leverage the entangled state already present in the \( X \) gate sequence, namely 
$
|\psi_{\mathrm{Bell}}\rangle = \frac{1}{\sqrt{2}}(\alpha_\downarrow\left |\downarrow_1\uparrow_2\right\rangle + \alpha_\uparrow\left |\uparrow_1\downarrow_2\right\rangle)
$
as shown in Figure~\ref{fig_X_rydberg}. We then incorporate conditional tunneling that depends on the state of the target qubit at site 4 and the bit-flipped version of the control qubit at site 2, along with two additional candidate qubits at sites 5 and 6 initialized in the state \( \left |\uparrow_5\downarrow_6\right\rangle \). This functionality is enabled using the two- and three-site occupation-controlled tunneling gates introduced in Sec.~\ref{sec_controlled_tunneling}.

Figure~\ref{fig_X_rydberg}(b) illustrates the key steps of the entangling protocol used to implement a bias-preserving $\mathrm{CNOT}$ gate. The sequence proceeds as follows:  
The first step involves state-dependent tunneling such that an atom in state \( \left |\uparrow\right\rangle \) tunnels between sites 2 and 4 only if the spins on those two sites are anti-aligned, i.e., in configurations \( \left | \downarrow_2\uparrow_4\right\rangle \) or \( \left | \uparrow_2\downarrow_4\right\rangle \). This is achieved by sequentially applying the two-site controlled tunneling gates \( C_{2,4}^{\downarrow, \uparrow} T_{2,4}^{\uparrow} \) and \( C_{2,4}^{\uparrow, \downarrow} T_{2,4}^{\downarrow} \), which respectively enable tunneling of the \( \left |\uparrow\right\rangle \) and \( \left |\downarrow\right\rangle \) states from site 4 to site 2 under the corresponding anti-aligned spin conditions.

Now, the three-site gate $C_{2,4}^{\neg(\downarrow, \downarrow)} T_{4,5}^\uparrow$ is applied to move the  \( \left |\uparrow\right\rangle \) state from site 5 to site 4, but only if the atom pair at sites 2 and 4 is not in the state  \( \left |\downarrow_2\downarrow_4\right\rangle \). 
The gate $C_{5,6}^{\neg(\uparrow, \downarrow)} T_{5,6}^\downarrow$ is applied next. This enables tunneling of the $|0\rangle$ ($\downarrow$) state from site 6 to site 5, conditioned on the pair of atoms at those sites not being in the configuration \( \left |\uparrow_5\downarrow_6\right\rangle \). As a result, this operation modifies sites 5 as the output state of the target qubit, enabling completion of state transformation required for the $\mathrm{CNOT}$ operation, i.e. $\left |q_c^{(1)} q_t^{(5)} \right\rangle = \mathrm{CNOT}\left |q_c^{(1)} q_t^{(4)} \right\rangle$.
After the operation has been performed, atoms located at auxiliary sites 0, 2, 3, 4 and 6 are no longer needed and are discarded similarly to the disentagling operation for the $X$ protocol.

\subsection{Error Mechanisms\label{sec_atoms_errors}}

To maintain a strong noise bias during quantum operations, replacement-type gates are designed to predominantly conserve the nature of the dominant error channels. However, achieving robust suppression of undesired error processes also requires careful qubit engineering at the hardware level. In this work, we addressed this by utilizing stretched nuclear spin states of neutral atom qubits with a large nuclear spin. Specifically, the qubit states are chosen to reside at the extreme ends of the magnetic quantum number spectrum in the ground-state manifold. Due to angular momentum selection rules, this configuration inherently forbids direct transitions between the two qubit states, providing intrinsic protection against unwanted couplings.

This advantageous property can extend to the metastable and Rydberg manifolds as well. In the metastable manifold, where atoms are temporarily excited for transport, only one of the qubit states is mainly affected due to the state-selective nature of the process. Consequently, imperfections in transport tend to primarily impact a single computational basis state, which can help maintain noise asymmetry. Similarly, in the Rydberg manifold, used for implementing entangling gates, spontaneous emission events are constrained by selection rules to decay within the same stretched subspace (i.e., conserving the sign of $M_I$). This can lead to a preferential preservation of the original error bias, even in the presence of decoherence mechanisms such as photon scattering.

However, not all operations necessarily preserve this asymmetry. For instance, imperfections in the pick-and-drop process during storage-to-transport conversion, or phase errors in the entangling components of conditional tunneling gates, can introduce unwanted couplings between candidate qubits, ultimately leading to bias-breaking errors. Mitigating such effects requires these operations to be implemented with high fidelity. Although the system exhibits characteristics that are supportive of bias-preserving protocols, this compatibility is not absolute and depends critically on specific implementation details, which lie beyond the scope of this work.

On top of Pauli errors, the leakage error in neutral-atom platforms could substantially degrade the performance of quantum computational tasks, even if it is not dominant. This error could originate from atom loss or any leakage out of the computational subspace. Here, we outline several mitigation techniques that have been verified in experiments. In particular, state-selective imaging is a well-established technique that can be used to infer atom loss and leakage \cite{Wu2019, Norica2023}. In this approach, by leveraging the internal structure of the atom, both qubit states undergo independent imaging, hence detecting the leftover population out of the computational subspace, which in turn can be used to discard faulty simulation runs.  
Additionally, such measurements with single-site resolution allow for detecting the location of the leakage error, also known as leakage-to-erasure conversion \cite{Scholl2023, Ma2023}. This enhances the gate performance conditioned on applying erasure checks at each step of our gate protocols. Note that the erasure conversion can also be done at a circuit level by transferring information about whether the qubit is present onto an auxiliary atom, e.g. see \cite{Chow2024}.

Accurately simulating the performance of these gates requires detailed noise models and hardware specifications backed by yet unavailable experimental data. As a rough estimate, we perform a back-of-the-envelope calculation of their fidelities. We assume an infidelity of $10^{-3}$ for entangling gates and measurements, and $10^{-4}$ for single-site rotations and atomic movements, values that are challenging but potentially achievable in the near future. Under these assumptions, the tunneling gate achieves an estimated fidelity of $\sim 99.95\%$, and the occupation-independent tunneling gate $\sim 99.78\%$. Consequently, the overall fidelities of the noise-biased $X$ gate and the $\mathrm{CNOT}$ gate, including the disentangling operations, are expected to reach approximately $98.89\%$ and $96.27\%$, respectively. In this estimate, we conservatively assume a worst-case scenario in which the disentangling operation requires the same gate resources as the entangling step. 

Another key characteristic is the gate duration. We conservatively assume $\SI{600}{\micro\second}$ for the tunneling gate and $200$–$\SI{500}{\nano\second}$ for the entangling gates, making the overall timescales of replacement-type gates primarily limited by tunneling operations: approximately $\SI{4.8}{\milli\second}$ for the $X$ gate and roughly three times that for the $\mathrm{CNOT}$ gate. While the measurement process could also take a few milliseconds, we exclude the measurement time associated with the disentangling operation from these estimates, as it can be parallelized with the subsequent Clifford gates. Although the duration of the replacement-type gates for the proposed scheme with neutral atoms is relatively long, they remain well below the coherence time of nuclear-spin qubits in neutral atoms. Moreover, since the outputs of these gates are encoded on qubits different from the inputs, discarded atoms can be replaced with fresh ones, enabling coherent operations that extend beyond the coherence time of individual qubits. Recent experiments have successfully demonstrated such coherent atom replacement in large arrays \cite{Chiu2025,Li2025,Muniz2025}.

\section{Replacement-Type Gates in Practice\label{sec_practice}}

Based on the properties of the hardware implementations of replacement-type gates in Secs.~\ref{sec_spins}, \ref{sec_atoms}, in this Section we outline advantageous use cases. Compared to conventional gates, replacement-type gates can be expected to have a lower fidelity, require an overhead in qubit number, and may have a longer duration, especially in the case of neutral atoms. Thus, at first glance, our proposed gate may appear inferior to established techniques. For NISQ applications, where a small computation is performed within the lifetime of physical qubits and only few qubits are available, replacement-type gates are not favorable.

However, their noise-bias-preserving quality can outweigh these disadvantages in QEC. On a platform whose native two-qubit gate is a noise-biased $\mathrm{CPhase}$ gate, like spin or neutral atom qubits, one promising option for QEC is the concatenation of two codes $C_1$ and $C_2$~\cite{PhysRevA.78.052331}. The classical inner code $C_1$ protects against the dominant dephasing. The resulting weaker, more balanced noise can be efficiently suppressed by the outer code $C_2$. However, to maintain the protection of $C_1$, the encoded gates and syndrome measurements in $C_2$ must be constructed from the fundamentally noise-biased physical operations.

A gadget for a $\mathrm{CNOT}$ gate protected by $C_1$ is realized using a variant of state teleportation~\cite{PhysRevA.78.052331}. This method requires two auxiliary logical qubits and several stabilizer measurements. The measurements should be repeated $r$ times with fresh measurement qubits for fault tolerance. If $C_1$ is a repetition code of distance $d$, additional $2d + 4r$ qubits are required, and $21r$ $\mathrm{CZ}$ gates and $4r$ qubit readouts, with a depth of $18 + 3(r-1)$ $\mathrm{CZ}$ gates and at least one measurement. Additionally, layers of shuttling are likely to be necessary to achieve the required connectivity.

This massive overhead contrasts with a transversal noise-bias-preserving replacement-type $\mathrm{CNOT}$ -- $d$ pairwise $\mathrm{CNOT}$s in parallel -- relying on $6d$ additional qubits. Serializing the $\mathrm{CNOT}$ gates allows to reduce the overhead in qubit number at the cost of a deeper circuit because the auxiliary qubits of the replacement-type gate can be reset and reused. Thus, depending on $d$ and $r$, the replacement-type gate can be expected to require fewer qubits and have a higher total fidelity. Depending on the amounts of serialization and shuttling, transversal replacement-type gates can be faster than the standard gadget.

Another option for QEC with biased-noise qubits are Clifford-deformed topological codes such as the XZZX surface code~\cite{BonillaAtaides2021}. In an XZZX surface code qubit realized on a square grid of $d\times (d+1)$ physical qubits, the logical failure probability is $\bar P \propto (p / \sqrt{\eta})^{d/2}$, where $p$ is the total error probability and $\eta$ is the factor by which dephasing occurs more frequently than bit errors~\cite{BonillaAtaides2021}. Using replacement-type gates scales the qubit number up by a factor of $k$. The maximal $k = 4$ corresponds to parallel gates on all (pairs of) qubits, it can be reduced by serialization and reuse of auxiliary qubits. Comparing replacement-type gates with error rate $p_r$ and bias $\eta \gg 1$ to non-bias-preserving gates with error rate $p_s$ and $\eta = 1$, we find that replacement-type gates achieve a better protection with fewer physical qubits if $\log(p_s) / \log(p_r / \sqrt{\eta}) \lesssim 1 / \sqrt{k}$.

Thus, despite the overhead of replacement-type gates and their lower fidelity, in the context of QEC they can allow for the same protection as standard gates with a lower total overhead. We note that with the fidelities predicted here, replacement-type gates would be already close or beyond the break-even point.

Finally, we comment on the serialization of replacement-type gates. Their duration may seem prohibitive. However, as pointed out in Secs.~\ref{sec_spins_qpt} and \ref{sec_atoms_errors}, the effects of motional narrowing and the coherent replacement of qubits can be expected to extend the coherence of the qubits beyond the lifetime of a single, idle qubit. We further emphasize that all other qubits waiting for the completion of a replacement-type gate can be subjected to a suitable DD protocol. To this end it may be beneficial to combine noise-bias-preserving replacement-type two-qubit gates with conventional high-fidelity single qubit gates.

\section{Conclusion and Outlook\label{sec_conclusion}}

In this work, we introduced the paradigm of replacement-type gates, which are designed for manipulating the qubit state in an extended Hilbert space instead of Bloch sphere rotations, therefore facilitating approximate noise bias preservation. Replacement-type gates are an option for qubits encoded in the internal states of quantum particles and rely on conditionally replacing the input particles with a superposition of candidate particles at an output site. We discussed the abstract concept of the two constituents of a replacement-type gate: a rearrangement operation which entangles the state of the input qubits with the configuration of a register of mobile candidate qubits, and a disentangling operation, which discards the original input qubits and all candidate qubits not in dedicated output sites by appropriate measurements.

As a concrete example, we presented protocols for replacement-type gates for Loss--DiVincenzo spin qubits, built from simple primitives which are either experimentally demonstrated or close to available techniques. The recent demonstration of charge-driven logic in spin qubits is a promising first step towards a replacement-type gates~\cite{Jones2025}. Through a numerical evaluation we confirm the validity of the rearrangement operation, we find that the gates are strongly noise biased towards phase errors, and we identify the remaining sources of bit errors. The fidelity estimated with typical experimental parameters is lower than for comparable conventional gates, but our promising results warrant future theoretical or experimental studies aiming to find the optimal range of microscopic parameters, such as the magnetic gradient, or to optimize the shape of the detuning ramps.

In fact, even with parameters reported in recent publications we find that the replacement-type gate is close to the threshold of certain QEC codes which exploit the noise bias, like the XZZX surface code~\cite{BonillaAtaides2021} and have a duration comparable to conventional gates. Furthermore, the intrinsic noise bias preservation of the replacement-type quantum gate can make the preparation, distillation and injection of magic states for fault-tolerant quantum computing obsolete and thus remove a significant overhead in qubit numbers. The replacement-type gate itself has a dynamic overhead in qubit numbers since a limited number of auxiliary qubits is required only during the gate and can be reused afterwards. Especially for shuttling-based architectures relying on dedicated interaction zones, replacement-type gates are therefore an interesting option for simplifying QEC. Considering the extremely high fidelity of conventional single qubit gates~\cite{Wu2025highfidelitygate} the combination of near-perfect conventional single qubit gates and strongly noise-biased replacement-type two-qubit gates may prove optimal.

To demonstrate the versatility of the replacement-type gate paradigm, we also develop an example realization for neutral atom qubits in optical tweezers. The discussion of the error sources suggests that a strong noise bias can be expected in this case as well, although some of the required operations here are currently hypothetical and no experimental reference is available. The formulation of an accurate error model and the assessment of gate fidelity and noise bias should be addressed in further research.

Finally, we study the balance between the overhead required for approximately noise-bias-preserving replacement-type gates and the overhead for QEC. Even with a lower fidelity, the noise bias can allow for a lower total overhead compared to standard gates without sacrificing protection. This encouraging result points out the possible role of our proposal in future fault-tolerant quantum computing architectures.

In further studies, special attention should be paid to the small but non-vanishing probability of charge errors, which may need tailored QEC codes~\cite{Chow2024,Stricker2020}. Future research should also investigate the interplay of repeated qubit replacement with time-correlated noise, such as $1/f$ charge noise, and the consequences for QEC. Another open topic is the development of protocols for other hardware platforms. Possible candidates that could support replacement-type gates are shuttling-based ion traps, floating electrons on cryogenic substrates~\cite{PhysRevApplied.20.054022,Jennings2024}, or molecular tweezer arrays~\cite{Cairncross2021, Holland2023, Langen2024}. Additional degrees of freedom of trapped atoms~\cite{Huie2025} or the vibration and rotation of a trapped molecule may allow for additional flexibility or may reduce the overhead. Another beneficial research direction is the design of chip layouts and blueprints that are optimized to accommodate the overhead of initializing and discarding auxiliary qubits within the limitations of specific hardware platforms.

%%%%%%%%%%%%%%%%%%%%%%%%%%%%%%%%%%%%%%%%%%%%%%%%%%%%%%%%%%%%%%%%%
%%%%%%%%%%%%%%%%%%%%%%%% Acknowledgments %%%%%%%%%%%%%%%%%%%%%%%%%%%%%
%%%%%%%%%%%%%%%%%%%%%%%%%%%%%%%%%%%%%%%%%%%%%%%%%%%%%%%%%%%%%%%%%

\begin{acknowledgments}
We thank Anette Messinger, Michael Schuler, Federico Dominguez, Philipp Aumann, Frederik Lohof and Christophe Goeller for helpful discussions.
This study was supported by the Austrian Research Promotion Agency (FFG Project
No. FO999909249, FFG Basisprogramm) as well as funded in part by the Austrian Research Promotion Agency (FFG Project No. 884444, QFTE 2020), NextGenerationEU via FFG and Quantum Austria (FFG
Project No. FO999896208) and the Horizon Europe programme HORIZON-
CL4-2022-QUANTUM-02-SGA via the project 101113690 (PASQuanS2.1). Moreover, this publication has received funding by the Austrian Science Fund (FWF) SFB BeyondC Project No. F7108-N38, as well as funding within the QuantERA II Programme that is supported by the European Union’s Horizon 2020 research and innovation programme under Grant Agreement No. 101017733. For the purpose of open access, the authors have applied a CC BY public copyright license to any Author Accepted Manuscript version arising from this submission. Furthermore, funding is acknowledged by the German Federal Ministry for Education and Research within ATIQ (Project No. 13N16127) and MuniQC-Atoms (Project No. 13N16080).
\end{acknowledgments}

%% ----------------- Bibliography --------------------%%

\bibliography{literature}

@article{PhysRevA.57.120,
  title = {Quantum computation with quantum dots},
  author = {Loss, Daniel and DiVincenzo, David P.},
  journal = {Phys. Rev. A},
  volume = {57},
  issue = {1},
  pages = {120--126},
  numpages = {0},
  year = {1998},
  month = {Jan},
  publisher = {American Physical Society},
  doi = {10.1103/PhysRevA.57.120},
  url = {https://link.aps.org/doi/10.1103/PhysRevA.57.120}
}

@article{RevModPhys.95.025003,
  title = {Semiconductor spin qubits},
  author = {Burkard, Guido and Ladd, Thaddeus D. and Pan, Andrew and Nichol, John M. and Petta, Jason R.},
  journal = {Rev. Mod. Phys.},
  volume = {95},
  issue = {2},
  pages = {025003},
  numpages = {58},
  year = {2023},
  month = {Jun},
  publisher = {American Physical Society},
  doi = {10.1103/RevModPhys.95.025003},
  url = {https://link.aps.org/doi/10.1103/RevModPhys.95.025003}
}

@article{RevModPhys.79.1217,
  title = {Spins in few-electron quantum dots},
  author = {Hanson, R. and Kouwenhoven, L. P. and Petta, J. R. and Tarucha, S. and Vandersypen, L. M. K.},
  journal = {Rev. Mod. Phys.},
  volume = {79},
  issue = {4},
  pages = {1217--1265},
  numpages = {0},
  year = {2007},
  month = {Oct},
  publisher = {American Physical Society},
  doi = {10.1103/RevModPhys.79.1217},
  url = {https://link.aps.org/doi/10.1103/RevModPhys.79.1217}
}

@article{RevModPhys.85.961,
  title = {Silicon quantum electronics},
  author = {Zwanenburg, Floris A. and Dzurak, Andrew S. and Morello, Andrea and Simmons, Michelle Y. and Hollenberg, Lloyd C. L. and Klimeck, Gerhard and Rogge, Sven and Coppersmith, Susan N. and Eriksson, Mark A.},
  journal = {Rev. Mod. Phys.},
  volume = {85},
  issue = {3},
  pages = {961--1019},
  numpages = {0},
  year = {2013},
  month = {Jul},
  publisher = {American Physical Society},
  doi = {10.1103/RevModPhys.85.961},
  url = {https://link.aps.org/doi/10.1103/RevModPhys.85.961}
}

@article{Taylor2005,
	author = {Taylor, J. M. and Engel, H. -A. and D{\"u}r, W. and Yacoby, A. and Marcus, C. M. and Zoller, P. and Lukin, M. D.},
	date = {2005/12/01},
	doi = {10.1038/nphys174},
	isbn = {1745-2481},
	journal = {Nat. Phys.},
	number = {3},
	pages = {177--183},
	title = {Fault-tolerant architecture for quantum computation using electrically controlled semiconductor spins},
	url = {https://doi.org/10.1038/nphys174},
	volume = {1},
	year = {2005},
}

@article{Noiri2022,
	author = {Noiri, Akito and Takeda, Kenta and Nakajima, Takashi and Kobayashi, Takashi and Sammak, Amir and Scappucci, Giordano and Tarucha, Seigo},
	date = {2022/09/30},
	doi = {10.1038/s41467-022-33453-z},
	isbn = {2041-1723},
	journal = {Nat. Commun.},
	number = {1},
	pages = {5740},
	title = {A shuttling-based two-qubit logic gate for linking distant silicon quantum processors},
	url = {https://doi.org/10.1038/s41467-022-33453-z},
	volume = {13},
	year = {2022},
}

@article{Kuenne2024,
	author = {K{\"u}nne, Matthias and Willmes, Alexander and Oberl{\"a}nder, Max and Gorjaew, Christian and Teske, Julian D. and Bhardwaj, Harsh and Beer, Max and Kammerloher, Eugen and Otten, Ren{\'e} and Seidler, Inga and Xue, Ran and Schreiber, Lars R. and Bluhm, Hendrik},
	date = {2024/06/11},
	doi = {10.1038/s41467-024-49182-4},
	isbn = {2041-1723},
	journal = {Nat. Communs.},
	number = {1},
	pages = {4977},
	title = {The SpinBus architecture for scaling spin qubits with electron shuttling},
	url = {https://doi.org/10.1038/s41467-024-49182-4},
	volume = {15},
	year = {2024},
}

@article{PhysRevB.110.075302,
  title = {Scalable parity architecture with a shuttling-based spin qubit processor},
  author = {Ginzel, Florian and Fellner, Michael and Ertler, Christian and Schreiber, Lars R. and Bluhm, Hendrik and Lechner, Wolfgang},
  journal = {Phys. Rev. B},
  volume = {110},
  issue = {7},
  pages = {075302},
  numpages = {16},
  year = {2024},
  month = {Aug},
  publisher = {American Physical Society},
  doi = {10.1103/PhysRevB.110.075302},
  url = {https://link.aps.org/doi/10.1103/PhysRevB.110.075302}
}

@article{Koppens2006,
	author = {Koppens, F. H. L. and Buizert, C. and Tielrooij, K. J. and Vink, I. T. and Nowack, K. C. and Meunier, T. and Kouwenhoven, L. P. and Vandersypen, L. M. K.},
	date = {2006/08/01},
	doi = {10.1038/nature05065},
	isbn = {1476-4687},
	journal = {Nature},
	number = {7104},
	pages = {766--771},
	title = {Driven coherent oscillations of a single electron spin in a quantum dot},
	url = {https://doi.org/10.1038/nature05065},
	volume = {442},
	year = {2006},
}

@article{Pioro-Ladriere2008,
	author = {Pioro-Ladri{\`e}re, M. and Obata, T. and Tokura, Y. and Shin, Y. -S. and Kubo, T. and Yoshida, K. and Taniyama, T. and Tarucha, S.},
	date = {2008/10/01},
	doi = {10.1038/nphys1053},
	isbn = {1745-2481},
	journal = {Nat. Phys.},
	number = {10},
	pages = {776--779},
	title = {Electrically driven single-electron spin resonance in a slanting Zeeman field},
	url = {https://doi.org/10.1038/nphys1053},
	volume = {4},
	year = {2008},
}

@article{PhysRevB.74.165319,
  title = {Electric-dipole-induced spin resonance in quantum dots},
  author = {Golovach, Vitaly N. and Borhani, Massoud and Loss, Daniel},
  journal = {Phys. Rev. B},
  volume = {74},
  issue = {16},
  pages = {165319},
  numpages = {10},
  year = {2006},
  month = {Oct},
  publisher = {American Physical Society},
  doi = {10.1103/PhysRevB.74.165319},
  url = {https://link.aps.org/doi/10.1103/PhysRevB.74.165319}
}

@article{Petta2005,
author = {J. R. Petta  and A. C. Johnson  and J. M. Taylor  and E. A. Laird  and A. Yacoby  and M. D. Lukin  and C. M. Marcus  and M. P. Hanson  and A. C. Gossard},
title = {Coherent Manipulation of Coupled Electron Spins in Semiconductor Quantum Dots},
journal = {Science},
volume = {309},
number = {5744},
pages = {2180-2184},
year = {2005},
doi = {10.1126/science.1116955},
URL = {https://www.science.org/doi/abs/10.1126/science.1116955},
eprint = {https://www.science.org/doi/pdf/10.1126/science.1116955},
}

@article{PhysRevLett.103.160503,
  title = {Rapid Single-Shot Measurement of a Singlet-Triplet Qubit},
  author = {Barthel, C. and Reilly, D. J. and Marcus, C. M. and Hanson, M. P. and Gossard, A. C.},
  journal = {Phys. Rev. Lett.},
  volume = {103},
  issue = {16},
  pages = {160503},
  numpages = {4},
  year = {2009},
  month = {Oct},
  publisher = {American Physical Society},
  doi = {10.1103/PhysRevLett.103.160503},
  url = {https://link.aps.org/doi/10.1103/PhysRevLett.103.160503}
}

@article{PhysRevB.102.195418,
  title = {Spin shuttling in a silicon double quantum dot},
  author = {Ginzel, Florian and Mills, Adam R. and Petta, Jason R. and Burkard, Guido},
  journal = {Phys. Rev. B},
  volume = {102},
  issue = {19},
  pages = {195418},
  numpages = {13},
  year = {2020},
  month = {Nov},
  publisher = {American Physical Society},
  doi = {10.1103/PhysRevB.102.195418},
  url = {https://link.aps.org/doi/10.1103/PhysRevB.102.195418}
}

@article{PRXQuantum.4.020305,
  title = {Blueprint of a Scalable Spin Qubit Shuttle Device for Coherent Mid-Range Qubit Transfer in Disordered ${\text{Si/SiGe/SiO}}_{2}$},
  author = {Langrock, Veit and Krzywda, Jan A. and Focke, Niels and Seidler, Inga and Schreiber, Lars R. and Cywi\ifmmode \acute{n}\else \'{n}\fi{}ski, \L{}ukasz},
  journal = {PRX Quantum},
  volume = {4},
  issue = {2},
  pages = {020305},
  numpages = {35},
  year = {2023},
  month = {Apr},
  publisher = {American Physical Society},
  doi = {10.1103/PRXQuantum.4.020305},
  url = {https://link.aps.org/doi/10.1103/PRXQuantum.4.020305}
}

@article{Struck2023,
	author = {Struck, Tom and Volmer, Mats and Visser, Lino and Offermann, Tobias and Xue, Ran and Tu, Jhih-Sian and Trellenkamp, Stefan and Cywi{\'n}ski, {\L}ukasz and Bluhm, Hendrik and Schreiber, Lars R.},
	date = {2024/02/13},
	doi = {10.1038/s41467-024-45583-7},
	journal = {Nat. Commun.},
	number = {1},
	pages = {1325},
	title = {Spin-EPR-pair separation by conveyor-mode single electron shuttling in {Si/SiGe}},
	url = {https://doi.org/10.1038/s41467-024-45583-7},
	volume = {15},
	year = {2024},
}

@article{PRXQuantum.2.030331,
  title = {Enhanced Spin Coherence while Displacing Electron in a Two-Dimensional Array of Quantum Dots},
  author = {Mortemousque, Pierre-Andr\'e and Jadot, Baptiste and Chanrion, Emmanuel and Thiney, Vivien and B\"auerle, Christopher and Ludwig, Arne and Wieck, Andreas D. and Urdampilleta, Matias and Meunier, Tristan},
  journal = {PRX Quantum},
  volume = {2},
  issue = {3},
  pages = {030331},
  numpages = {12},
  year = {2021},
  month = {Aug},
  publisher = {American Physical Society},
  doi = {10.1103/PRXQuantum.2.030331},
  url = {https://link.aps.org/doi/10.1103/PRXQuantum.2.030331}
}

@article{vanDiepen2021,
	author = {van Diepen, Cornelis J. and Hsiao, Tzu-Kan and Mukhopadhyay, Uditendu and Reichl, Christian and Wegscheider, Werner and Vandersypen, Lieven M. K.},
	date = {2021/01/04},
	doi = {10.1038/s41467-020-20388-6},
	isbn = {2041-1723},
	journal = {Nat. Commun.},
	number = {1},
	pages = {77},
	title = {Electron cascade for distant spin readout},
	url = {https://doi.org/10.1038/s41467-020-20388-6},
	volume = {12},
	year = {2021},
}

@misc{Fattal2025,
      title={Radio frequency single electron transmission spectroscopy of a semiconductor
{S}i/{S}i{G}e quantum dot}, 
      author={I. Fattal and J. Van Damme and B. Raes and C. Godfrin and G. Jaliel and K. Chen and T. VanCaekenberghe and A. Loenders and S. Kubicek and S. Massar and Y. Canvel and J. Jussot and Y. Shimura and R. Loo and D. Wan and M. Mongillo and and K. De Greve},
      year={2025},
      eprint={2504.05016},
      archivePrefix={arXiv},
      primaryClass={cond-mat.mes-hall}
}

@article{Connors2022,
	author = {Connors, Elliot J. and Nelson, J. and Edge, Lisa F. and Nichol, John M.},
	date = {2022/02/17},
	doi = {10.1038/s41467-022-28519-x},
	isbn = {2041-1723},
	journal = {Nat. Commun.},
	number = {1},
	pages = {940},
	title = {Charge-noise spectroscopy of Si/SiGe quantum dots via dynamically-decoupled exchange oscillations},
	url = {https://doi.org/10.1038/s41467-022-28519-x},
	volume = {13},
	year = {2022},
}

@article{PhysRevApplied.13.024019,
  title = {Rapid High-Fidelity Spin-State Readout in $\mathrm{Si}$/$\mathrm{Si}$-$\mathrm{Ge}$ Quantum Dots via rf Reflectometry},
  author = {Connors, Elliot J. and Nelson, JJ and Nichol, John M.},
  journal = {Phys. Rev. Appl.},
  volume = {13},
  issue = {2},
  pages = {024019},
  numpages = {9},
  year = {2020},
  month = {Feb},
  publisher = {American Physical Society},
  doi = {10.1103/PhysRevApplied.13.024019},
  url = {https://link.aps.org/doi/10.1103/PhysRevApplied.13.024019}
}

@article{Takeda2016,
author = {Kenta Takeda  and Jun Kamioka  and Tomohiro Otsuka  and Jun Yoneda  and Takashi Nakajima  and Matthieu R. Delbecq  and Shinichi Amaha  and Giles Allison  and Tetsuo Kodera  and Shunri Oda  and Seigo Tarucha },
title = {A fault-tolerant addressable spin qubit in a natural silicon quantum dot},
journal = {Sci. Adv.},
volume = {2},
number = {8},
pages = {e1600694},
year = {2016},
doi = {10.1126/sciadv.1600694},
URL = {https://www.science.org/doi/abs/10.1126/sciadv.1600694},
eprint = {https://www.science.org/doi/pdf/10.1126/sciadv.1600694},
}

@article{PhysRevLett.110.046805,
  title = {Dispersive Readout of a Few-Electron Double Quantum Dot with Fast rf Gate Sensors},
  author = {Colless, J. I. and Mahoney, A. C. and Hornibrook, J. M. and Doherty, A. C. and Lu, H. and Gossard, A. C. and Reilly, D. J.},
  journal = {Phys. Rev. Lett.},
  volume = {110},
  issue = {4},
  pages = {046805},
  numpages = {5},
  year = {2013},
  month = {Jan},
  publisher = {American Physical Society},
  doi = {10.1103/PhysRevLett.110.046805},
  url = {https://link.aps.org/doi/10.1103/PhysRevLett.110.046805}
}

@article{Fujisawa2002,
	author = {Fujisawa, Toshimasa and Austing, David Guy and Tokura, Yasuhiro and Hirayama, Yoshiro and Tarucha, Seigo},
	date = {2002/09/01},
	doi = {10.1038/nature00976},
	isbn = {1476-4687},
	journal = {Nature},
	number = {6904},
	pages = {278--281},
	title = {Allowed and forbidden transitions in artificial hydrogen and helium atoms},
	url = {https://doi.org/10.1038/nature00976},
	volume = {419},
	year = {2002},
}

@article{PhysRevB.80.041301,
  title = {Pauli spin blockade in the presence of strong spin-orbit coupling},
  author = {Danon, J. and Nazarov, Yu. V.},
  journal = {Phys. Rev. B},
  volume = {80},
  issue = {4},
  pages = {041301},
  numpages = {4},
  year = {2009},
  month = {Jul},
  publisher = {American Physical Society},
  doi = {10.1103/PhysRevB.80.041301},
  url = {https://link.aps.org/doi/10.1103/PhysRevB.80.041301}
}

@article{PhysRevB.82.155312,
  title = {Quantum dot spin qubits in silicon: Multivalley physics},
  author = {Culcer, Dimitrie and Cywi\ifmmode \acute{n}\else \'{n}\fi{}ski, \L{}ukasz and Li, Qiuzi and Hu, Xuedong and Das Sarma, S.},
  journal = {Phys. Rev. B},
  volume = {82},
  issue = {15},
  pages = {155312},
  numpages = {20},
  year = {2010},
  month = {Oct},
  publisher = {American Physical Society},
  doi = {10.1103/PhysRevB.82.155312},
  url = {https://link.aps.org/doi/10.1103/PhysRevB.82.155312}
}

@article{PhysRevB.82.155424,
  title = {Spin-valley blockade in carbon nanotube double quantum dots},
  author = {P\'alyi, Andr\'as and Burkard, Guido},
  journal = {Phys. Rev. B},
  volume = {82},
  issue = {15},
  pages = {155424},
  numpages = {14},
  year = {2010},
  month = {Oct},
  publisher = {American Physical Society},
  doi = {10.1103/PhysRevB.82.155424},
  url = {https://link.aps.org/doi/10.1103/PhysRevB.82.155424}
}

@article{PhysRevB.80.201404,
  title = {Hyperfine-induced valley mixing and the spin-valley blockade in carbon-based quantum dots},
  author = {P\'alyi, Andr\'as and Burkard, Guido},
  journal = {Phys. Rev. B},
  volume = {80},
  issue = {20},
  pages = {201404},
  numpages = {4},
  year = {2009},
  month = {Nov},
  publisher = {American Physical Society},
  doi = {10.1103/PhysRevB.80.201404},
  url = {https://link.aps.org/doi/10.1103/PhysRevB.80.201404}
}

@article{Wittig2005,
	author = {Wittig, Curt},
	date = {2005/05/01},
	doi = {10.1021/jp040627u},
	journal = {J. Phys. Chem. B},
	month = {05},
	number = {17},
	pages = {8428--8430},
	publisher = {American Chemical Society},
	title = {The Landau-Zener Formula},
	volume = {109},
	year = {2005},
	url = {https://pubs.acs.org/doi/10.1021/jp040627u},
}

@article{PhysRevA.53.4288,
  title = {Landau-Zener model: Effects of finite coupling duration},
  author = {Vitanov, N. V. and Garraway, B. M.},
  journal = {Phys. Rev. A},
  volume = {53},
  issue = {6},
  pages = {4288--4304},
  numpages = {0},
  year = {1996},
  month = {Jun},
  publisher = {American Physical Society},
  doi = {10.1103/PhysRevA.53.4288},
  url = {https://link.aps.org/doi/10.1103/PhysRevA.53.4288}
}

@article{PhysRevApplied.18.054090,
  title = {Quantum Control of Hole Spin Qubits in Double Quantum Dots},
  author = {Fern\'andez-Fern\'andez, D. and Ban, Yue and Platero, G.},
  journal = {Phys. Rev. Appl.},
  volume = {18},
  issue = {5},
  pages = {054090},
  numpages = {11},
  year = {2022},
  month = {Nov},
  publisher = {American Physical Society},
  doi = {10.1103/PhysRevApplied.18.054090},
  url = {https://link.aps.org/doi/10.1103/PhysRevApplied.18.054090}
}

@misc{Ventura2024,
      title={Quantum geometric protocols for fast high-fidelity adiabatic state transfer}, 
      author={Christian Ventura Meinersen and Stefano Bosco and Maximilian Rimbach-Russ},
      year={2024},
      eprint={2409.03084},
      archivePrefix={arXiv},
      primaryClass={quant-ph}
}

@article{PhysRevB.101.035303,
  title = {Adiabatic electron charge transfer between two quantum dots in presence of $1/f$ noise},
  author = {Krzywda, Jan A. and Cywi\ifmmode \acute{n}\else \'{n}\fi{}ski, \L{}ukasz},
  journal = {Phys. Rev. B},
  volume = {101},
  issue = {3},
  pages = {035303},
  numpages = {12},
  year = {2020},
  month = {Jan},
  publisher = {American Physical Society},
  doi = {10.1103/PhysRevB.101.035303},
  url = {https://link.aps.org/doi/10.1103/PhysRevB.101.035303}
}

@article{PhysRevB.104.075439,
  title = {Interplay of charge noise and coupling to phonons in adiabatic electron transfer between quantum dots},
  author = {Krzywda, Jan A. and Cywi\ifmmode \acute{n}\else \'{n}\fi{}ski, \L{}ukasz},
  journal = {Phys. Rev. B},
  volume = {104},
  issue = {7},
  pages = {075439},
  numpages = {21},
  year = {2021},
  month = {Aug},
  publisher = {American Physical Society},
  doi = {10.1103/PhysRevB.104.075439},
  url = {https://link.aps.org/doi/10.1103/PhysRevB.104.075439}
}

@misc{Stehouwer2025,
      title={Engineering {G}e profiles in {S}i/{S}i{G}e heterostructures for increased valley splitting}, 
      author={Lucas E. A. Stehouwer and Merrit P. Losert and Maia Rigot and Davide Degli Esposti and Sara Martí-Sánchez and Maximillian Rimbach-Russ and Jordi Arbiol and Mark Friesen and Giordano Scappucci},
      year={2025},
      eprint={2505.22295},
      archivePrefix={arXiv},
      primaryClass={cond-mat.mes-hall}
}

@article{Lima_2023,
doi = {10.1088/2633-4356/acd743},
url = {https://dx.doi.org/10.1088/2633-4356/acd743},
year = {2023},
month = {may},
publisher = {IOP Publishing},
volume = {3},
number = {2},
pages = {025004},
author = {Lima, Jonas R F and Burkard, Guido},
title = {Interface and electromagnetic effects in the valley splitting of Si quantum dots},
journal = {Mater. quantum technol.}}

@article{PhysRevB.104.085406,
  title = {Valley splittings in {S}i/{S}i{G}e quantum dots with a germanium spike in the silicon well},
  author = {McJunkin, Thomas and MacQuarrie, E. R. and Tom, Leah and Neyens, S. F. and Dodson, J. P. and Thorgrimsson, Brandur and Corrigan, J. and Ercan, H. Ekmel and Savage, D. E. and Lagally, M. G. and Joynt, Robert and Coppersmith, S. N. and Friesen, Mark and Eriksson, M. A.},
  journal = {Phys. Rev. B},
  volume = {104},
  issue = {8},
  pages = {085406},
  numpages = {11},
  year = {2021},
  month = {Aug},
  publisher = {American Physical Society},
  doi = {10.1103/PhysRevB.104.085406},
  url = {https://link.aps.org/doi/10.1103/PhysRevB.104.085406}
}

@article{McJunkin2022,
	author = {McJunkin, Thomas and Harpt, Benjamin and Feng, Yi and Losert, Merritt P. and Rahman, Rajib and Dodson, J. P. and Wolfe, M. A. and Savage, D. E. and Lagally, M. G. and Coppersmith, S. N. and Friesen, Mark and Joynt, Robert and Eriksson, M. A.},
	date = {2022/12/15},
	doi = {10.1038/s41467-022-35510-z},
	journal = {Nat. Commun.},
	number = {1},
	pages = {7777},
	title = {SiGe quantum wells with oscillating Ge concentrations for quantum dot qubits},
	url = {https://doi.org/10.1038/s41467-022-35510-z},
	volume = {13},
	year = {2022}
}

@article{PhysRevB.106.085304,
  title = {Enhanced valley splitting in {S}i layers with oscillatory {G}e concentration},
  author = {Feng, Yi and Joynt, Robert},
  journal = {Phys. Rev. B},
  volume = {106},
  issue = {8},
  pages = {085304},
  numpages = {11},
  year = {2022},
  month = {Aug},
  publisher = {American Physical Society},
  doi = {10.1103/PhysRevB.106.085304},
  url = {https://link.aps.org/doi/10.1103/PhysRevB.106.085304}
}

@article{PaqueletWuetz2022,
	author = {Paquelet Wuetz, Brian and Losert, Merritt P. and Koelling, Sebastian and Stehouwer, Lucas E. A. and Zwerver, Anne-Marije J. and Philips, Stephan G. J. and M{\k a}dzik, Mateusz T. and Xue, Xiao and Zheng, Guoji and Lodari, Mario and Amitonov, Sergey V. and Samkharadze, Nodar and Sammak, Amir and Vandersypen, Lieven M. K. and Rahman, Rajib and Coppersmith, Susan N. and Moutanabbir, Oussama and Friesen, Mark and Scappucci, Giordano},
	date = {2022/12/13},
	doi = {10.1038/s41467-022-35458-0},
	journal = {Nat. Commun.},
	number = {1},
	pages = {7730},
	title = {Atomic fluctuations lifting the energy degeneracy in {S}i/{S}i{G}e quantum dots},
	url = {https://doi.org/10.1038/s41467-022-35458-0},
	volume = {13},
	year = {2022},
}

@article{PhysRevLett.110.086804,
  title = {Coherent Adiabatic Spin Control in the Presence of Charge Noise Using Tailored Pulses},
  author = {Ribeiro, Hugo and Burkard, Guido and Petta, J. R. and Lu, H. and Gossard, A. C.},
  journal = {Phys. Rev. Lett.},
  volume = {110},
  issue = {8},
  pages = {086804},
  numpages = {5},
  year = {2013},
  month = {Feb},
  publisher = {American Physical Society},
  doi = {10.1103/PhysRevLett.110.086804},
  url = {https://link.aps.org/doi/10.1103/PhysRevLett.110.086804}
}

@article{Takeda2024,
	author = {Takeda, Kenta and Noiri, Akito and Nakajima, Takashi and Camenzind, Leon C. and Kobayashi, Takashi and Sammak, Amir and Scappucci, Giordano and Tarucha, Seigo},
	date = {2024/02/13},
	doi = {10.1038/s41534-024-00813-0},
	journal = {npj Quant. Inf.},
	number = {1},
	pages = {22},
	title = {Rapid single-shot parity spin readout in a silicon double quantum dot with fidelity exceeding 99{\%}},
	url = {https://doi.org/10.1038/s41534-024-00813-0},
	volume = {10},
	year = {2024},
}

@article{PhysRevB.111.115305,
  title = {Decoherence of electron spin qubit during transfer between two semiconductor quantum dots at low magnetic fields},
  author = {Krzywda, Jan A. and Cywi\ifmmode \acute{n}\else \'{n}\fi{}ski, \L{}ukasz},
  journal = {Phys. Rev. B},
  volume = {111},
  issue = {11},
  pages = {115305},
  numpages = {21},
  year = {2025},
  month = {Mar},
  publisher = {American Physical Society},
  doi = {10.1103/PhysRevB.111.115305},
  url = {https://link.aps.org/doi/10.1103/PhysRevB.111.115305}
}

@article{PhysRevB.102.125406,
  title = {Simulated coherent electron shuttling in silicon quantum dots},
  author = {Buonacorsi, Brandon and Shaw, Benjamin and Baugh, Jonathan},
  journal = {Phys. Rev. B},
  volume = {102},
  issue = {12},
  pages = {125406},
  numpages = {11},
  year = {2020},
  month = {Sep},
  publisher = {American Physical Society},
  doi = {10.1103/PhysRevB.102.125406},
  url = {https://link.aps.org/doi/10.1103/PhysRevB.102.125406}
}

@article{PRXQuantum.4.030303,
  title = {Shuttling an Electron Spin through a Silicon Quantum Dot Array},
  author = {Zwerver, A.M.J. and Amitonov, S.V. and de Snoo, S.L. and M\k{a}dzik, M.T. and Rimbach-Russ, M. and Sammak, A. and Scappucci, G. and Vandersypen, L.M.K.},
  journal = {PRX Quantum},
  volume = {4},
  issue = {3},
  pages = {030303},
  numpages = {11},
  year = {2023},
  month = {Jul},
  publisher = {American Physical Society},
  doi = {10.1103/PRXQuantum.4.030303},
  url = {https://link.aps.org/doi/10.1103/PRXQuantum.4.030303}
}

@article{PhysRevB.110.235305,
  title = {Characterization of individual charge fluctuators in {S}i/{S}i{G}e quantum dots},
  author = {Ye, Feiyang and Ellaboudy, Ammar and Albrecht, Dylan and Vudatha, Rohith and Jacobson, N. Tobias and Nichol, John M.},
  journal = {Phys. Rev. B},
  volume = {110},
  issue = {23},
  pages = {235305},
  numpages = {12},
  year = {2024},
  month = {Dec},
  publisher = {American Physical Society},
  doi = {10.1103/PhysRevB.110.235305},
  url = {https://link.aps.org/doi/10.1103/PhysRevB.110.235305}
}

@article{PhysRevB.100.165305,
  title = {Low-frequency charge noise in {S}i/{S}i{G}e quantum dots},
  author = {Connors, Elliot J. and Nelson, JJ and Qiao, Haifeng and Edge, Lisa F. and Nichol, John M.},
  journal = {Phys. Rev. B},
  volume = {100},
  issue = {16},
  pages = {165305},
  numpages = {6},
  year = {2019},
  month = {Oct},
  publisher = {American Physical Society},
  doi = {10.1103/PhysRevB.100.165305},
  url = {https://link.aps.org/doi/10.1103/PhysRevB.100.165305}
}

@misc{Choi2024,
      title={Ballast charges for semiconductor spin qubits}, 
      author={Yujun Choi and John M. Nichol and Edwin Barnes},
      year={2024},
      eprint={2405.14027},
      archivePrefix={arXiv},
      primaryClass={cond-mat.mes-hall},
      url={https://arxiv.org/abs/2405.14027}, 
}

@article{PhysRevA.109.032433,
  title = {Tailoring quantum error correction to spin qubits},
  author = {Het\'enyi, Bence and Wootton, James R.},
  journal = {Phys. Rev. A},
  volume = {109},
  issue = {3},
  pages = {032433},
  numpages = {22},
  year = {2024},
  month = {Mar},
  publisher = {American Physical Society},
  doi = {10.1103/PhysRevA.109.032433},
  url = {https://link.aps.org/doi/10.1103/PhysRevA.109.032433}
}

@misc{Matsumoto2025,
      title={Two-qubit logic and teleportation with mobile spin qubits in silicon}, 
      author={Yuta Matsumoto and Maxim De Smet and Larysa Tryputen and Sander L. de Snoo and Sergey V. Amitonov and Amir Sammak and Maximilian Rimbach-Russ and Giordano Scappucci and Lieven M. K. Vandersypen},
      year={2025},
      eprint={2503.15434},
      archivePrefix={arXiv},
      primaryClass={quant-ph},
      url={https://arxiv.org/abs/2503.15434}, 
}

@article{Mills2022,
    author = {A. R. Mills  and C. R. Guinn  and M. J. Gullans  and A. J. Sigillito  and M. M. Feldman  and E. Nielsen  and J. R. Petta },
    title = {Two-qubit silicon quantum processor with operation fidelity exceeding 99\%},
    journal = {Sci. Adv.},
    volume = {8},
    number = {14},
    pages = {eabn5130},
    year = {2022},
    doi = {10.1126/sciadv.abn5130},
    URL = {https://www.science.org/doi/abs/10.1126/sciadv.abn5130},
}

@article{Xue2022,
	author = {Xue, X. and Russ, M. and Samkharadze, N. and Undseth, B. and Sammak, A. and Scappucci, G. and Vandersypen, L. M. K.},
	date = {2022/01/01},
	doi = {10.1038/s41586-021-04273-w},
	journal = {Nature},
	number = {7893},
	pages = {343--347},
	title = {Quantum logic with spin qubits crossing the surface code threshold},
	url = {https://doi.org/10.1038/s41586-021-04273-w},
	volume = {601},
	year = {2022},
}

@article{PaqueletWuetz2023,
	author = {Paquelet Wuetz, Brian and Degli Esposti, Davide and Zwerver, Anne-Marije J. and Amitonov, Sergey V. and Botifoll, Marc and Arbiol, Jordi and Sammak, Amir and Vandersypen, Lieven M. K. and Russ, Maximilian and Scappucci, Giordano},
	date = {2023/03/13},
	doi = {10.1038/s41467-023-36951-w},
	isbn = {2041-1723},
	journal = {Nat. Commun.},
	number = {1},
	pages = {1385},
	title = {Reducing charge noise in quantum dots by using thin silicon quantum wells},
	url = {https://doi.org/10.1038/s41467-023-36951-w},
	volume = {14},
	year = {2023},
}

@misc{RojasArias2024,
      title={Spatial noise correlations beyond nearest-neighbor in $^{28}${S}i/{S}i{G}e spin qubits}, 
      author={Juan S. Rojas-Arias and Yohei Kojima and Kenta Takeda and Peter Stano, Takashi Nakajima and Jun Yoneda and Akito Noiri and Takashi Kobayashi and Daniel Loss and Seigo Tarucha},
      year={2023},
      eprint={2302.11717},
      archivePrefix={arXiv},
      primaryClass={cond-mat.mes-hall},
      url={https://arxiv.org/abs/2302.11717}, 
}

@article{PhysRevX.13.041015,
  title = {Hotter is Easier: Unexpected Temperature Dependence of Spin Qubit Frequencies},
  author = {Undseth, Brennan and Pietx-Casas, Oriol and Raymenants, Eline and Mehmandoost, Mohammad and M\k{a}dzik, Mateusz T. and Philips, Stephan G. J. and de Snoo, Sander L. and Michalak, David J. and Amitonov, Sergey V. and Tryputen, Larysa and Wuetz, Brian Paquelet and Fezzi, Viviana and Esposti, Davide Degli and Sammak, Amir and Scappucci, Giordano and Vandersypen, Lieven M. K.},
  journal = {Phys. Rev. X},
  volume = {13},
  issue = {4},
  pages = {041015},
  numpages = {18},
  year = {2023},
  month = {Oct},
  publisher = {American Physical Society},
  doi = {10.1103/PhysRevX.13.041015},
  url = {https://link.aps.org/doi/10.1103/PhysRevX.13.041015}
}

@misc{Unseld2024,
      title={Baseband control of single-electron silicon spin qubits in two dimensions}, 
      author={Florian K. Unseld and Brennan Undseth and Eline Raymenants and Yuta Matsumoto and Saurabh Karwal and Oriol Pietx-Casas and Alexander S. Ivlev and Marcel Meyer and Amir Sammak and Menno Veldhorst and Giordano Scappucci and Lieven M. K. Vandersypen},
      year={2024},
      eprint={2412.05171},
      archivePrefix={arXiv},
      primaryClass={cond-mat.mes-hall},
      url={https://arxiv.org/abs/2412.05171}, 
}

@misc{Wu2025highfidelitygate,
      title={Simultaneous High-Fidelity Single-Qubit Gates in a Spin Qubit Array}, 
      author={Yi-Hsien Wu and Leon C. Camenzind and Patrick B{\"u}tler and Ik Kyeong Jin and Akito Noiri and Kenta Takeda and Takashi Nakajima and Takashi Kobayashi and Giordano Scappucci and Hsi-Sheng Goan and Seigo Tarucha},
      year={2025},
      eprint={2507.11918},
      archivePrefix={arXiv},
      primaryClass={quant-ph},
      url={https://arxiv.org/abs/2507.11918}, 
}

@misc{Jones2025,
      title={Mid-circuit logic executed in the qubit layer of a quantum processor}, 
      author={Cameron Jones and Piper Wysocki and MengKe Feng and Gerardo A. Paz-Silva and Corey I. Ostrove and Tuomo Tanttu and Kenneth M. Rudinger and Samuel K. Bartee and Kevin Young and Fay E. Hudson and Wee Han Lim and Nikolay V. Abrosimov and Hans-Joachim Pohl and Michael L. W. Thewalt and Robin Blume-Kohout and Andrew S. Dzurak and Andre Saraiva and Arne Laucht and Chih Hwan Yang},
      year={2025},
      eprint={2512.12648},
      archivePrefix={arXiv},
      primaryClass={quant-ph},
      url={https://arxiv.org/abs/2512.12648v1}, 
}

@article{ion_trap_review,
    author = {Bruzewicz, Colin D. and Chiaverini, John and McConnell, Robert and Sage, Jeremy M.},
    title = {Trapped-ion quantum computing: Progress and challenges},
    journal = {Appl. Phys. Rev.},
    year = {2019},
    url = {https://doi.org/10.1063/1.5088164},
    volume = {6},
    issue = {2},
    pages = {021314},
    doi = {10.1063/1.5088164}
}

@article{PhysRevX.13.041052,
  title = {A Race-Track Trapped-Ion Quantum Processor},
  author = {Moses, S. A. and Baldwin, C. H. and Allman, M. S. and Ancona, R. and Ascarrunz, L. and Barnes, C. and Bartolotta, J. and Bjork, B. and Blanchard, P. and Bohn, M. and Bohnet, J. G. and Brown, N. C. and Burdick, N. Q. and Burton, W. C. and Campbell, S. L. and Campora, J. P. and Carron, C. and Chambers, J. and Chan, J. W. and Chen, Y. H. and Chernoguzov, A. and Chertkov, E. and Colina, J. and Curtis, J. P. and Daniel, R. and DeCross, M. and Deen, D. and Delaney, C. and Dreiling, J. M. and Ertsgaard, C. T. and Esposito, J. and Estey, B. and Fabrikant, M. and Figgatt, C. and Foltz, C. and Foss-Feig, M. and Francois, D. and Gaebler, J. P. and Gatterman, T. M. and Gilbreth, C. N. and Giles, J. and Glynn, E. and Hall, A. and Hankin, A. M. and Hansen, A. and Hayes, D. and Higashi, B. and Hoffman, I. M. and Horning, B. and Hout, J. J. and Jacobs, R. and Johansen, J. and Jones, L. and Karcz, J. and Klein, T. and Lauria, P. and Lee, P. and Liefer, D. and Lu, S. T. and Lucchetti, D. and Lytle, C. and Malm, A. and Matheny, M. and Mathewson, B. and Mayer, K. and Miller, D. B. and Mills, M. and Neyenhuis, B. and Nugent, L. and Olson, S. and Parks, J. and Price, G. N. and Price, Z. and Pugh, M. and Ransford, A. and Reed, A. P. and Roman, C. and Rowe, M. and Ryan-Anderson, C. and Sanders, S. and Sedlacek, J. and Shevchuk, P. and Siegfried, P. and Skripka, T. and Spaun, B. and Sprenkle, R. T. and Stutz, R. P. and Swallows, M. and Tobey, R. I. and Tran, A. and Tran, T. and Vogt, E. and Volin, C. and Walker, J. and Zolot, A. M. and Pino, J. M.},
  journal = {Phys. Rev. X},
  volume = {13},
  issue = {4},
  pages = {041052},
  numpages = {25},
  year = {2023},
  month = {Dec},
  publisher = {American Physical Society},
  doi = {10.1103/PhysRevX.13.041052},
  url = {https://link.aps.org/doi/10.1103/PhysRevX.13.041052}
}

@article{PRXQuantum.4.040313,
  title = {How to Wire a $1000$-Qubit Trapped-Ion Quantum Computer},
  author = {Malinowski, M. and Allcock, D.T.C. and Ballance, C.J.},
  journal = {PRX Quantum},
  volume = {4},
  issue = {4},
  pages = {040313},
  numpages = {21},
  year = {2023},
  month = {Oct},
  publisher = {American Physical Society},
  doi = {10.1103/PRXQuantum.4.040313},
  url = {https://link.aps.org/doi/10.1103/PRXQuantum.4.040313}
}

@article{Lekitsch2017,
author = {Bjoern Lekitsch  and Sebastian Weidt  and Austin G. Fowler  and Klaus Mølmer  and Simon J. Devitt  and Christof Wunderlich  and Winfried K. Hensinger },
title = {Blueprint for a microwave trapped ion quantum computer},
journal = {Sci. Adv.},
volume = {3},
number = {2},
pages = {e1601540},
year = {2017},
doi = {10.1126/sciadv.1601540},
URL = {https://www.science.org/doi/abs/10.1126/sciadv.1601540},
eprint = {https://www.science.org/doi/pdf/10.1126/sciadv.1601540},
}

@article{PhysRevLett.117.220501,
  title = {Trapped-Ion Quantum Logic with Global Radiation Fields},
  author = {Weidt, S. and Randall, J. and Webster, S. C. and Lake, K. and Webb, A. E. and Cohen, I. and Navickas, T. and Lekitsch, B. and Retzker, A. and Hensinger, W. K.},
  journal = {Phys. Rev. Lett.},
  volume = {117},
  issue = {22},
  pages = {220501},
  numpages = {6},
  year = {2016},
  month = {Nov},
  publisher = {American Physical Society},
  doi = {10.1103/PhysRevLett.117.220501},
  url = {https://link.aps.org/doi/10.1103/PhysRevLett.117.220501}
}

@article{Chuang01111997,
author = {Isaac L. Chuang and M. A. Nielsen},
title = {Prescription for experimental determination of the dynamics of a quantum black box},
journal = {J. Mod. Opt.},
volume = {44},
number = {11-12},
pages = {2455--2467},
year = {1997},
publisher = {Taylor \& Francis},
doi = {10.1080/09500349708231894},
URL = {https://www.tandfonline.com/doi/abs/10.1080/09500349708231894},
}

@article{PhysRevLett.78.390,
  title = {Complete Characterization of a Quantum Process: The Two-Bit Quantum Gate},
  author = {Poyatos, J. F. and Cirac, J. I. and Zoller, P.},
  journal = {Phys. Rev. Lett.},
  volume = {78},
  issue = {2},
  pages = {390--393},
  numpages = {0},
  year = {1997},
  month = {Jan},
  publisher = {American Physical Society},
  doi = {10.1103/PhysRevLett.78.390},
  url = {https://link.aps.org/doi/10.1103/PhysRevLett.78.390}
}

@article{PhysRevLett.93.080502,
  title = {Quantum Process Tomography of a Controlled-NOT Gate},
  author = {O'Brien, J. L. and Pryde, G. J. and Gilchrist, A. and James, D. F. V. and Langford, N. K. and Ralph, T. C. and White, A. G.},
  journal = {Phys. Rev. Lett.},
  volume = {93},
  issue = {8},
  pages = {080502},
  numpages = {4},
  year = {2004},
  month = {Aug},
  publisher = {American Physical Society},
  doi = {10.1103/PhysRevLett.93.080502},
  url = {https://link.aps.org/doi/10.1103/PhysRevLett.93.080502}
}

@book{NielsenChuang,
    author = {M. A. Nielsen and I. L. Chuang},
    title = {Quantum Computation and Quantum Information},
    publisher = {Cambrige University Press},
    address = {Cambridge},
    year = {2000}
}

@article{PhysRevX.9.041053,
  title = {Repetition Cat Qubits for Fault-Tolerant Quantum Computation},
  author = {Guillaud, J\'er\'emie and Mirrahimi, Mazyar},
  journal = {Phys. Rev. X},
  volume = {9},
  issue = {4},
  pages = {041053},
  numpages = {23},
  year = {2019},
  month = {Dec},
  publisher = {American Physical Society},
  doi = {10.1103/PhysRevX.9.041053},
  url = {https://link.aps.org/doi/10.1103/PhysRevX.9.041053}
}

@article{mirrahimi2014dynamically,
  title={Dynamically protected cat-qubits: a new paradigm for universal quantum computation},
  author={Mirrahimi, Mazyar and Leghtas, Zaki and Albert, Victor V and Touzard, Steven and Schoelkopf, Robert J and Jiang, Liang and Devoret, Michel H},
  journal={New Journal of Physics},
  volume={16},
  number={4},
  pages={045014},
  year={2014},
  publisher={IOP Publishing},
  doi = {10.1088/1367-2630/16/4/045014},
  url = {https://dx.doi.org/10.1088/1367-2630/16/4/045014}
}

@article{chamberland2022building,
  title = {Building a Fault-Tolerant Quantum Computer Using Concatenated Cat Codes},
  author = {Chamberland, Christopher and Noh, Kyungjoo and Arrangoiz-Arriola, Patricio and Campbell, Earl T. and Hann, Connor T. and Iverson, Joseph and Putterman, Harald and Bohdanowicz, Thomas C. and Flammia, Steven T. and Keller, Andrew and Refael, Gil and Preskill, John and Jiang, Liang and Safavi-Naeini, Amir H. and Painter, Oskar and Brand\~ao, Fernando G.S.L.},
  journal = {PRX Quantum},
  volume = {3},
  issue = {1},
  pages = {010329},
  numpages = {117},
  year = {2022},
  month = {Feb},
  publisher = {American Physical Society},
  doi = {10.1103/PRXQuantum.3.010329},
  url = {https://link.aps.org/doi/10.1103/PRXQuantum.3.010329}
}

@article{puri2020bias,
  title={Bias-preserving gates with stabilized cat qubits},
  author={Puri, Shruti and St-Jean, Lucas and Gross, Jonathan A and Grimm, Alexander and Frattini, Nicholas E and Iyer, Pavithran S and Krishna, Anirudh and Touzard, Steven and Jiang, Liang and Blais, Alexandre and Flammia, Steven T and Girvin, S M},
  journal={Science advances},
  volume={6},
  number={34},
  pages={eaay5901},
  year={2020},
  publisher={American Association for the Advancement of Science},
  doi={10.1126/sciadv.aay590},
  url={https://www.science.org/doi/10.1126/sciadv.aay5901}
}

@article{putterman2025hardware,
  title={Hardware-efficient quantum error correction via concatenated bosonic qubits},
  author={Putterman, Harald and Noh, Kyungjoo and Hann, Connor T and MacCabe, Gregory S and Aghaeimeibodi, Shahriar and Patel, Rishi N and Lee, Menyoung and Jones, William M and Moradinejad, Hesam and Rodriguez, Roberto and others},
  journal={Nature},
  volume={638},
  number={8052},
  pages={927--934},
  year={2025},
  publisher={Nature Publishing Group UK London},
  doi={10.1038/s41586-025-08642-7},
  url={https://www.nature.com/articles/s41586-025-08642-7#citeas}
}

@article{ruiz2025ldpc,
  title={LDPC-cat codes for low-overhead quantum computing in 2D},
  author={Ruiz, Diego and Guillaud, J{\'e}r{\'e}mie and Leverrier, Anthony and Mirrahimi, Mazyar and Vuillot, Christophe},
  journal={Nature Communications},
  volume={16},
  number={1},
  pages={1040},
  year={2025},
  publisher={Nature Publishing Group UK London},
  url={https://www.nature.com/articles/s41467-025-56298-8}
}

@article{Messinger2024,
  title = {Fault-tolerant quantum computing with the parity code and biased-noise qubits},
  author = {Messinger, Anette and Torggler, Valentin and Klaver, Berend and Fellner, Michael and Lechner, Wolfgang},
  journal = {Phys. Rev. Appl.},
  volume = {23},
  issue = {4},
  pages = {044032},
  numpages = {11},
  year = {2025},
  month = {Apr},
  publisher = {American Physical Society},
  doi = {10.1103/PhysRevApplied.23.044032},
  url = {https://link.aps.org/doi/10.1103/PhysRevApplied.23.044032}
}

@article{PhysRevA.52.R2493,
  title = {Scheme for reducing decoherence in quantum computer memory},
  author = {Shor, Peter W.},
  journal = {Phys. Rev. A},
  volume = {52},
  issue = {4},
  pages = {R2493--R2496},
  numpages = {0},
  year = {1995},
  month = {Oct},
  publisher = {American Physical Society},
  doi = {10.1103/PhysRevA.52.R2493},
  url = {https://link.aps.org/doi/10.1103/PhysRevA.52.R2493}
}

@article{PhysRevA.54.1098,
  title = {Good quantum error-correcting codes exist},
  author = {Calderbank, A. R. and Shor, Peter W.},
  journal = {Phys. Rev. A},
  volume = {54},
  issue = {2},
  pages = {1098--1105},
  numpages = {0},
  year = {1996},
  month = {Aug},
  publisher = {American Physical Society},
  doi = {10.1103/PhysRevA.54.1098},
  url = {https://link.aps.org/doi/10.1103/PhysRevA.54.1098}
}

@article{PhysRevLett.77.793,
  title = {Error Correcting Codes in Quantum Theory},
  author = {Steane, A. M.},
  journal = {Phys. Rev. Lett.},
  volume = {77},
  issue = {5},
  pages = {793--797},
  numpages = {0},
  year = {1996},
  month = {Jul},
  publisher = {American Physical Society},
  doi = {10.1103/PhysRevLett.77.793},
  url = {https://link.aps.org/doi/10.1103/PhysRevLett.77.793}
}

@article{KITAEV20032,
title = {Fault-tolerant quantum computation by anyons},
journal = {Ann. Phys},
volume = {303},
number = {1},
pages = {2-30},
year = {2003},
issn = {0003-4916},
doi = {https://doi.org/10.1016/S0003-4916(02)00018-0},
url = {https://www.sciencedirect.com/science/article/pii/S0003491602000180},
author = {A.Yu. Kitaev},
}

@article{PhysRevA.86.032324,
  title = {Surface codes: Towards practical large-scale quantum computation},
  author = {Fowler, Austin G. and Mariantoni, Matteo and Martinis, John M. and Cleland, Andrew N.},
  journal = {Phys. Rev. A},
  volume = {86},
  issue = {3},
  pages = {032324},
  numpages = {48},
  year = {2012},
  month = {Sep},
  publisher = {American Physical Society},
  doi = {10.1103/PhysRevA.86.032324},
  url = {https://link.aps.org/doi/10.1103/PhysRevA.86.032324}
}

@article{Bravyi2024,
	author = {Bravyi, Sergey and Cross, Andrew W. and Gambetta, Jay M. and Maslov, Dmitri and Rall, Patrick and Yoder, Theodore J.},
	doi = {10.1038/s41586-024-07107-7},
	journal = {Nature},
	number = {8005},
	pages = {778--782},
	title = {High-threshold and low-overhead fault-tolerant quantum memory},
	url = {https://doi.org/10.1038/s41586-024-07107-7},
	volume = {627},
	year = {2024},
}

@misc{Jayashankar2022,
      title={Quantum Error Correction: Noise-adapted Techniques and Applications}, 
      author={Akshaya Jayashankar and Prabha Mandayam},
      year={2024},
      eprint={2208.00365},
      archivePrefix={arXiv},
      primaryClass={quant-ph},
      url={https://arxiv.org/abs/2208.00365}, 
}

@misc{Wu2025,
      title={Bias-tailored single-shot quantum LDPC codes}, 
      author={Shixin Wu and Todd A. Brun and Daniel A. Lidar},
      year={2025},
      eprint={2507.02239},
      archivePrefix={arXiv},
      primaryClass={quant-ph},
      url={https://arxiv.org/abs/2507.02239}, 
}

@article{PRXQuantum.4.030338,
  title = {Tailoring Three-Dimensional Topological Codes for Biased Noise},
  author = {Huang, Eric and Pesah, Arthur and Chubb, Christopher T. and Vasmer, Michael and Dua, Arpit},
  journal = {PRX Quantum},
  volume = {4},
  issue = {3},
  pages = {030338},
  numpages = {35},
  year = {2023},
  month = {Sep},
  publisher = {American Physical Society},
  doi = {10.1103/PRXQuantum.4.030338},
  url = {https://link.aps.org/doi/10.1103/PRXQuantum.4.030338}
}

@article{Tiurev2023correctingnon,
  doi = {10.22331/q-2023-09-26-1123},
  url = {https://doi.org/10.22331/q-2023-09-26-1123},
  title = {Correcting non-independent and non-identically distributed errors with surface codes},
  author = {Tiurev, Konstantin and Derks, Peter-Jan H. S. and Roffe, Joschka and Eisert, Jens and Reiner, Jan-Michael},
  journal = {{Quantum}},
  issn = {2521-327X},
  publisher = {{Verein zur F{\"{o}}rderung des Open Access Publizierens in den Quantenwissenschaften}},
  volume = {7},
  pages = {1123},
  month = sep,
  year = {2023}
}

@article{Roffe2023biastailoredquantum,
  doi = {10.22331/q-2023-05-15-1005},
  url = {https://doi.org/10.22331/q-2023-05-15-1005},
  title = {Bias-tailored quantum {LDPC} codes},
  author = {Roffe, Joschka and Cohen, Lawrence Z. and Quintavalle, Armanda O. and Chandra, Daryus and Campbell, Earl T.},
  journal = {{Quantum}},
  issn = {2521-327X},
  publisher = {{Verein zur F{\"{o}}rderung des Open Access Publizierens in den Quantenwissenschaften}},
  volume = {7},
  pages = {1005},
  month = may,
  year = {2023}
}

@article{BonillaAtaides2021,
	author = {Bonilla Ataides, J. Pablo and Tuckett, David K. and Bartlett, Stephen D. and Flammia, Steven T. and Brown, Benjamin J.},
	date = {2021/04/12},
	doi = {10.1038/s41467-021-22274-1},
	isbn = {2041-1723},
	journal = {Nature Communications},
	number = {1},
	pages = {2172},
	title = {The XZZX surface code},
	url = {https://doi.org/10.1038/s41467-021-22274-1},
	volume = {12},
	year = {2021},
}

@article{PhysRevA.78.052331,
  title = {Fault-tolerant quantum computation against biased noise},
  author = {Aliferis, Panos and Preskill, John},
  journal = {Phys. Rev. A},
  volume = {78},
  issue = {5},
  pages = {052331},
  numpages = {9},
  year = {2008},
  month = {Nov},
  publisher = {American Physical Society},
  doi = {10.1103/PhysRevA.78.052331},
  url = {https://link.aps.org/doi/10.1103/PhysRevA.78.052331}
}

@article{PhysRevA.92.062309,
  title = {Reducing the overhead for quantum computation when noise is biased},
  author = {Webster, Paul and Bartlett, Stephen D. and Poulin, David},
  journal = {Phys. Rev. A},
  volume = {92},
  issue = {6},
  pages = {062309},
  numpages = {8},
  year = {2015},
  month = {Dec},
  publisher = {American Physical Society},
  doi = {10.1103/PhysRevA.92.062309},
  url = {https://link.aps.org/doi/10.1103/PhysRevA.92.062309}
}

@article{PhysRevA.71.022316,
  title = {Universal quantum computation with ideal Clifford gates and noisy ancillas},
  author = {Bravyi, Sergey and Kitaev, Alexei},
  journal = {Phys. Rev. A},
  volume = {71},
  issue = {2},
  pages = {022316},
  numpages = {14},
  year = {2005},
  month = {Feb},
  publisher = {American Physical Society},
  doi = {10.1103/PhysRevA.71.022316},
  url = {https://link.aps.org/doi/10.1103/PhysRevA.71.022316}
}

@article{Gidney2019efficientmagicstate,
  doi = {10.22331/q-2019-04-30-135},
  url = {https://doi.org/10.22331/q-2019-04-30-135},
  title = {Efficient magic state factories with a catalyzed {$|CCZ\rangle$} to {$2|T\rangle$} transformation},
  author = {Gidney, Craig and Fowler, Austin G.},
  journal = {{Quantum}},
  issn = {2521-327X},
  publisher = {{Verein zur F{\"{o}}rderung des Open Access Publizierens in den Quantenwissenschaften}},
  volume = {3},
  pages = {135},
  month = apr,
  year = {2019}
}

@article{Preskill2018quantumcomputingin,
  doi = {10.22331/q-2018-08-06-79},
  url = {https://doi.org/10.22331/q-2018-08-06-79},
  title = {Quantum {C}omputing in the {NISQ} era and beyond},
  author = {Preskill, John},
  journal = {{Quantum}},
  issn = {2521-327X},
  publisher = {{Verein zur F{\"{o}}rderung des Open Access Publizierens in den Quantenwissenschaften}},
  volume = {2},
  pages = {79},
  month = aug,
  year = {2018}
}

@article{Preskill1998,
  doi = {10.1098/rspa.1998.0167},
  url = {http://doi.org/10.1098/rspa.1998.0167},
  title = {Reliable quantum computers},
  author = {Preskill, John},
  journal = {Proc. R. Soc. Lond. A.},
  volume = {454},
  pages = {385-410},
  year = {1998}
}

@article{PhysRevApplied.20.054022,
  title = {Blueprint for quantum computing using electrons on helium},
  author = {Kawakami, Erika and Chen, Jiabao and Benito, M\'onica and Konstantinov, Denis},
  journal = {Phys. Rev. Appl.},
  volume = {20},
  issue = {5},
  pages = {054022},
  numpages = {26},
  year = {2023},
  month = {Nov},
  publisher = {American Physical Society},
  doi = {10.1103/PhysRevApplied.20.054022},
  url = {https://link.aps.org/doi/10.1103/PhysRevApplied.20.054022}
}

@article{Jennings2024,
  title = {Quantum computing using floating electrons on cryogenic substrates: Potential and challenges},
  author = {Jennings, A. and Zhou, X. and Grytsenko, I. and Kawakami, E.},
  journal = {Appl. Phys. Lett.},
  volume = {124},
  issue = {12},
  pages = {120501},
  year = {2024},
  doi = {10.1063/5.0179700},
  url = {https://doi.org/10.1063/5.0179700}
}

@article{Stricker2020,
	author = {Stricker, Roman and Vodola, Davide and Erhard, Alexander and Postler, Lukas and Meth, Michael and Ringbauer, Martin and Schindler, Philipp and Monz, Thomas and M{\"u}ller, Markus and Blatt, Rainer},
	date = {2020/09/01},
	doi = {10.1038/s41586-020-2667-0},
	isbn = {1476-4687},
	journal = {Nature},
	number = {7824},
	pages = {207--210},
	title = {Experimental deterministic correction of qubit loss},
	url = {https://doi.org/10.1038/s41586-020-2667-0},
	volume = {585},
	year = {2020},
}

@misc{Huie2025,
      title={Three-qubit encoding in ytterbium-171 atoms for simulating 1+1{D QCD}}, 
      author={William Huie and Cianan Conefrey-Shinozaki and Zhubing Jia and Patrick Draper and and Jacob P. Covey},
      year={2025},
      eprint={2507.18426},
      archivePrefix={arXiv},
      primaryClass={quant-ph},
      url={https://arxiv.org/abs/2507.18426}, 
}

@article{PhysRevX.13.041013,
  title = {High-Threshold Codes for Neutral-Atom Qubits with Biased Erasure Errors},
  author = {Sahay, Kaavya and Jin, Junlan and Claes, Jahan and Thompson, Jeff D. and Puri, Shruti},
  journal = {Phys. Rev. X},
  volume = {13},
  issue = {4},
  pages = {041013},
  numpages = {12},
  year = {2023},
  month = {Oct},
  publisher = {American Physical Society},
  doi = {10.1103/PhysRevX.13.041013},
  url = {https://link.aps.org/doi/10.1103/PhysRevX.13.041013}
}

@article{PhysRevX.12.021049,
  title = {Hardware-Efficient, Fault-Tolerant Quantum Computation with {R}ydberg Atoms},
  author = {Cong, Iris and Levine, Harry and Keesling, Alexander and Bluvstein, Dolev and Wang, Sheng-Tao and Lukin, Mikhail D.},
  journal = {Phys. Rev. X},
  volume = {12},
  issue = {2},
  pages = {021049},
  numpages = {31},
  year = {2022},
  month = {Jun},
  publisher = {American Physical Society},
  doi = {10.1103/PhysRevX.12.021049},
  url = {https://link.aps.org/doi/10.1103/PhysRevX.12.021049}
}

@article{Zache2023,
  doi = {10.22331/q-2023-10-16-1140},
  url = {https://doi.org/10.22331/q-2023-10-16-1140},
  title = {Fermion-qudit quantum processors for simulating lattice gauge theories with matter},
  author = {Zache, Torsten V. and Gonz{\'{a}}lez-Cuadra, Daniel and Zoller, Peter},
  journal = {{Quantum}},
  issn = {2521-327X},
  publisher = {{Verein zur F{\"{o}}rderung des Open Access Publizierens in den Quantenwissenschaften}},
  volume = {7},
  pages = {1140},
  month = oct,
  year = {2023}
}

@article{Gonzalez2023,
  title={Fermionic quantum processing with programmable neutral atom arrays},
  author = {D. Gonz{\'a}lez-Cuadra  and D. Bluvstein  and M. Kalinowski  and R. Kaubruegger  and N. Maskara  and P. Naldesi  and T. V. Zache  and A. M. Kaufman  and M. D. Lukin  and H. Pichler  and B. Vermersch  and Jun Ye  and P. Zoller },
  journal = {Proc. Natl. Acad. Sci.},
  volume = {120},
  number = {35},
  pages = {e2304294120},
  year = {2023},
  doi = {10.1073/pnas.2304294120},
  URL = {https://www.pnas.org/doi/abs/10.1073/pnas.2304294120},
}

@article{Gonzalez2022,
  title = {Hardware Efficient Quantum Simulation of Non-Abelian Gauge Theories with Qudits on Rydberg Platforms},
  author = {Gonz\'alez-Cuadra, Daniel and Zache, Torsten V. and Carrasco, Jose and Kraus, Barbara and Zoller, Peter},
  journal = {Phys. Rev. Lett.},
  volume = {129},
  issue = {16},
  pages = {160501},
  numpages = {8},
  year = {2022},
  month = {Oct},
  publisher = {American Physical Society},
  doi = {10.1103/PhysRevLett.129.160501},
  url = {https://link.aps.org/doi/10.1103/PhysRevLett.129.160501}
}

@article{Sharma2021,
  title = {Driven-dissipative control of cold atoms in tilted optical lattices},
  author = {Sharma, Vaibhav and Mueller, Erich J.},
  journal = {Phys. Rev. A},
  volume = {103},
  issue = {4},
  pages = {043322},
  numpages = {10},
  year = {2021},
  month = {Apr},
  publisher = {American Physical Society},
  doi = {10.1103/PhysRevA.103.043322},
  url = {https://link.aps.org/doi/10.1103/PhysRevA.103.043322}
}

@article{Barnes2022,
  title={Assembly and coherent control of a register of nuclear spin qubits},
  author={Barnes, Katrina and Battaglino, Peter and Bloom, Benjamin J and Cassella, Kayleigh and Coxe, Robin and Crisosto, Nicole and King, Jonathan P and Kondov, Stanimir S and Kotru, Krish and Larsen, Stuart C and others},
  journal={Nat. Commun.},
  volume={13},
  number={1},
  pages={2779},
  year={2022},
  url={https://www.nature.com/articles/s41467-022-29977-z},
  doi={10.1038/s41467-022-29977-z},
  publisher={Nature Publishing Group UK London}
}

@article{Cooper2018,
  title = {Alkaline-Earth Atoms in Optical Tweezers},
  author = {Cooper, Alexandre and Covey, Jacob P. and Madjarov, Ivaylo S. and Porsev, Sergey G. and Safronova, Marianna S. and Endres, Manuel},
  journal = {Phys. Rev. X},
  volume = {8},
  issue = {4},
  pages = {041055},
  numpages = {19},
  year = {2018},
  month = {Dec},
  publisher = {American Physical Society},
  doi = {10.1103/PhysRevX.8.041055},
  url = {https://link.aps.org/doi/10.1103/PhysRevX.8.041055}
}

@article{Lis2023,
  title = {Midcircuit Operations Using the omg Architecture in Neutral Atom Arrays},
  author = {Lis, Joanna W. and Senoo, Aruku and McGrew, William F. and R\"onchen, Felix and Jenkins, Alec and Kaufman, Adam M.},
  journal = {Phys. Rev. X},
  volume = {13},
  issue = {4},
  pages = {041035},
  numpages = {22},
  year = {2023},
  month = {Nov},
  publisher = {American Physical Society},
  doi = {10.1103/PhysRevX.13.041035},
  url = {https://link.aps.org/doi/10.1103/PhysRevX.13.041035}
}

@article{Daley2011,
  title={Quantum computing and quantum simulation with {group-II} atoms},
  author={Daley, Andrew J},
  journal={Quant. Inf. Process},
  volume={10},
  pages={865--884},
  year={2011},
  publisher={Springer},
  doi={10.1007/s11128-011-0293-3},
  url={https://doi.org/10.1007/s11128-011-0293-3},
}

@article{Jenkins2022,
  title = {Ytterbium Nuclear-Spin Qubits in an Optical Tweezer Array},
  author = {Jenkins, Alec and Lis, Joanna W. and Senoo, Aruku and McGrew, William F. and Kaufman, Adam M.},
  journal = {Phys. Rev. X},
  volume = {12},
  issue = {2},
  pages = {021027},
  numpages = {21},
  year = {2022},
  month = {May},
  publisher = {American Physical Society},
  doi = {10.1103/PhysRevX.12.021027},
  url = {https://link.aps.org/doi/10.1103/PhysRevX.12.021027}
}

@article{Ma2022,
  title = {Universal Gate Operations on Nuclear Spin Qubits in an Optical Tweezer Array of $^{171}\mathrm{Yb}$ Atoms},
  author = {Ma, Shuo and Burgers, Alex P. and Liu, Genyue and Wilson, Jack and Zhang, Bichen and Thompson, Jeff D.},
  journal = {Phys. Rev. X},
  volume = {12},
  issue = {2},
  pages = {021028},
  numpages = {12},
  year = {2022},
  month = {May},
  publisher = {American Physical Society},
  doi = {10.1103/PhysRevX.12.021028},
  url = {https://link.aps.org/doi/10.1103/PhysRevX.12.021028}
}

@article{Chow2024,
  title = {Circuit-Based Leakage-to-Erasure Conversion in a Neutral-Atom Quantum Processor},
  author = {Chow, Matthew N. H. and Buchemmavari, Vikas and Omanakuttan, Sivaprasad and Little, Bethany J. and Pandey, Saurabh and Deutsch, Ivan H. and Jau, Yuan-Yu},
  journal = {PRX Quantum},
  volume = {5},
  issue = {4},
  pages = {040343},
  numpages = {14},
  year = {2024},
  month = {Dec},
  publisher = {American Physical Society},
  doi = {10.1103/PRXQuantum.5.040343},
  url = {https://link.aps.org/doi/10.1103/PRXQuantum.5.040343}
}

@article{Ma2023,
  title={High-fidelity gates and mid-circuit erasure conversion in an atomic qubit},
  author={Ma, Shuo and Liu, Genyue and Peng, Pai and Zhang, Bichen and Jandura, Sven and Claes, Jahan and Burgers, Alex P and Pupillo, Guido and Puri, Shruti and Thompson, Jeff D},
  journal={Nature},
  volume={622},
  number={7982},
  pages={279--284},
  year={2023},
  publisher={Nature Publishing Group UK London},
  url={https://doi.org/10.1038/s41586-023-06438-1},
  doi={10.1038/s41586-023-06438-1},
}

@article{Scholl2023,
  title={Erasure conversion in a high-fidelity {R}ydberg quantum simulator},
  author={Scholl, Pascal and Shaw, Adam L and Tsai, Richard Bing-Shiun and Finkelstein, Ran and Choi, Joonhee and Endres, Manuel},
  journal={Nature},
  volume={622},
  number={7982},
  pages={273--278},
  year={2023},
  publisher={Nature Publishing Group UK London},
  url={https://doi.org/10.1038/s41586-023-06516-4},
  doi={10.1038/s41586-023-06516-4},
}

@article{Norica2023,
  title = {Midcircuit Qubit Measurement and Rearrangement in a $^{171}\mathrm{Yb}$ Atomic Array},
  author = {Norcia, M. A. and Cairncross, W. B. and Barnes, K. and Battaglino, P. and Brown, A. and Brown, M. O. and Cassella, K. and Chen, C.-A. and Coxe, R. and Crow, D. and Epstein, J. and Griger, C. and Jones, A. M. W. and Kim, H. and Kindem, J. M. and King, J. and Kondov, S. S. and Kotru, K. and Lauigan, J. and Li, M. and Lu, M. and Megidish, E. and Marjanovic, J. and McDonald, M. and Mittiga, T. and Muniz, J. A. and Narayanaswami, S. and Nishiguchi, C. and Notermans, R. and Paule, T. and Pawlak, K. A. and Peng, L. S. and Ryou, A. and Smull, A. and Stack, D. and Stone, M. and Sucich, A. and Urbanek, M. and van de Veerdonk, R. J. M. and Vendeiro, Z. and Wilkason, T. and Wu, T.-Y. and Xie, X. and Zhang, X. and Bloom, B. J.},
  journal = {Phys. Rev. X},
  volume = {13},
  issue = {4},
  pages = {041034},
  numpages = {12},
  year = {2023},
  month = {Nov},
  publisher = {American Physical Society},
  doi = {10.1103/PhysRevX.13.041034},
  url = {https://link.aps.org/doi/10.1103/PhysRevX.13.041034}
}

@article{Pagano2019,
author = {Pagano, Guido and Scazza, Francesco and Foss-Feig, Michael},
title = {Fast and Scalable Quantum Information Processing with Two-Electron Atoms in Optical Tweezer Arrays},
journal = {Adv. Quant Technol.},
volume = {2},
number = {3-4},
pages = {1800067},
doi = {https://doi.org/10.1002/qute.201800067},
url = {https://advanced.onlinelibrary.wiley.com/doi/abs/10.1002/qute.201800067},
year = {2019}
}

@incollection{Grimm2000,
title = {Optical Dipole Traps for Neutral Atoms},
editor = {Benjamin Bederson and Herbert Walther},
series = {Advances In Atomic, Molecular, and Optical Physics},
publisher = {Academic Press},
volume = {42},
pages = {95-170},
year = {2000},
issn = {1049-250X},
doi = {https://doi.org/10.1016/S1049-250X(08)60186-X},
url = {https://www.sciencedirect.com/science/article/pii/S1049250X0860186X},
author = {Rudolf Grimm and Matthias Weidemüller and Yurii B. Ovchinnikov},
}

@article{Heinz2020,
  title = {State-Dependent Optical Lattices for the Strontium Optical Qubit},
  author = {Heinz, A. and Park, A. J. and \ifmmode \check{S}\else \v{S}\fi{}anti\ifmmode \acute{c}\else \'{c}\fi{}, N. and Trautmann, J. and Porsev, S. G. and Safronova, M. S. and Bloch, I. and Blatt, S.},
  journal = {Phys. Rev. Lett.},
  volume = {124},
  issue = {20},
  pages = {203201},
  numpages = {7},
  year = {2020},
  month = {May},
  publisher = {American Physical Society},
  doi = {10.1103/PhysRevLett.124.203201},
  url = {https://link.aps.org/doi/10.1103/PhysRevLett.124.203201}
}

@article{Jandura2022,
  title={Time-optimal two-and three-qubit gates for {R}ydberg atoms},
  author={Jandura, Sven and Pupillo, Guido},
  journal={Quantum},
  volume={6},
  pages={712},
  year={2022},
  publisher={Verein zur F{\"o}rderung des Open Access Publizierens in den Quantenwissenschaften},
  url={https://doi.org/10.22331/q-2022-05-13-712},
  doi={10.22331/q-2022-05-13-712},
}

@article{Safronova2015,
  title = {Extracting transition rates from zero-polarizability spectroscopy},
  author = {Safronova, M. S. and Zuhrianda, Z. and Safronova, U. I. and Clark, Charles W.},
  journal = {Phys. Rev. A},
  volume = {92},
  issue = {4},
  pages = {040501},
  numpages = {5},
  year = {2015},
  month = {Oct},
  publisher = {American Physical Society},
  doi = {10.1103/PhysRevA.92.040501},
  url = {https://link.aps.org/doi/10.1103/PhysRevA.92.040501}
}

@article{Pagano2022,
  title = {Error budgeting for a controlled-phase gate with strontium-88 {R}ydberg atoms},
  author = {Pagano, Alice and Weber, Sebastian and Jaschke, Daniel and Pfau, Tilman and Meinert, Florian and Montangero, Simone and B\"uchler, Hans Peter},
  journal = {Phys. Rev. Res.},
  volume = {4},
  issue = {3},
  pages = {033019},
  numpages = {10},
  year = {2022},
  month = {Jul},
  publisher = {American Physical Society},
  doi = {10.1103/PhysRevResearch.4.033019},
  url = {https://link.aps.org/doi/10.1103/PhysRevResearch.4.033019}
}

@article{Kazemi2025,
  title = {Multiqubit parity gates for Rydberg atoms in various configurations},
  author = {Kazemi, Javad and Schuler, Michael and Ertler, Christian and Lechner, Wolfgang},
  journal = {Phys. Rev. Res.},
  volume = {7},
  issue = {3},
  pages = {033269},
  numpages = {16},
  year = {2025},
  month = {Sep},
  publisher = {American Physical Society},
  doi = {10.1103/56qk-rmsz},
  url = {https://link.aps.org/doi/10.1103/56qk-rmsz}
}

@article{Cao2024,
  title={Multi-qubit gates and {S}chr{\"o}dinger cat states in an optical clock},
  author={Cao, Alec and Eckner, William J and Lukin Yelin, Theodor and Young, Aaron W and Jandura, Sven and Yan, Lingfeng and Kim, Kyungtae and Pupillo, Guido and Ye, Jun and Darkwah Oppong, Nelson and others},
  journal={Nature},
  volume={634},
  number={8033},
  pages={315--320},
  year={2024},
  publisher={Nature Publishing Group UK London},
  url={https://doi.org/10.1038/s41586-024-07913-z},
  doi={10.1038/s41586-024-07913-z},
}

@article{Evered2023,
  title={High-fidelity parallel entangling gates on a neutral-atom quantum computer},
  author={Evered, Simon J and Bluvstein, Dolev and Kalinowski, Marcin and Ebadi, Sepehr and Manovitz, Tom and Zhou, Hengyun and Li, Sophie H and Geim, Alexandra A and Wang, Tout T and Maskara, Nishad and others},
  journal={Nature},
  volume={622},
  number={7982},
  pages={268--272},
  year={2023},
  publisher={Nature Publishing Group UK London},
  url={https://doi.org/10.1038/s41586-023-06481-y},
  doi={10.1038/s41586-023-06481-y},
}

@misc{Schuckert2024,
      title={Fermion-qubit fault-tolerant quantum computing}, 
      author={Alexander Schuckert and Eleanor Crane and Alexey V. Gorshkov and Mohammad Hafezi and Michael J. Gullans},
      year={2024},
      eprint={2411.08955},
      archivePrefix={arXiv},
      primaryClass={quant-ph},
      url={https://arxiv.org/abs/2411.08955}, 
}

@article{
Shaw2025,
author = {Adam L. Shaw  and Pascal Scholl  and Ran Finkelstein  and Richard Bing-Shiun Tsai  and Joonhee Choi  and Manuel Endres },
title = {Erasure cooling, control, and hyperentanglement of motion in optical tweezers},
journal = {Science},
volume = {388},
number = {6749},
pages = {845-849},
year = {2025},
doi = {10.1126/science.adn2618},
URL = {https://www.science.org/doi/abs/10.1126/science.adn2618},
}

@article{Morgado2021,
  title={Quantum simulation and computing with {R}ydberg-interacting qubits},
  author={Morgado, M and Whitlock, S},
  journal={AVS Quantum Sci.},
  volume={3},
  number={2},
  year={2021},
  publisher={AIP Publishing},
  doi={10.1116/5.0036562},
  url={https://doi.org/10.1116/5.0036562},
}

@article{Henriet2020,
  title={Quantum computing with neutral atoms},
  author={Henriet, Lo{\"\i}c and Beguin, Lucas and Signoles, Adrien and Lahaye, Thierry and Browaeys, Antoine and Reymond, Georges-Olivier and Jurczak, Christophe},
  journal={Quantum},
  volume={4},
  pages={327},
  year={2020},
  publisher={Verein zur F{\"o}rderung des Open Access Publizierens in den Quantenwissenschaften},
  doi={10.22331/q-2020-09-21-327},
  url={	https://doi.org/10.22331/q-2020-09-21-327},
}

@article{Bluvstein2022,
  title={A quantum processor based on coherent transport of entangled atom arrays},
  author={Bluvstein, Dolev and Levine, Harry and Semeghini, Giulia and Wang, Tout T and Ebadi, Sepehr and Kalinowski, Marcin and Keesling, Alexander and Maskara, Nishad and Pichler, Hannes and Greiner, Markus and others},
  journal={Nature},
  volume={604},
  number={7906},
  pages={451--456},
  year={2022},
  publisher={Nature Publishing Group UK London},
  url={https://doi.org/10.1038/s41586-022-04592-6},
  doi={10.1038/s41586-022-04592-6},
}

@misc{Bluvstein2025,
      title={Architectural mechanisms of a universal fault-tolerant quantum computer}, 
      author={Dolev Bluvstein and Alexandra A. Geim and Sophie H. Li and Simon J. Evered and J. Pablo Bonilla Ataides and Gefen Baranes and Andi Gu and Tom Manovitz and Muqing Xu and Marcin Kalinowski and Shayan Majidy and Christian Kokail and Nishad Maskara and Elias C. Trapp and Luke M. Stewart and Simon Hollerith and Hengyun Zhou and Michael J. Gullans and Susanne F. Yelin and Markus Greiner and Vladan Vuletic and Madelyn Cain and Mikhail D. Lukin},
      year={2025},
      eprint={2506.20661},
      archivePrefix={arXiv},
      primaryClass={quant-ph},
      url={https://arxiv.org/abs/2506.20661}, 
}

@article{Finkelstein2024,
  title={Universal quantum operations and ancilla-based read-out for tweezer clocks},
  author={Finkelstein, Ran and Tsai, Richard Bing-Shiun and Sun, Xiangkai and Scholl, Pascal and Direkci, Su and Gefen, Tuvia and Choi, Joonhee and Shaw, Adam L and Endres, Manuel},
  journal={Nature},
  volume={634},
  number={8033},
  pages={321--327},
  year={2024},
  publisher={Nature Publishing Group UK London},
  doi={10.1038/s41586-024-08005-8},
  url={https://doi.org/10.1038/s41586-024-08005-8},
}

@article{Pagano2024,
  title = {Optimal control transport of neutral atoms in optical tweezers at finite temperature},
  author = {Pagano, Alice and Jaschke, Daniel and Weiss, Werner and Montangero, Simone},
  journal = {Phys. Rev. Res.},
  volume = {6},
  issue = {3},
  pages = {033282},
  numpages = {9},
  year = {2024},
  month = {Sep},
  publisher = {American Physical Society},
  doi = {10.1103/PhysRevResearch.6.033282},
  url = {https://link.aps.org/doi/10.1103/PhysRevResearch.6.033282}
}

@article{Hwang2025,
author = {Sunhwa Hwang and Hansub Hwang and Kangjin Kim and Andrew Byun and Kangheun Kim and Seokho Jeong and Maynardo Pratama Soegianto and Jaewook Ahn},
journal = {Opt. quantum},
number = {1},
pages = {64--71},
publisher = {Optica Publishing Group},
title = {Fast and reliable atom transport by optical tweezers},
volume = {3},
month = {Feb},
year = {2025},
url = {https://opg.optica.org/opticaq/abstract.cfm?URI=opticaq-3-1-64},
doi = {10.1364/OPTICAQ.546797},
}

@article{Wu2019,
  title={Stern--Gerlach detection of neutral-atom qubits in a state-dependent optical lattice},
  author={Wu, Tsung-Yao and Kumar, Aishwarya and Giraldo, Felipe and Weiss, David S},
  journal={Nat. Phys.},
  volume={15},
  number={6},
  pages={538--542},
  year={2019},
  publisher={Nature Publishing Group UK London},
  doi={10.1038/s41567-019-0478-8},
  url={https://doi.org/10.1038/s41567-019-0478-8},
}

@article{
Holland2023,
author = {Connor M. Holland  and Yukai Lu  and Lawrence W. Cheuk },
title = {On-demand entanglement of molecules in a reconfigurable optical tweezer array},
journal = {Science},
volume = {382},
number = {6675},
pages = {1143-1147},
year = {2023},
doi = {10.1126/science.adf4272},
URL = {https://www.science.org/doi/abs/10.1126/science.adf4272},
}

@article{Cairncross2021,
  title = {Assembly of a Rovibrational Ground State Molecule in an Optical Tweezer},
  author = {Cairncross, William B. and Zhang, Jessie T. and Picard, Lewis R. B. and Yu, Yichao and Wang, Kenneth and Ni, Kang-Kuen},
  journal = {Phys. Rev. Lett.},
  volume = {126},
  issue = {12},
  pages = {123402},
  numpages = {6},
  year = {2021},
  month = {Mar},
  publisher = {American Physical Society},
  doi = {10.1103/PhysRevLett.126.123402},
  url = {https://link.aps.org/doi/10.1103/PhysRevLett.126.123402}
}

@article{Langen2024,
  title={Quantum state manipulation and cooling of ultracold molecules},
  author={Langen, Tim and Valtolina, Giacomo and Wang, Dajun and Ye, Jun},
  journal={Nature Physics},
  volume={20},
  number={5},
  pages={702--712},
  year={2024},
  publisher={Nature Publishing Group UK London},
  doi={10.1038/s41567-024-02423-1},
  url={https://doi.org/10.1038/s41567-024-02423-1}
}

@article{Yang2025,
  title={Minute-scale {S}chr{\"o}dinger-cat state of spin-5/2 atoms},
  author={Yang, YA and Luo, W-T and Zhang, J-L and Wang, S-Z and Zou, Chang-Ling and Xia, T and Lu, Z-T},
  journal={Nature Photonics},
  volume={19},
  number={1},
  pages={89--94},
  year={2025},
  url={https://doi.org/10.1038/s41566-024-01555-3},
  doi={10.1038/s41566-024-01555-3},
  publisher={Nature Publishing Group UK London}
}

@misc{Lu2025,
      title={Astigmatism-free 3D Optical Tweezer Control for Rapid Atom Rearrangement}, 
      author={Yue-Hui Lu and Nathan Song and Tai Xiang and Jacquelyn Ho and Tsai-Chen Lee and Zhenjie Yan and Dan M. Stamper-Kurn},
      year={2025},
      eprint={2510.11451},
      archivePrefix={arXiv},
      primaryClass={physics.optics},
      url={https://arxiv.org/abs/2510.11451}, 
}

@misc{Muniz2025,
      title={Repeated ancilla reuse for logical computation on a neutral atom quantum computer}, 
      author={J. A. Muniz and D. Crow and H. Kim and J. M. Kindem and W. B. Cairncross and A. Ryou and T. C. Bohdanowicz and C. -A. Chen and Y. Ji and A. M. W. Jones and E. Megidish and C. Nishiguchi and M. Urbanek and L. Wadleigh and T. Wilkason and D. Aasen and K. Barnes and J. M. Bello-Rivas and I. Bloomfield and G. Booth and A. Brown and M. O. Brown and K. Cassella and G. Cowan and J. Epstein and M. Feldkamp and C. Griger and Y. Hassan and A. Heinz and E. Halperin and T. Hofler and F. Hummel and M. Jaffe and E. Kapit and K. Kotru and J. Lauigan and J. Marjanovic and M. Meredith and M. McDonald and R. Morshead and S. Narayanaswami and K. A. Pawlak and K. L. Pudenz and D. Rodríguez Pérez and P. Sabharwal and J. Simon and A. Smull and M. Sorensen and D. T. Stack and M. Stone and L. Taneja and R. J. M. van de Veerdonk and Z. Vendeiro and R. T. Weverka and K. White and T. -Y. Wu and X. Xie and E. Zalys-Geller and X. Zhang and J. King and B. J. Bloom and M. A. Norcia},
      year={2025},
      eprint={2506.09936},
      archivePrefix={arXiv},
      primaryClass={quant-ph},
      url={https://arxiv.org/abs/2506.09936}, 
}

@misc{Li2025,
      title={Fast, continuous and coherent atom replacement in a neutral atom qubit array}, 
      author={Yiyi Li and Yicheng Bao and Michael Peper and Chenyuan Li and Jeff D. Thompson},
      year={2025},
      eprint={2506.15633},
      archivePrefix={arXiv},
      primaryClass={quant-ph},
      url={https://arxiv.org/abs/2506.15633}, 
}

@article{Chiu2025,
  title={Continuous operation of a coherent 3,000-qubit system},
  author={Chiu, Neng-Chun and Trapp, Elias C and Guo, Jinen and Abobeih, Mohamed H and Stewart, Luke M and Hollerith, Simon and Stroganov, Pavel L and Kalinowski, Marcin and Geim, Alexandra A and Evered, Simon J and others},
  journal={Nature},
  pages={1--3},
  year={2025},
  url={https://doi.org/10.1038/s41586-025-09596-6},
  doi={10.1038/s41586-025-09596-6},
  publisher={Nature Publishing Group UK London}
}

\end{document}